\documentclass[preprint,prd,aps,showpacs,groupedaddress,superscriptaddress]{revtex4-1}
\usepackage{amsmath}
\usepackage{amssymb}
\usepackage{mathtools}
\usepackage{graphicx}
\usepackage{subfig}
\usepackage{bm}
\usepackage{color}
\usepackage{soul}
\usepackage{placeins}
\usepackage{mathrsfs}
\usepackage{hyperref}
\usepackage{slashed}
\usepackage{breqn}
\usepackage{etoolbox}
\usepackage{multirow}
\usepackage{mdframed}
\usepackage{float}

\makeatletter
\preto\maketitle{%
	\begingroup\lccode`~=`,
	\lowercase{\endgroup
		\let\saved@breqn@active@comma~
		\let~}\active@comma 
}
\appto\maketitle{%
	\begingroup\lccode`~=`,
	\lowercase{\endgroup
		\let~}\saved@breqn@active@comma 
}
\makeatother

\begin{document}
	
\title{Domain Wall Fermion QCD with the Exact One Flavor Algorithm}

\author{C.~Jung}
\affiliation{Department of Physics, Brookhaven National Laboratory, Upton, NY 11973, USA}
\author{C.~Kelly}
\author{R.D.~Mawhinney}
\author{D.J.~Murphy}
\affiliation{Department of Physics, Columbia University, New York, NY 10027, USA}


\begin{abstract}
Lattice QCD calculations including the effects of one or more non-degenerate sea quark flavors are conventionally performed using the Rational Hybrid Monte Carlo (RHMC) algorithm, which computes the square root of the determinant of $\mathscr{D}^{\dagger} \mathscr{D}$, where $\mathscr{D}$ is the Dirac operator. The special case of two degenerate quark flavors with the same mass is described directly by the determinant of $\mathscr{D}^{\dagger} \mathscr{D}$ --- in particular, no square root is necessary --- enabling a variety of algorithmic developments, which have driven down the cost of simulating the light (up and down) quarks in the isospin-symmetric limit of equal masses. As a result, the relative cost of single quark flavors --- such as the strange or charm --- computed with RHMC has become more expensive. This problem is even more severe in the context of our measurements of the $\Delta I = 1/2$ $K \rightarrow \pi \pi$ matrix elements on lattice ensembles with $G$-parity boundary conditions, since $G$-parity is associated with a doubling of the number of quark flavors described by $\mathscr{D}$, and thus RHMC is needed for the isospin-symmetric light quarks as well. In this paper we report on our implementation of the exact one flavor algorithm (EOFA) introduced by the TWQCD collaboration for simulations including single flavors of domain wall quarks. We have developed a new preconditioner for the EOFA Dirac equation, which both reduces the cost of solving the Dirac equation and allows us to re-use the bulk of our existing high-performance code. Coupling these improvements with careful tuning of our integrator, the time per accepted trajectory in the production of our 2+1 flavor $G$-parity ensembles with physical pion and kaon masses has been decreased by a factor of 4.2.
\end{abstract}

\maketitle

\section{Introduction}
\label{sec:intro}
Lattice QCD simulations are typically performed using variants of the hybrid Monte Carlo (HMC) algorithm, which includes the effects of dynamical sea quarks through the determinant of a fermion matrix evaluated by stochastically sampling a discretized QCD path integral. Conventional simulations choose the Hermitian fermion matrix $\mathcal{M} = \mathscr{D}^{\dagger} \mathscr{D}$ rather than the lattice Dirac operator $\mathcal{M} = \mathscr{D}$, since the latter, in general, has a complex spectrum, and is thus less amenable to standard numerical algorithms. While $\mathscr{D}$ describes a single quark flavor, $\mathscr{D}^{\dagger} \mathscr{D}$ describes two degenerate quark flavors with the same mass. As a result the standard HMC algorithm naturally describes the light (up and down) quarks in the isospin-symmetric limit $m_{u} = m_{d}$ considered in most lattice calculations. Simulations including single quark flavors (such as the strange or charm) are typically performed by taking an overall square root of the determinant of $\mathcal{M} = \mathscr{D}^{\dagger} \mathscr{D}$, leading to the rational hybrid Monte Carlo (RHMC) algorithm. While RHMC has found widespread usage in the lattice QCD community, RHMC calculations are typically more expensive than HMC calculations for the same input quark mass, in part because many of the techniques which have been developed to accelerate HMC simulations of degenerate quark flavor pairs are not applicable to RHMC. 

A number of recent developments in the HMC algorithm used by the RBC/UKQCD collaboration have driven down the cost of simulating degenerate pairs of isospin-symmetric quarks with the same mass. These developments include: extensive force tuning via Hasenbush mass preconditioning \cite{Hasenbusch:2002ai}, the zM\"{o}bius domain wall fermion action \cite{greg_thesis}, reduced $L_{s}$ approximations to the light quark determinant \cite{greg_thesis}, and the use of implicitly restarted, mixed-precision defect correction methods in the conjugate gradient algorithm \footnote{We elaborate on the details of our defect correction solver in Section \ref{sec:optimizations}.}. In Table \ref{tab:charm_ensemble_timings} we list timings for a recent large-scale calculation which utilizes these techniques.
\begin{table}[!ht]
	\centering
	\begin{tabular}{ccc}
		\hline
		\hline
		\rule{0cm}{0.4cm}\textbf{Action Component} & \multicolumn{2}{c}{\textbf{Timings}} \\
		Gauge & 5970 s & 12.0\% \\
		Light Quarks & 19600 s & 39.4\% \\
		Strange and Charm Quarks & 24200 s & 48.6\% \\
		\hline
		\rule{0cm}{0.4cm}Total & 49770 s & --- \\ 
		\hline
		\hline
	\end{tabular}
	\caption{Timings for one HMC trajectory of RBC/UKQCD's $80^{2} \times 96 \times 192 \times 32$ $N_{f}=2+1+1$ ensemble with physical quark masses and $a^{-1} \approx 3$ GeV on a 12,288-node Blue Gene/Q partition \cite{greg_thesis}.}
	\label{tab:charm_ensemble_timings}
\end{table}
We now find that the single-flavor strange and charm quark determinants, which we simulate using the RHMC algorithm, are collectively the most expensive part of the calculation. To address this, we have turned to exploring TWQCD's recently proposed exact one flavor algorithm (EOFA), which allows for simulating single quark flavors without the need for RHMC \cite{Chen:2014hyy}. Preliminary results have suggested that EOFA simulations can outperform RHMC simulations, both in terms of computer time and a reduced memory footprint, while producing exactly the same physics \cite{Chen:2014bbc,Murphy:2016ywx}.  

The RBC/UKQCD collaboration's ongoing efforts to probe direct $CP$-violation in $K \rightarrow \pi \pi$ decays provide a second motivation for exploring EOFA. The collaboration has recently reported the first calculation of the $\Delta I = 1/2$ $K \rightarrow \pi \pi$ decay amplitude with physical kinematics in Ref.~\cite{Bai:2015nea}, which, when combined with previous results for the $\Delta I = 3/2$ amplitude \cite{Blum:2015ywa} determines the Standard Model $CP$-violating parameters $\epsilon$ and $\epsilon'$ entirely from first principles. An important ingredient in this calculation was the introduction of $G$-parity boundary conditions for the quark fields \cite{Kelly:2013ana,Bai:2015nea}: since the pion is $G$-parity odd, the pion momenta are quantized along $G$-parity directions as
\begin{equation}
p_{\pi}^{i} = \frac{\left(2 n_{i} + 1 \right) \pi}{L}, \quad n_{i} \in \mathbb{Z},
\end{equation}
allowing the ensemble parameters to be tuned such that the $K \rightarrow \pi \pi$ decay has both physical kinematics and the final pions in the ground state. Since the $G$-parity transformation $G = C e^{i \pi I_{y}}$ is the product of charge conjugation and a $180^{\circ}$ isospin rotation about the $y$-axis --- at the lattice boundary the light quark doublet transforms as $(u,d) \mapsto (\overline{d},-\overline{u})$ --- the $G$-parity Dirac operator inherently describes two quark flavors. 
The standard lattice technique for obtaining a Hermitian, positive-definite fermion matrix --- by taking the square of the Dirac operator, $\mathcal{M} = \mathscr{D}^{\dagger} \mathscr{D}$ --- results in a theory with four degenerate quark flavors on a $G$-parity ensemble, and a square root is required to reduce to a two-flavor simulation.
Describing the light quark pair on a $G$-parity ensemble is a particularly attractive target for EOFA, since many of the techniques we use to accelerate the calculation of the light quark determinant for ensembles with periodic boundary conditions --- including defect correction solvers, the forecasted force gradient integrator \cite{Yin:2011np}, and Hasenbusch mass preconditioning --- are not applicable or of limited utility for RHMC simulations, but are expected to perform well in the context of EOFA. More generally, since there is no straightforward way to start the multishift conjugate gradient solver used for RHMC with a nonzero initial guess, techniques which rely on forecasting or restarting the solver are not applicable.

In this work we discuss the RBC/UKQCD collaboration's implementation and tests of the exact one flavor algorithm, as well as the use of EOFA in generating gauge field configurations for our ongoing first-principles calculation of the ratio of Standard Model parameters $\epsilon'/\epsilon$ from $\Delta I = 1/2$ $K \rightarrow \pi \pi$ decays with $G$-parity boundary conditions. We have independently implemented EOFA in the Columbia Physics System (CPS), BAGEL fermion sparse matrix library (BFM), and the Grid data parallel C++ QCD library (Grid), for Shamir and M\"{o}bius domain wall fermions, with periodic, anti-periodic, and $G$-parity boundary conditions. We will demonstrate in the following sections that a significant improvement over the RHMC algorithm in terms of wall clock time is indeed possible with EOFA after introducing a variety of preconditioning and tuning techniques. Early work in this direction was presented at the 34th International Symposium on Lattice Field Theory \cite{Murphy:2016ywx}; here we will elaborate on the details and discuss our first large-scale EOFA calculation. 
\FloatBarrier

\section{The Exact One Flavor Algorithm}
\label{sec:eofa}
The exact one flavor algorithm was developed by the TWQCD collaboration and used to enable efficient simulations of single quark flavors on GPU clusters, where memory usage is a significant constraint. In Ref.~\cite{Ogawa:2009ex} the authors discuss their construction of a positive-definite pseudofermion action describing a single flavor of Wilson or domain wall quark, and elaborate on the details of this construction in Ref.~\cite{Chen:2014hyy}. The key is their observation that a ratio of determinants of domain wall fermion (DWF) Dirac operators can be factorized as
\begin{equation}
\label{eqn:EOFA_factorization}
\det \left( \frac{\mathscr{D}(m_{1})}{\mathscr{D}(m_{2})} \right) = \frac{1}{\det \left( \mathcal{M}_{L} \right)} \cdot \frac{1}{\det \left( \mathcal{M}_{R} \right)},
\end{equation}
with $\mathcal{M}_{L}$ and $\mathcal{M}_{R}$ Hermitian and positive-definite. In a subsequent paper the authors benchmark EOFA against RHMC for $N_{f} = 1$ and $N_{f} = 2+1$ lattice QCD simulations, and demonstrate a number of advantages of the EOFA formalism \cite{Chen:2014bbc}. These include substantial reductions in the pseudofermion force and in the memory footprint of the algorithm, since, in the context of EOFA, inversions of the Dirac operator can be performed using the ordinary conjugate gradient (CG) algorithm rather than the multishift CG used for RHMC. They ultimately find that they are able to generate HMC trajectories 15-20\% faster using EOFA rather than RHMC after retuning their integration scheme to take advantage of these properties. More recently, TWQCD has used EOFA to generate $N_{f} = 2 + 1 + 1$ domain wall fermion ensembles with dynamical strange and charm quarks \cite{Chen:2017kxr}.  

We note that the construction of the exact one flavor pseudofermion action has been detailed by TWQCD in Ref.~\cite{Ogawa:2009ex,Chen:2014hyy} and summarized in our own formalism in Ref.~\cite{Murphy:2016ywx}. We will not repeat this discussion here, other than to give a brief overview and to introduce the notation used in this work. We write the 5D M\"{o}bius domain wall fermion (MDWF) operator $\mathscr{D}_{\rm DWF}$ in terms of the 4D Wilson Dirac operator $D_{W}$ and 5D hopping matrix $L_{ss'}$ as
\begin{equation}
\label{eqn:Ddwf}
\begin{aligned}
\left( \mathscr{D}_{\rm DWF} \right)_{xx',ss'} &= \big( \left( c + d \right) \left( D_{W} \right)_{xx'} + \delta_{xx'} \big) \delta_{ss'} + \big( \left( c - d \right) \left( D_{W} \right)_{xx'} - \delta_{xx'} \big) L_{ss'} \\
\left( D_{W} \right)_{xx'} &= \left( 4 - M_{5} \right) \delta_{xx'} - \frac{1}{2} \sum_{\mu} \left[ \left( 1 - \gamma_{\mu} \right) U_{\mu}(x) \delta_{x+\hat{\mu},x'} + \left( 1 + \gamma_{\mu} \right) U_{\mu}^{\dagger}(x') \delta_{x-\hat{\mu},x'} \right] \\
L_{ss'} &= \left( L_{+} \right)_{ss'} P_{+} + \left( L_{-} \right)_{ss'} P_{-} \\
\end{aligned}
\end{equation}
with
\begin{equation}
\left( L_{+} \right)_{ss'} = \left( L_{-} \right)_{s's} = \begin{dcases} \begin{array}{cc} - m \delta_{L_{s}-1,s'}, & s = 0 \\ \delta_{s-1,s'}, & 1 \leq s \leq L_{s} - 1 \end{array} \end{dcases} .
\end{equation}
Here $x$ and $s$ are spacetime indices in the 4D bulk and along the fifth dimension, respectively, with $L_{s}$ denoting the total number of $s$ sites, $P_{\pm} = ( 1 \pm \gamma_{5} ) / 2$ denoting the chiral projection operators, and $(R_{5})_{ss'} \equiv \delta_{s,Ls-1-s'}$ denoting the operator which performs a reflection in the fifth dimension. We recover four-dimensional quark fields $q$ and $\overline{q}$ with definite chiralities from the five-dimensional quark fields $\psi$ and $\overline{\psi}$ described by $\mathscr{D}_{\rm DWF}$ at the boundaries of the fifth dimension
\begin{equation}
\setlength\arraycolsep{10pt}
\begin{array}{cc}
q_{R} = P_{+} \psi_{L_{s}-1} & q_{L} = P_{-} \psi_{0} \\
\overline{q}_{R} = \overline{\psi}_{L_{s}-1} P_{-} & \overline{q}_{L} = \overline{\psi}_{0} P_{+} \\
\end{array} .
\end{equation}
Green's functions constructed from $q$ and $\overline{q}$ approximate continuum QCD arbitrarily well in the limit of vanishing lattice spacing and infinite 5D spacetime volume. The tunable parameters in Eqn.~\eqref{eqn:Ddwf} are the domain wall parameter $M_{5}$, the bare quark mass $m$, and the M\"{o}bius scale $\alpha = 2 c$; the parameter $d$ is fixed at $d=1/2$. 
DWF with the Shamir kernel is recovered from the more general M\"{o}bius operator in the limit $\alpha \rightarrow 1$. For more detail regarding our MDWF formalism we direct the reader to Ref.~\cite{Blum:2014tka}.

The construction of the exact one flavor action for domain wall fermions begins by factorizing the MDWF Dirac operator as \footnote{We emphasize that the operator we call $\mathscr{D}_{\rm EOFA}$ is merely an alternative notation for TWQCD's rescaled Dirac operator $D_{T}$ \cite{Chen:2014hyy}.}
\begin{equation}
\label{eqn:eofa_dop}
\mathscr{D}_{\rm DWF} = \mathscr{D}_{\rm EOFA} \cdot \widetilde{\mathscr{D}},
\end{equation}
with
\begin{equation}
\label{eqn:eofa_dop_def}
\begin{aligned}
\big( \mathscr{D}_{\rm EOFA} \big)_{xx',ss'} &\equiv \left( D_{W} \right)_{xx'} \delta_{ss'} + \delta_{xx'} \left(M_{+}\right)_{ss'} P_{+} + \delta_{xx'} \left(M_{-}\right)_{ss'} P_{-}  \\ 
\big( \widetilde{\mathscr{D}} \big)_{ss'} &\equiv d \left( \delta_{ss'} - L_{ss'} \right)+ c \left( \delta_{ss'} + L_{ss'} \right)
\end{aligned} .
\end{equation}
The operator $\widetilde{\mathscr{D}}$ relating $\mathscr{D}_{\rm DWF}$ and $\mathscr{D}_{\rm EOFA}$ has no dependence on the gauge field, so we are free to replace $\mathscr{D}_{\rm DWF}$ with $\mathscr{D}_{\rm EOFA}$ in Eqn.~\eqref{eqn:EOFA_factorization} without modifying physical observables described by a properly normalized path integral. In fact, it can be shown analytically using the explicit form of $\widetilde{\mathscr{D}}$ listed in Appendix \ref{appendix:eofa_operators} that
\begin{equation}
\label{eqn:dtilde_det}
\det \big( \widetilde{\mathscr{D}} \big) = \left( \left( c + d \right)^{L_{s}} + m \left( c - d \right)^{L_{s}} \right)^{12 V},
\end{equation}
where $V = L^{3} T$ is the 4D spacetime volume. This substitution facilitates the construction of a proper action since the operator $\gamma_{5} R_{5} \mathscr{D}_{\rm EOFA}$ is manifestly Hermitian for any choice of the M\"{o}bius scale $\alpha$, whereas $\mathscr{D}_{\rm DWF}$ satisfies a less trivial $\gamma_{5}$-Hermiticity condition when $\alpha \ne 1$ \cite{Brower:2004xi}. However, this comes at the cost of substantially more expensive inversions, since $\mathscr{D}_{\rm EOFA}$ is dense in $ss'$ whereas $\mathscr{D}_{\rm DWF}$ has a well-known tridiagonal block structure.

After introducing $\mathscr{D}_{\rm EOFA}$, TWQCD's construction proceeds by applying the Schur identity
\begin{equation}
\det \left[ \left( \begin{array}{cc} A & B \\ C & D \end{array} \right) \right] = \det \left( A \right) \det \left( D - C A^{-1} B \right) = \det \left( D \right) \det \left( A - B D^{-1} C \right)
\end{equation}
to $\mathscr{D}_{\rm EOFA}$, treated as a $2 \times 2$ block matrix in its spinor indices, and rearranging terms to arrive at the right-hand side of Eqn.~\eqref{eqn:EOFA_factorization}. Crucially, factors of $\gamma_{5} R_{5}$ can be freely inserted under the determinant to replace $\mathscr{D}_{\rm EOFA}$ with the Hermitian operator $H \equiv \gamma_{5} R_{5} \mathscr{D}_{\rm EOFA}$, since $\det ( \gamma_{5} ) = \det ( R_{5} ) = 1$. The final form of the exact one flavor pseudofermion action is $S_{\rm EOFA} = \phi^{\dagger} \mathcal{M}_{\rm EOFA} \phi$, with
\begin{equation}
\label{eqn:eofa_action}
\mathcal{M}_{\rm EOFA} \equiv 1 - k P_{-} \Omega_{-}^{\dagger} \left[ H(m_{1}) \right]^{-1} \Omega_{-} P_{-} + k P_{+} \Omega_{+}^{\dagger} \left[ H(m_{2}) - \Delta_{+} \left( m_{1}, m_{2} \right) P_{+} \right]^{-1} \Omega_{+} P_{+}.
\end{equation}
In Appendix \ref{appendix:eofa_operators} we collect explicit expressions for $k$, $\Omega_{\pm}$, $\Delta_{\pm}$, $\mathscr{D}_{\rm EOFA}$, and $\widetilde{\mathscr{D}}$ for Shamir and M\"{o}bius DWF, since, to the authors' knowledge, these expressions have not previously appeared in the literature. In Ref.~\cite{Chen:2014hyy} these operators are constructed recursively for the more general case of DWF with weights $\rho_{s} = c \omega_{s} + d$ and $\sigma_{s} = c \omega_{s} - d$ that are allowed to vary along the fifth dimension, subject to the constraint that $\omega_{s}$ is reflection-symmetric in $s$. Shamir and M\"{o}bius DWF are simpler, special cases with $\omega_{s} = 1$.
\FloatBarrier

\section{Summary of Ensembles Used in This Work}
\label{sec:ensembles}
The properties of the lattices used in this work are summarized in Tables \ref{tab:ensembles} and \ref{tab:ensembles_measured}. In all cases we use the Iwasaki gauge action (I) \cite{IWASAKI1984449}, and on some ensembles supplement this with the dislocation suppressing determinant ratio (DSDR) \cite{PhysRevD.74.034512,Renfrew:2009wu}; we abbreviate the combined action including both terms as ``ID''. The additional DSDR term is designed to suppress the dislocations of the gauge field associated with tunneling between topological sectors, thereby reducing the degree of residual chiral symmetry breaking. For strong coupling simulations, where these dislocations occur frequently, the DSDR term reduces the costs associated with light quark masses while still maintaining good topological sampling. We simulate $N_{f} = 2+1$ quark flavors using domain wall fermions, with either the Shamir (DWF) \cite{KAPLAN1992342,SHAMIR199390} or M\"{o}bius (MDWF) \cite{BROWER2005686,BROWER2006191,Brower:2012vk} kernel. Finally, on ensembles marked ``-G'' we use $G$-parity boundary conditions in one or more of the spatial directions. 
\begin{table}[!h]
	\centering
	\resizebox{\linewidth}{!}{
		\setlength{\tabcolsep}{8pt}
		\begin{tabular}{cccccccc}
			\hline
			\hline
			\rule{0cm}{0.4cm}Ensemble & Action & $\beta$ & $L^{3} \times T \times L_{s}$ & M\"{o}bius Scale & $G$-Parity B.C. & $a m_{l}$ & $a m_{h}$ \\
			\hline
			\rule{0cm}{0.4cm}16I & DWF + I & 2.13 & $16^{3} \times 32 \times 16$ & --- & --- & 0.01 & 0.032 \\
			\rule{0cm}{0.4cm}16I-G & DWF + I & 2.13 & $16^{3} \times 32 \times 16$ & --- & $x$ & 0.01 & 0.032 \\
			\rule{0cm}{0.4cm}16ID-G & MDWF + ID & 1.75 & $16^{3} \times 32 \times 8$ & 2.00 & $x$,$y$,$z$ & 0.01 & 0.045 \\
			\rule{0cm}{0.4cm}24ID & MDWF + ID & 1.633 & $24^{3} \times 64 \times 24$ & 4.00 & --- & 0.00789 & 0.085 \\
			\rule{0cm}{0.4cm}32ID-G & MDWF + ID & 1.75 & $32^{3} \times 64 \times 12$ & 2.67 & $x$,$y$,$z$ & 0.0001 & 0.045 \\
			\hline
			\hline
		\end{tabular}
	}
	\caption{Summary of ensembles and input parameters used in this work. Here $\beta$ is the gauge coupling, $L^{3} \times T \times L_{s}$ is the lattice volume decomposed into the length of the spatial ($L$), temporal ($T$), and fifth $(L_{s}$) dimensions, and $a m_{l}$ and $a m_{h}$ are the bare, input light and heavy quark masses. On the 16I-G, 16ID-G, and 32ID-G ensembles $G$-parity boundary conditions are applied to the fermion fields at one or more of the spatial boundaries of the lattice; otherwise periodic boundary conditions are applied, and in all cases antiperiodic boundary conditions are used along the temporal direction.}
	\label{tab:ensembles}
\end{table}

The 16I ensemble was first generated and used to study light meson spectroscopy with domain wall fermions in Ref.~\cite{Allton:2007hx}. The 16I-G ensemble is identical to the 16I ensemble except for the boundary conditions along the $x$-direction, which have been changed from periodic to $G$-parity. Likewise, the parameters of the 16ID-G ensemble have been chosen based on a series of $\beta = 1.75$ DSDR ensembles generated in Ref.~\cite{Arthur:2012yc}, but have $G$-parity boundary conditions in all three spatial directions. Collectively, these three lattices are used as inexpensive, small-volume test ensembles with unphysical, heavy pion masses to perform cross-checks of the EOFA algorithm and its implementation in the BFM and CPS code libraries. The larger 24ID \cite{coarse_ensembles} and 32ID-G \cite{PhysRevLett.115.212001} ensembles have physical pion masses and are currently being generated as part of production RBC/UKQCD calculations.
\begin{table}[!h]
	\centering
	\begin{tabular}{cccc}
		\hline
		\hline
		\rule{0cm}{0.4cm}Ensemble & $L$ (fm) & $a^{-1}$ (GeV) & $m_{\pi}$ (MeV) \\
		\hline
		\rule{0cm}{0.4cm}16I & 1.95(5) & 1.62(4) & 400(11) \\
		\rule{0cm}{0.4cm}16I-G & 1.95(5) & 1.62(4) & 388(14) \\
		\rule{0cm}{0.4cm}16ID-G & 2.29(1) & 1.378(7) & 575(11) \\
		\rule{0cm}{0.4cm}24ID & 4.82(19) & 0.981(39) & 137.1(5.5) \\
		\rule{0cm}{0.4cm}32ID-G & 4.57(2) & 1.378(7) & 143.1(2.0) \\
		\hline
		\hline
	\end{tabular}
	\caption{Summary of spatial volumes, lattice cutoffs, and pion masses in physical units for the ensembles used in this work. All values for the 16I and 32ID-G ensembles are from Refs.~\cite{Allton:2007hx} and \cite{Bai:2015nea}, respectively. On the 16I-G (16ID-G) ensemble we assume the lattice cutoff is the same as the 16I (32ID-G) ensemble since the same action and value of $\beta$ has been used. The pion masses on the 16I-G and 16ID-G ensembles have been extracted using the fitted value of the lowest energy pion states from Table \ref{tab:repro_test_spectra_gparity} and the continuum dispersion relation. Finally, the determination of the lattice scale for the 24ID ensemble was performed in Ref.~\cite{Boyle:2015exm}, and the determination of the pion mass in Ref.~\cite{coarse_ensembles}.}
	\label{tab:ensembles_measured}
\end{table}

\section{Hybrid Monte Carlo with EOFA}
\label{sec:HMC_with_EOFA}
In lattice QCD correlation functions are computed in terms of a discretized Euclidean path integral
\begin{equation}
\label{eqn:euclidean_pi}
\left\langle \mathscr{O}_{1} \cdots \mathscr{O}_{n} \right\rangle = \frac{1}{\mathcal{Z}} \int \mathscr{D} U \left( \prod_{f} \mathscr{D} \psi_{f} \mathscr{D} \overline{\psi}\,\hspace{-0.9mm}_{f} \right) \left( \mathscr{O}_{1}[U] \cdots \mathscr{O}_{n}[U] \right) e^{-S[U,\overline{\psi}\,\hspace{-0.9mm}_{f},\psi_{f}]}.
\end{equation}
Here $U$ is the gauge field, $\psi_{f}$ is the quark field associated with flavor $f$, and $S[U,\overline{\psi}\,\hspace{-0.9mm}_{f},\psi_{f}]$ is the action, which decomposes into a sum of contributions from the gauge field, fermions, and possibly other terms (e.g. the dislocation suppressing determinant ratio). To avoid having to deal with anticommuting Grassman variables in a computer, dynamical fermion flavors are integrated out and then reintroduced in terms of bosonic ``pseudofermion'' fields $\phi$ as
\begin{equation}
\label{eqn:ferm_det}
\frac{1}{\mathcal{Z}} \int \mathscr{D} \psi \mathscr{D} \overline{\psi} \, e^{-\overline{\psi} M \psi} = \det \left( M \right) = \frac{1}{\det \left( M^{-1} \right)} = \frac{1}{\mathcal{Z}} \int \mathscr{D} \phi \mathscr{D} \phi^{\dagger} \, e^{-\phi^{\dagger} M^{-1} \phi},
\end{equation}
provided $M$ is positive-definite. While pseudofermions can be represented straightforwardly in a computer, they come at the cost of applications of $M^{-1}$ rather than $M$, which is not typically available in an explicit form. Even after discretization the integration in Eqn.~\eqref{eqn:euclidean_pi} is far too expensive to perform directly due to the enormous number of degrees of freedom on a typical lattice. Instead, Monte Carlo techniques are used to ergodically sample a sequence of representative configurations of the gauge field $\{ U_{i} \}$, for which 
\begin{equation}
\left\langle \mathscr{O}_{1} \cdots \mathscr{O}_{n} \right\rangle \approx \frac{1}{N} \sum_{i=1}^{N} \mathscr{O}_{1}(U_{i}) \cdots \mathscr{O}_{n}(U_{i}).
\end{equation} 
The standard Monte Carlo technique used in modern lattice QCD calculations is known as the Hybrid Monte Carlo (HMC) algorithm.

HMC generates a Markov chain of gauge field configurations $\{ U_{i} \}$ by evolving a Hamiltonian system in unphysical Molecular Dynamics (MD) ``time''. This Hamiltonian system is constructed by treating $U_{\mu}(x)$ as a generalized coordinate, introducing an $\mathfrak{su}(3)$-valued conjugate momentum $\pi_{\mu}(x)$, and forming the standard Hamiltonian
\begin{equation}
\label{eqn:hmc_h}
H = \frac{1}{2} \pi^{2} + S(U).
\end{equation}
The associated equations of motion
\begin{equation}
\label{eqn:hmc_eom}
\begin{dcases}
\partial_{\tau} U_{\mu}(x) = \pi_{\mu}(x) U_{\mu}(x) \\
\partial_{\tau} \pi_{\mu}(x) = - T^{a} \partial_{x,\mu}^{a} S(U)
\end{dcases}
\end{equation}
can then be integrated using numerical integration techniques. The integration is performed over intervals of length $\Delta \tau$ --- referred to as a single MD trajectory --- as a sequence of $N$ small steps $\delta \tau$, with $N = \Delta \tau / \delta \tau$. Finite precision integration errors are corrected stochastically with a Metropolis accept/reject step: after every $N$ integration steps by $\delta \tau$ the total change in the Hamiltonian $\Delta H$ is computed, and the current gauge field $U_{\mu}'(x)$ is accepted as the next configuration in the Markov chain with probability
\begin{equation}
P_{\rm accept} = \min \left( 1, e^{- \Delta H} \right).
\end{equation}
One can show that the resulting algorithm satisfies detailed balance provided the scheme used to numerically integrate Eqn.~\eqref{eqn:hmc_eom} is reversible \cite{Rothe:1492203}. Ergodicity is achieved by performing a heatbath step each time the integration is restarted to pick a new conjugate momentum $\pi_{\mu}(x)$, and thus a new trajectory in the phase space $\{ (U,\pi) \}$. HMC generates a sequence of gauge field configurations whose statistical independence is governed by the length of each MD trajectory, $\Delta \tau$. The number of MD trajectories separating statistically independent gauge field configurations is typically determined \textit{ex post facto} by examining the integrated autocorrelation times of representative physical observables.

The fermionic contribution to the Hamiltonian in Eqn.~\eqref{eqn:hmc_h} introduces a technical obstacle for the HMC algorithm since, generically, the lattice Dirac operator $\mathscr{D}$ has a complex spectrum. Replacing $\mathscr{D}$ with the Hermitian fermion matrix $M = \mathscr{D}^{\dagger} \mathscr{D}$ in Eqn.~\eqref{eqn:ferm_det} has a number of advantages. Most importantly, it allows $M^{-1}$ to be applied to pseudofermion vectors using the conjugate gradient algorithm, and it allows for a straightforward pseudofermion heatbath step: at the beginning of each MD trajectory a random Gaussian vector $\eta$ is drawn according to $P(\eta) \propto \exp(-\eta^{\dagger} \eta/2)$ and the initial pseudofermion field is seeded as $\phi = \mathscr{D} \eta$, ensuring that $\phi$ is correctly sampled as $P(\phi) \propto \exp(-\phi^{\dagger} M^{-1} \phi/2)$. However, the fermion matrix $M = \mathscr{D}^{\dagger} \mathscr{D}$ describes two degenerate quark flavors with the same mass. Single flavor simulations are typically performed by taking an overall square root of the fermion determinant, 
\begin{equation}
\det \left( \mathscr{D} \right) = \left[ \det \left( \mathscr{D}^{\dagger} \mathscr{D} \right) \right]^{1/2}.
\end{equation} 
In the pseudofermion formalism applications of the operator $( \mathscr{D}^{\dagger} \mathscr{D} )^{-1/2}$ are approximated by a matrix-valued function $f(\mathscr{D}^{\dagger} \mathscr{D})$, 
\begin{equation}
\label{eqn:rhmc_approx_pi}
\left[ \det \left( \mathscr{D}^{\dagger} \mathscr{D} \right) \right]^{1/2} = \frac{1}{\mathcal{Z}} \int \mathscr{D} \phi \mathscr{D} \phi^{\dagger} e^{-\phi^{\dagger} ( \mathscr{D}^{\dagger} \mathscr{D} )^{-1/2} \phi} \simeq \frac{1}{\mathcal{Z}} \int \mathscr{D} \phi \mathscr{D} \phi^{\dagger} e^{-\phi^{\dagger} f ( \mathscr{D}^{\dagger} \mathscr{D} ) \phi}
\end{equation}
where $f(x)$ is a suitably constructed approximation to the inverse square root, valid over the spectral range of $\mathscr{D}^{\dagger} \mathscr{D}$. Variants of the HMC algorithm which construct $f$ from different classes of functions have been proposed and used in the literature; the most common is the rational HMC (RHMC) algorithm \cite{Clark:2006wq}, where
\begin{equation}
\label{eqn:rat_approx}
f(x) = \alpha_{0} + \sum_{k=1}^{N} \frac{\alpha_{k}}{\beta_{k} + x}
\end{equation}
is a rational function. While rational functions are in many ways a good choice --- they are economical in the sense that the inverse square root can usually be well-approximated by a modest number of terms, and the multishift CG algorithm can be used to efficiently invert $( \mathscr{D}^{\dagger} \mathscr{D} + \beta_{k} )$ for all $k$ simultaneously --- the additional complexity of evaluating $f(\mathscr{D}^{\dagger} \mathscr{D})$ and the associated molecular dynamics pseudofermion force makes single flavor RHMC simulations significantly more costly than degenerate two flavor HMC simulations at the same bare quark mass. This additional cost can be largely attributed to the significant linear algebra overhead associated with multishift CG.

EOFA provides an alternative construction of a single-flavor pseudofermion action through Eqn.~\eqref{eqn:EOFA_factorization}: a ratio of fermion determinants can be factorized as a product of two determinants, each of which involves an operator which is Hermitian and positive-definite. This product can then be represented as a path integral over a bosonic pseudofermion field with a two-term action (Eqn.~\eqref{eqn:eofa_action})
\begin{equation}
\label{eqn:eofa_pi}
\det \left( \frac{\mathscr{D}_{\rm EOFA}(m_{1})}{\mathscr{D}_{\rm EOFA}(m_{2})} \right) = \frac{1}{\mathcal{Z}} \int \mathscr{D} \phi \mathscr{D} \phi^{\dagger} e^{-\phi^{\dagger} \mathcal{M}_{\rm EOFA} \phi},
\end{equation} 
leading to an algorithm which is ``exact'' in the sense that it avoids the numerical approximations required to implement the square root in RHMC (Eqn.~\eqref{eqn:rhmc_approx_pi}) and related HMC variants. EOFA is also expected to be somewhat faster than RHMC, since there is no rational approximation entering into evaluations of the Hamiltonian or the pseudofermion force, eliminating the overhead associated with multishift CG. In the remainder of this section we elaborate on the details of the action, heatbath step, and pseudofermion force entering into the Hamiltonian equations of motion (Eqn.~\eqref{eqn:hmc_eom}) for HMC with EOFA.

\subsection{Action}

\begin{figure}[!h]
\centering
\subfloat{\includegraphics[width=0.85\linewidth]{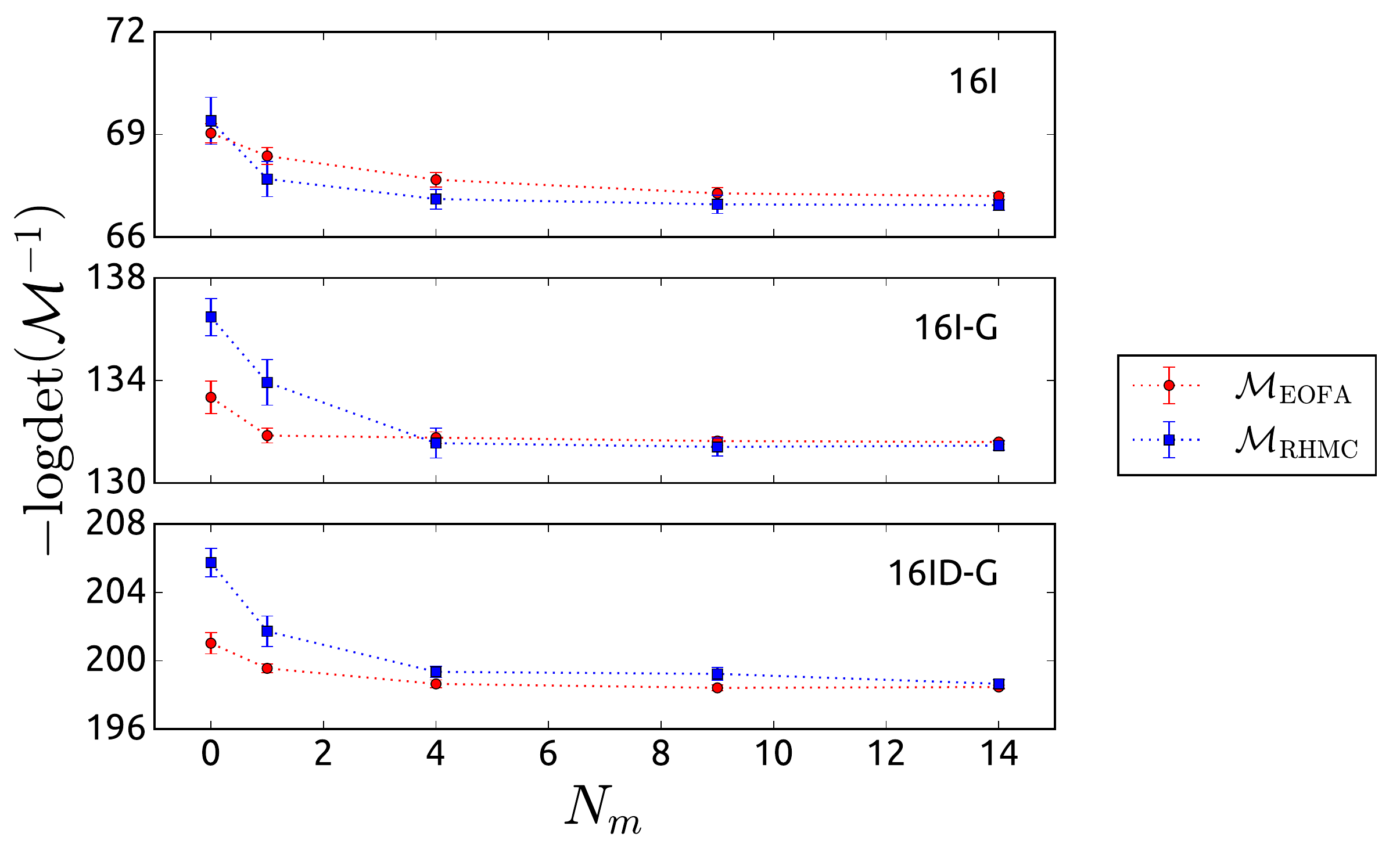}} \\
\subfloat{\includegraphics[width=0.475\linewidth]{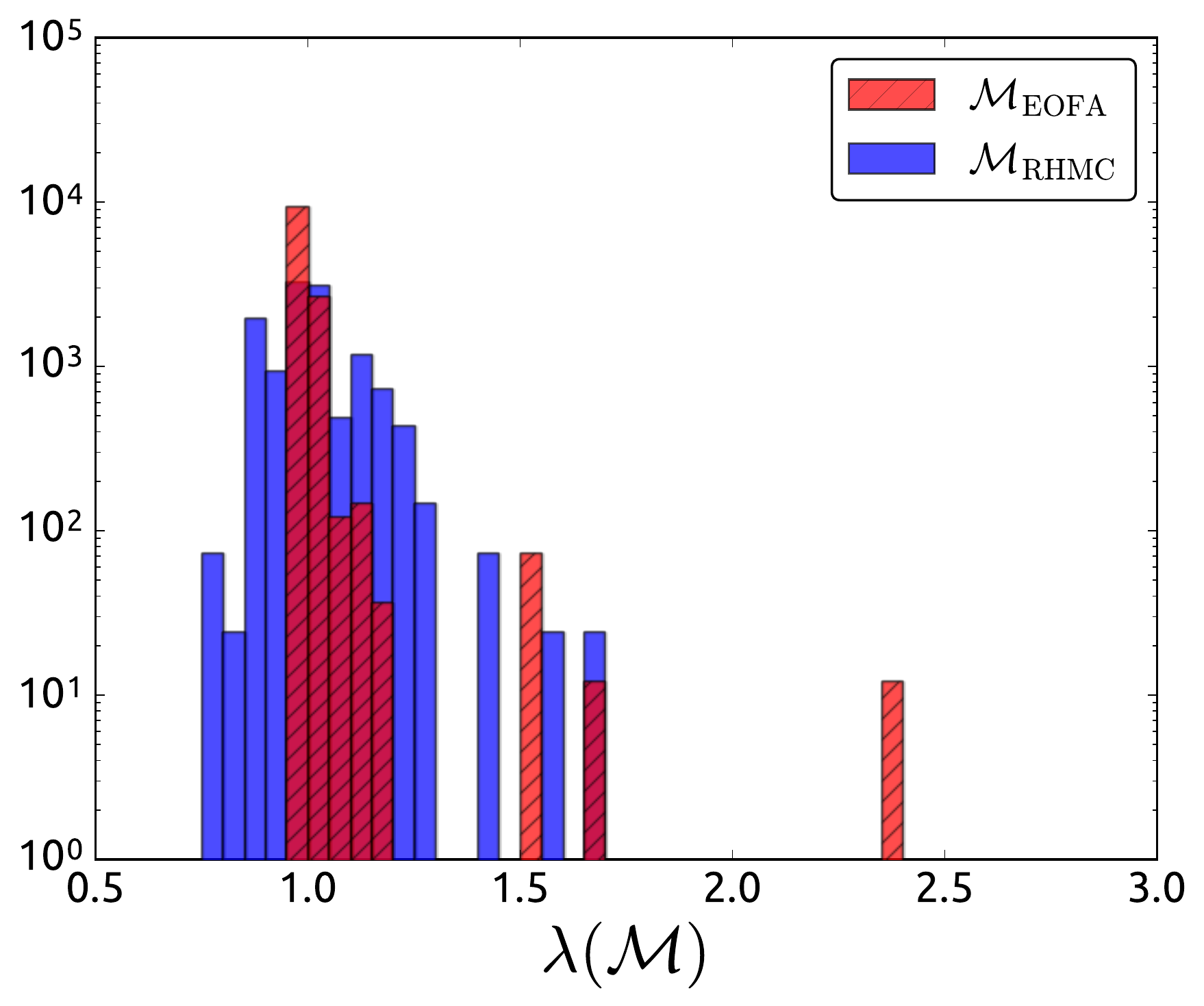}} \quad
\subfloat{\includegraphics[width=0.475\linewidth]{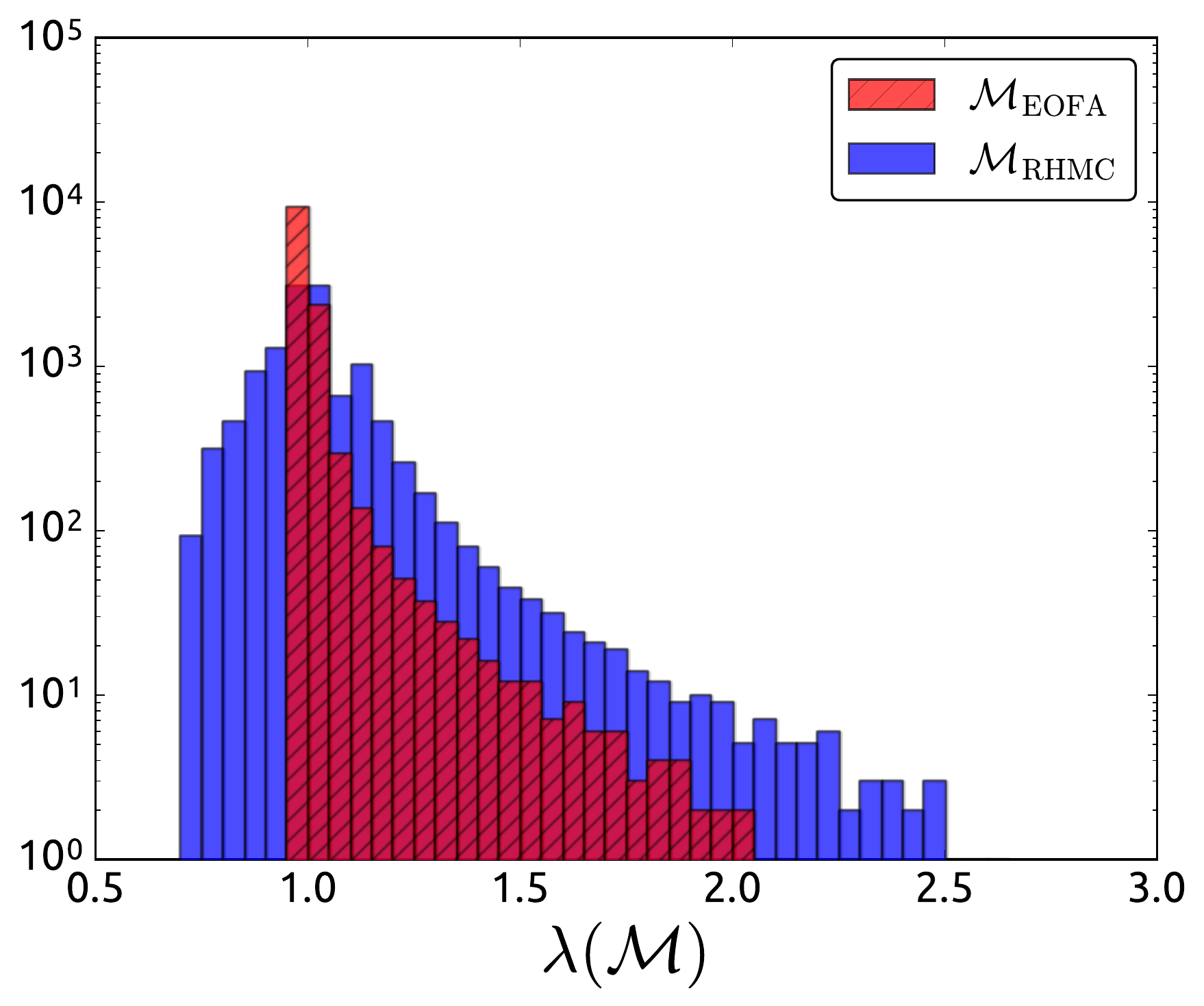}}
\caption{Top: log determinants of the EOFA and RHMC actions as a function of the number of intermediate masses ($N_{m}$) used to compute Eqn.~\eqref{eqn:det_ratio_prod}, computed on a single, thermalized configuration of the 16I, 16I-G, and 16ID-G ensembles. We set $(a m_{1}, a m_{2})$ to (0.032, 0.042), (0.032,0.042), and (0.045,0.055) on the 16I, 16I-G, and 16ID-G ensemble, respectively. We note that the error bars are purely statistical; for small $N_{m}$ there is a large, unaccounted systematic error associated with setting $\Sigma = 1$ in Eqns.~\eqref{eqn:pi_action_test} and \eqref{eqn:logdet_action_test}. Bottom: eigenvalue spectra of $\mathcal{M}_{\rm EOFA}$ and $\mathcal{M}_{\rm RHMC}$ on a $4^{5}$ lattice with $a m_{1} = 0.01$, $a m_{2} = 1.0$, and $a M_{5} = 1.8$. In the bottom left plot all of the gauge links are set to $U_{\mu}(x) = 1$ (\textit{i.e.} the free field limit); in the bottom right plot each gauge link is set to an independent, random $SU(3)$ matrix.}
\label{fig:eofa_action_test}
\end{figure}

The EOFA action (Eqn.~\eqref{eqn:eofa_action}) computes a ratio of determinants of $\mathscr{D}_{\rm EOFA}$ upon integrating out the pseudofermion fields (Eqn.~\eqref{eqn:eofa_pi}). This ratio can be related to the conventional determinant ratio computed by the RHMC algorithm through Eqns.~\eqref{eqn:eofa_dop} and \eqref{eqn:dtilde_det}, leading to the relationship
\begin{equation}
\label{eqn:eofa_action_test}
\det \left( \frac{\mathscr{D}_{\rm DWF}(m_{1})}{\mathscr{D}_{\rm DWF}(m_{2})} \right) = \left( \frac{ \left(c+d \right)^{L_{s}} + m_{1} \left( c - d \right)^{L_{s}} }{ \left(c+d \right)^{L_{s}} + m_{2} \left( c - d \right)^{L_{s}} } \right)^{12 V} \det \left( \frac{\mathscr{D}_{\rm EOFA}(m_{1})}{\mathscr{D}_{\rm EOFA}(m_{2})} \right).
\end{equation}
We use this relationship as a test of the equivalence of RHMC and EOFA, as well as our implementation of the EOFA action, by stochastically computing the left side of Eqn.~\eqref{eqn:eofa_action_test} with the RHMC action 
\begin{equation}
\mathcal{M}_{\rm RHMC} = \left[ \mathscr{D}_{\rm DWF}^{\dagger} \mathscr{D}_{\rm DWF}(m_{2}) \right]^{1/4} \left[ \mathscr{D}_{\rm DWF}^{\dagger} \mathscr{D}_{\rm DWF}(m_{1}) \right]^{-1/2} \left[ \mathscr{D}_{\rm DWF}^{\dagger} \mathscr{D}_{\rm DWF}(m_{2}) \right]^{1/4}
\end{equation}
and the right side with the EOFA action (Eqn.~\eqref{eqn:eofa_action}) on the same gauge field configuration. Observing that we can, in general, rewrite a determinant as
\begin{equation}
\label{eqn:pi_action_test}
\det \left( \mathcal{M}^{-1} \right) = \frac{1}{\mathcal{Z}} \int \mathscr{D} \phi \mathscr{D} \phi^{\dagger} e^{-\phi^{\dagger} \mathcal{M} \phi} = \frac{1}{\mathcal{Z}} \int \mathscr{D} \phi \mathscr{D} \phi^{\dagger} e^{-\frac{1}{2} \phi^{\dagger} \Sigma^{-1} \phi} e^{\phi^{\dagger} \left( \frac{1}{2} \Sigma^{-1} - \mathcal{M} \right) \phi}
\end{equation} 
suggests the following simple Monte Carlo integration scheme: we draw random pseudofermion vectors by independently sampling the real and imaginary parts of each component from the standard normal distribution $\mathcal{N}(\mu=0,\sigma=1)$, and compute the expectation value
\begin{equation}
\label{eqn:logdet_action_test}
-\mathrm{logdet} \left( \mathcal{M}^{-1} \right) \approx \left\langle \phi^{\dagger}_{i} \left( \mathcal{M} - \frac{1}{2} \Sigma^{-1} \right) \phi_{i} \right\rangle_{i},
\end{equation}
where the average is computed using the jackknife resampling technique. This will accurately approximate the true log determinant for finite, realistically calculable values of $N$ provided $m_{1}$ and $m_{2}$ are sufficiently close that the integrand is well-approximated by a Gaussian with unit variance. To address this latter systematic, we consider splitting Eqn.~\eqref{eqn:eofa_action_test} as a product of determinants
\begin{equation}
\label{eqn:det_ratio_prod}
\det \left( \frac{\mathscr{D}(m_{1})}{\mathscr{D}(m_{2})} \right) = \det \left( \frac{\mathscr{D}(m_{1})}{\mathscr{D}(m_{1}')} \right) \left[ \prod_{i=1}^{N_{m}} \det \left( \frac{\mathscr{D}(m_{i}')}{\mathscr{D}(m_{i+1}')} \right) \right] \det \left( \frac{\mathscr{D}(m_{N_{m}}')}{\mathscr{D}(m_{2})} \right)
\end{equation}
with equally-spaced intermediate masses
\begin{equation}
m_{i}' = m_{1} + \frac{m_{2} - m_{1}}{N_{m}+1} i, \quad i = 1, \ldots, N_{m},
\end{equation}
and study the dependence of the result on $N_{m}$ (this procedure is identical to the method introduced in Ref.~\cite{Aoki:2010dy} for computing quark mass reweighting factors). In the upper panel of Figure \ref{fig:eofa_action_test} we plot the log determinants of $\mathcal{M}_{\rm RHMC}^{-1}$ and $\mathcal{M}_{\rm EOFA}^{-1}$ as a function of $N_{m}$, with $N = 10$ stochastic evaluations, computed using a single thermalized trajectory of the 16I, 16I-G, and 16ID-G ensembles. For the case of the 16ID-G ensemble, which uses the M\"{o}bius DWF fermion action, we also include the overall constant multiplying the right side of Eqn.~\eqref{eqn:eofa_action_test} so that in all cases we are computing the same determinant ratio of $\mathscr{D}_{\rm DWF}$ using either action.

We observe, as expected, that both formalisms agree for sufficiently large $N_{m}$. Likewise, we observe that at sufficiently small $N_{m}$ the calculation generally becomes unreliable since we do not attempt to account for the systematic error associated with approximating the integrand of the path integral by a Gaussian with unit variance (\textit{i.e.}~setting $\Sigma = 1$ in Eqns.~\eqref{eqn:pi_action_test} and \eqref{eqn:logdet_action_test}.). In both cases ``sufficiently'' small or large $N_{m}$ is controlled by the size of the splitting between $m_{1}$ and $m_{2}$. We also observe that, for a given choice of $N_{m}$ and $N$, both the statistical and systematic errors of the determinant ratio computed via EOFA are suppressed relative to the errors of the determinant ratio computed via RHMC. We argue that the observed error suppression can be explained by comparing the spectrum of $\mathcal{M}_{\rm RHMC}$ to the spectrum of $\mathcal{M}_{\rm EOFA}$, which we plot in the lower panels of Figure \ref{fig:eofa_action_test} for a very small lattice volume ($4^{5}$) where the complete spectrum can be computed directly. While both operators have similar condition numbers, we find that most of the spectrum of the EOFA action is concentrated into a small interval $[1,1+\Delta]$ with $\Delta \sim \mathcal{O}(0.1)$, leading to an action which is easier to estimate stochastically. 

We propose that TWQCD's EOFA construction can be thought of as a kind of preconditioning which computes the same determinant ratio as RHMC but modifies the operator inside the determinant ($\mathcal{M}_{\rm RHMC}$), mapping its spectrum onto a more compact interval. This suggests an additional application of the EOFA formalism: quark mass reweighting factors can be computed substantially more cheaply using the EOFA action than using the RHMC action, especially at light quark masses, even if the ensemble was generated using RHMC. This could be useful, for example, to include the dynamical effects of isospin breaking in ensembles generated with isospin-symmetric up and down quarks.

\FloatBarrier
\subsection{Heatbath}

At the beginning of each HMC trajectory we wish to draw a random pseudofermion field $\phi$ according to the distribution $P(\phi) \propto \exp( -\phi^{\dagger} \mathcal{M}_{\rm EOFA} \phi )$. To do this, we first draw a random vector $\eta$ by independently sampling the real and imaginary parts of each component from the normal distribution with $\mu = 0$ and $\sigma^{2} = 1/2$, and then compute $\phi = \mathcal{M}_{\rm EOFA}^{-1/2} \eta$. As with the RHMC algorithm we approximate the inverse square root by an appropriately constructed rational function, but we stress that in the context of EOFA this rational approximation enters only into the heatbath and is not necessary to compute the EOFA action itself or the associated pseudofermion force. Naively applying a rational approximation with the form of Eqn.~\eqref{eqn:rat_approx} to the operator $\mathcal{M}_{\rm EOFA}$ results in
\begin{multline}
\mathcal{M}_{\rm EOFA}^{-1/2} \simeq \alpha_{0} + \sum_{k=1}^{N_{p}} \alpha_{l} \left[ \frac{1}{\gamma_{l}} - k P_{-} \Omega_{-}^{\dagger} \left[ H(m_{1}) \right]^{-1} \Omega_{-} P_{-} \right. \\
\left. + k P_{+} \Omega_{+}^{\dagger} \left[ H(m_{2}) - \Delta_{+}(m_{1},m_{2}) P_{+} \right]^{-1} \Omega_{+} P_{+} \right]^{-1},
\end{multline}
where we have defined $\gamma_{l} \equiv (1 + \beta_{l})^{-1}$. In this form, the nested inversions required to seed the heatbath would make EOFA prohibitively expensive. However, the Woodbury matrix identity
\begin{equation}
\left( A + B C D \right)^{-1} = A^{-1} - A^{-1} B \left( C^{-1} + D A^{-1} B \right)^{-1} D A^{-1}
\end{equation}
and the cancellation between cross-terms involving products of the chiral projection operators can be used to manipulate this expression into the equivalent form
\begin{multline}
\label{eqn:eofa_heatbath}
\mathcal{M}_{\rm EOFA}^{-1/2} \simeq \alpha_{0} + \sum_{k=1}^{N_{p}} \alpha_{l} \gamma_{l} \Big\{ 1 + k \gamma_{l} P_{-} \Omega_{-}^{\dagger} \left[ H(m_{1}) - \gamma_{l} \Delta_{-}(m_{1},m_{2}) P_{-} \right]^{-1} \Omega_{-} P_{-} \\
- k \gamma_{l} P_{+} \Omega_{+}^{\dagger} \left[ H(m_{2}) - \beta_{l} \gamma_{l} \Delta_{+}(m_{1},m_{2}) P_{+} \right]^{-1} \Omega_{+} P_{+} \Big\}.
\end{multline}
With this expression the EOFA heatbath step can be performed at the cost of $2 N_{p}$ CG inversions using a rational approximation with $N_{p}$ poles. Unlike the case of RHMC, multishift CG algorithms are not applicable to the EOFA heatbath since each of the $2 N_{p}$ operators in Eqn.~\eqref{eqn:eofa_heatbath} generates a different Krylov space. Furthermore, since the operators $\Delta_{\pm} P_{\pm}$ have a large number of zero modes and are therefore not invertible, there is no simple transformation by which this system can be recast into a form amenable to multishift CG.

\begin{figure}[!h]
\centering
\subfloat{\includegraphics[width=0.48\linewidth]{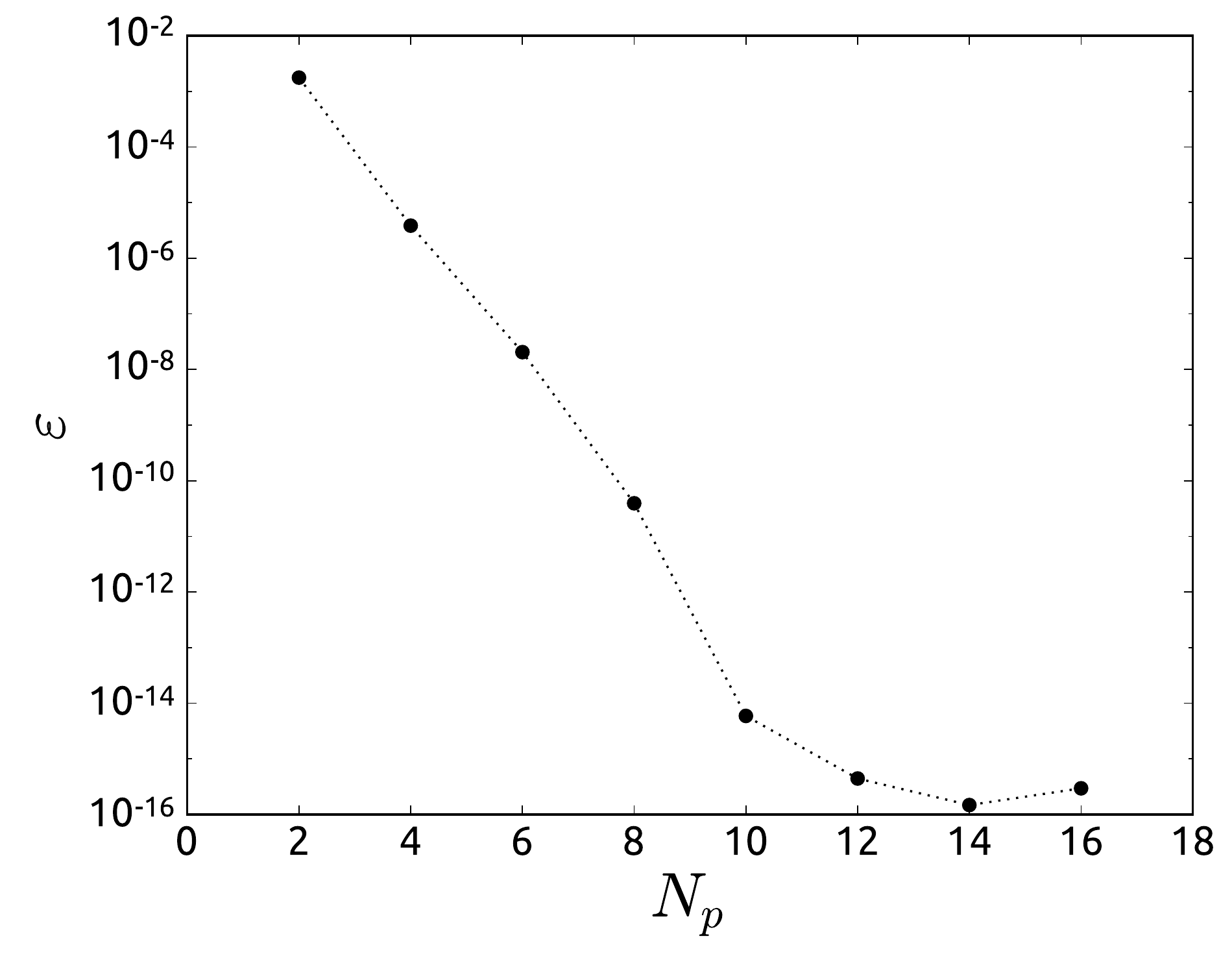}} \quad
\subfloat{\includegraphics[width=0.48\linewidth]{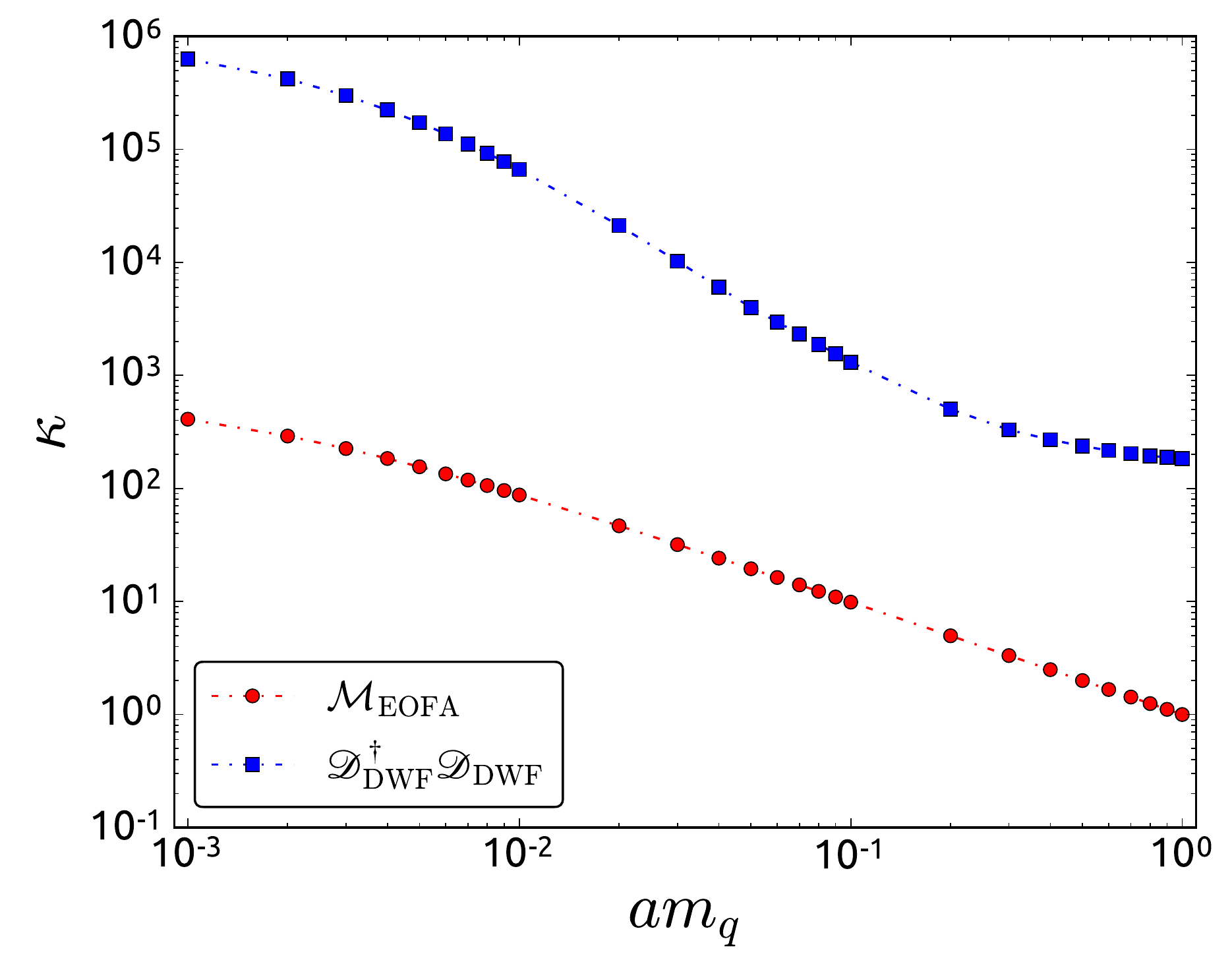}}
\caption{Left: relative error --- $\varepsilon$, defined by Eqn.~\eqref{eqn:heatbath_rel_err} --- in seeding the pseudofermion heatbath as a function of the number of poles in the rational approximation to the inverse square root ($N_{p}$), with $a m_{1} = 0.032$ set to the dynamical heavy quark mass, and a stopping residual of $10^{-10}$ for all CG inversions. Right: condition numbers of $\mathcal{M}_{\rm EOFA}$ and $\mathscr{D}_{\rm DWF}^{\dagger} \mathscr{D}_{\rm DWF}$ as a function of the bare input quark mass ($a m_{q}$); for $\mathcal{M}_{\rm EOFA}$ this is the numerator mass ($a m_{1} = a m_{q}$), while the denominator mass is fixed at $a m_{2} \equiv 1$. Both calculations were performed on a single, thermalized configuration of the 16I ensemble.}
\label{fig:heatbath_test}
\end{figure}

In the left panel of Figure \ref{fig:heatbath_test} we test Eqn.~\eqref{eqn:eofa_heatbath} on a single thermalized configuration of the 16I ensemble by computing the quantity
\begin{equation}
\label{eqn:heatbath_rel_err}
\varepsilon \equiv \frac{\left| \eta^{\dagger} \eta - \phi^{\dagger} \mathcal{M}_{\rm EOFA} \phi \right|}{\eta^{\dagger} \eta}
\end{equation} 
after seeding the pseudofermion field $\phi$ with a random Gaussian vector $\eta$. In exact arithmetic $\varepsilon = 0$; in practice it measures the relative error in the heatbath step arising from the choice of CG stopping conditions and rational approximation to the inverse square root. We repeat this calculation, varying the number of poles in the rational approximation but keeping the stopping conditions fixed, and observe that $\varepsilon$ reaches the limits of double-precision arithmetic even with a relatively modest number of poles compared to what is typically required to compute non-integer powers of $\mathscr{D}^{\dagger} \mathscr{D}$ accurately in the context of RHMC. In the right panel of Figure \ref{fig:heatbath_test} we demonstrate this explicitly by computing the condition numbers $\kappa = \lambda_{\rm max} / \lambda_{\min}$ of both operators as a function of the bare input quark mass. In Section \ref{sec:optimizations} we show how aggressive tuning of the rational approximation and stopping conditions, together with forecasting techniques for the initial CG guesses, can be combined to ameliorate the cost of the $2 N_{p}$ inversions required to apply $\mathcal{M}_{\rm EOFA}^{-1/2}$.

\subsection{Pseudofermion Force}

The pseudofermion force 
\begin{equation}
T^{a} \partial_{x,\mu}^{a} S_{f} (U) \equiv T^{a} \frac{d}{ds} S_{f} \left( e^{s T^{a}} U_{\mu}(x) \right) \Big|_{s=0}
\end{equation}
measures the back-reaction of the pseudofermions on the HMC system (Eqn.~\eqref{eqn:hmc_eom}) under an infinitesimal variation of the gauge field. In our notation $\{ T^{a} \}$ is a basis for the Lie algebra $\mathfrak{su}(3)$, with the CPS normalization convention
\begin{equation}
\mathrm{Tr} \left( T^{a} T^{b} \right) = - \frac{1}{2} \delta^{ab}.
\end{equation}
The EOFA pseudofermion force can be worked out explicitly by differentiating the EOFA action (Eqn.~\eqref{eqn:eofa_action}) and applying the matrix identity
\begin{equation}
\partial_{x} M^{-1} = - M^{-1} \left( \partial_{x} M \right) M^{-1},
\end{equation}
resulting in
\begin{equation}
\label{eqn:eofa_force}
T^{a} \partial_{x,\mu}^{a} S(U) = k T^{a} \left( \gamma_{5} R_{5} \chi_{1} \right)^{\dagger} \left( \partial_{x,\mu}^{a} D_{W} \right) \chi_{1} - k T^{a} \left( \gamma_{5} R_{5} \chi_{2} \right)^{\dagger} \left( \partial_{x,\mu}^{a} D_{W} \right) \chi_{2},
\end{equation}
with
\begin{equation}
\chi_{1} \equiv \left[ H(m_{1}) \right]^{-1} \Omega_{-} P_{-} \phi
\end{equation}
and
\begin{equation}
\chi_{2} \equiv \left[ H(m_{2}) - \Delta_{+}(m_{1},m_{2}) P_{+} \right]^{-1} \Omega_{+} P_{+} \phi.
\end{equation}
Standard manipulations can be used to write a more general Dirac bilinear as
\begin{align}
\begin{split}
a^{\dagger} \left( \partial_{x,\mu}^{a} D_{W} \right) b &= - \frac{k}{2} \sum_{x,s,\mu} \left[ a^{\dagger}(x,s) T^{a} U_{\mu}(x) \left( 1 - \gamma_{\mu} \right) b(x+\hat{\mu},s) \right. \\
&\hspace{5cm} \left. - a^{\dagger}(x+\hat{\mu},s) U^{\dagger}_{\mu}(x) T^{a} \left( 1 + \gamma_{\mu} \right) b(x) \right] \\
&= - \frac{k}{2} \sum_{x,s,\mu} \left[ U_{\mu}(x) \left( \mathop{\mathrm{Tr}}_{\mathrm{spin}} \left[ \left( 1 + \gamma_{\mu} \right) a(x+\hat{\mu},s) b^{\dagger}(x,s) \right] + \right. \right. \\
&\hspace{5cm}\left. \left. \mathop{\mathrm{Tr}}_{\mathrm{spin}} \left[ \left( 1 - \gamma_{\mu} \right) b(x+\hat{\mu},s) a^{\dagger}(x,s) \right] \right) \right],
\end{split}
\end{align}
allowing Eqn.~\eqref{eqn:eofa_force} to be efficiently computed locally in terms of a trace over spinor indices, at the cost of the two inversions required to form $\chi_{1}$ and $\chi_{2}$. Since Dirac bilinears of the form $a^{\dagger} \left( \partial_{x,\mu}^{a} D_{W} \right) b$ enter into the pseudofermion forces associated with many of the standard pseudofermion actions for Wilson and domain wall fermions --- including the RHMC action --- implementing the EOFA pseudofermion force requires little new code beyond what is required to implement the EOFA Hamiltonian. 
\FloatBarrier

\section{Small Volume Reproduction Tests}
\label{sec:16_cubed_reproduction_tests}
To further test our implementation of EOFA we have reproduced the 16I (16I-G, 16ID-G) ensemble using EOFA for the strange quark (light quarks) in place of RHMC. For these tests we have made no serious effort to tune EOFA for performance; we have simply checked that replacing RHMC with EOFA, but leaving all other details of the simulation fixed, has no discernible impact on physical observables such as the average plaquette, topological susceptibility, and low energy spectrum. 

\subsection{Ensemble Generation}

The details of the integrator parameters and nesting are summarized in Tables \ref{tab:test_int_parameters} and \ref{tab:test_int_nesting}, respectively. We use the abbreviations
\begin{equation}
\label{eqn:quo_action}
\mathrm{Quo}(m_{1},m_{2}) \equiv \det \left( \frac{\mathscr{D}_{\rm DWF}^{\dagger} \mathscr{D}_{\rm DWF}(m_{1})}{\mathscr{D}_{\rm DWF}^{\dagger} \mathscr{D}_{\rm DWF}(m_{2})} \right)
\end{equation}
and
\begin{equation}
\label{eqn:rat_quo_action}
\mathrm{RatQuo}_{1/n}(m_{1},m_{2}) \equiv \left[ \det \left( \frac{\mathscr{D}_{\rm DWF}^{\dagger} \mathscr{D}_{\rm DWF}(m_{1})}{\mathscr{D}_{\rm DWF}^{\dagger} \mathscr{D}_{\rm DWF}(m_{2})} \right) \right]^{1/n}
\end{equation}
to denote the quotient and rational quotient actions, and on the EOFA reproduction ensembles replace each instance of $\mathrm{RatQuo}_{1/2}(m_{1},m_{2})$ with the EOFA action (Eqn.~\eqref{eqn:eofa_action}) using the same mass parameters. The 16I and 16I-G EOFA reproduction runs were seeded with an ordered start --- \textit{i.e.} all gauge links were initially set to the unit matrix --- and evolved for 1500 and 2500 MD time units, respectively. For the 16ID-G ensemble the last RHMC configuration (MD trajectory 908) was used to seed the start of the EOFA reproduction run, and then evolved for an additional 500 MD time units.

\begin{table}[!h]
\centering
\begin{tabular}{cccccc}
\hline
\hline
\rule{0cm}{0.4cm}Ensemble & Integrator & $\delta \tau$ & $r_{\rm FG}$ & $r_{\rm MD}$ & $r_{\rm MC}$ \\
\hline
\rule{0cm}{0.4cm}16I & Force Gradient QPQPQ & 0.1000 & $10^{-7}$ & $10^{-8}$ & $10^{-10}$ \\
16I-G & Omelyan ($\lambda = 0.2$) & 0.2000 & --- & $10^{-8}$ & $10^{-10}$ \\
16ID-G & Force Gradient QPQPQ & 0.1667 & $10^{-6}$ & $10^{-7}$ & $10^{-10}$ \\
\hline
\hline
\end{tabular}
\caption{Basic integrator and HMC details for the generation of the 16I, 16I-G, and 16ID-G ensembles. We use nested Sexton-Weingarten integration schemes, detailed in Table \ref{tab:test_int_nesting}, with $\delta \tau$ the coarsest time step used to evolve the outermost level. We denote the CG stopping tolerances used for the force gradient forecasting, molecular dynamics, and Monte Carlo steps by $r_{\rm FG}$, $r_{\rm MD}$, and $r_{\rm MC}$, respectively.}
\label{tab:test_int_parameters}
\end{table}

\begin{table}[!h]
\centering
\begin{tabular}{cccc}
\hline
\hline
\rule{0cm}{0.4cm}Ensemble & Level & Action & Update \\
\hline
\rule{0cm}{0.4cm}\multirow{2}{*}{16I} & 1 & Quo(0.01,0.2) + Quo(0.2,1.0) + $\mathrm{RatQuo}_{1/2}$(0.032,1.0) & 4:1 \\
 & 2 & Gauge & 1:1 \\
\hline
\rule{0cm}{0.4cm}\multirow{3}{*}{16I-G} & 1 & $\mathrm{RatQuo}_{1/2}$(0.01,0.032) & 1:1 \\
& 2 & $\mathrm{RatQuo}_{1/4}$(0.032,1.0) + $\mathrm{RatQuo}_{1/4}$(0.032,1.0) + $\mathrm{RatQuo}_{1/4}$(0.032,1.0) & 8:1 \\
& 3 & Gauge & 1:1 \\
\hline
\rule{0cm}{0.4cm}\multirow{3}{*}{16ID-G} & 1 & $\mathrm{RatQuo}_{1/2}$(0.01,0.05) + $\mathrm{RatQuo}_{1/2}$(0.05,1.0) + $\mathrm{RatQuo}_{1/4}$(0.045,1.0) & 1:1 \\
& 2 & DSDR & 8:1 \\
& 3 & Gauge & 1:1 \\
\hline
\hline
\end{tabular}
\caption{Integrator layouts for the original RHMC runs. Here ``Quo'' is an abbreviation for the quotient action (Eqn.~\eqref{eqn:quo_action}) and ``$\mathrm{RatQuo}_{1/n}$'' is an abbreviation for the rational quotient action (Eqn.~\eqref{eqn:rat_quo_action}), with a rational function approximation used to apply $(\mathscr{D}^{\dagger} \mathscr{D})^{1/n}$ and its inverse. For the EOFA reproduction runs each instance of $\mathrm{RatQuo}_{1/2}$ is replaced by an EOFA determinant with the same masses (Eqn.~\eqref{eqn:eofa_action}), while all other ensemble and integrator details are left fixed. The notation A:B for the update scheme denotes the number of steps of the next innermost integrator level (A) per step of the current level (B).}
\label{tab:test_int_nesting}
\end{table}

\subsection{Basic Observables}

In Table \ref{tab:repro_test_basic_observables} we summarize results for the average plaquette $\langle P \rangle$, light quark and strange quark chiral $\langle \overline{\psi} \psi \rangle$ and pseudoscalar $\langle \overline{\psi} \gamma_{5} \psi \rangle$ condensates, and the topological susceptibility $\chi_{t} \equiv \langle Q^{2} \rangle / V$, computed on each ensemble; we observe statistically consistent results between the RHMC and EOFA ensembles in each case. Accompanying plots of the time evolution can be found in Section \ref{appendix:more_repro_plots}. The topological charge $Q$ has been measured using the 5Li discretization introduced in Ref.~\cite{DEFORCRAND1997409} after cooling the gauge fields with 20 steps of APE smearing \cite{ALBANESE1987163} using a smearing coefficient of 0.45. The ensemble averages were computed after binning over 50 (25) successive MD time units on the 16I and 16I-G (16ID-G) ensembles, where the bin size has been conservatively chosen based on the integrated autocorrelation times measured in Ref.~\cite{Allton:2007hx} for the 16I ensemble and Ref.~\cite{Arthur:2012yc} for a series of $\beta = 1.75$ DSDR ensembles. We expect that the runs produced for this study are too short to reliably compute integrated autocorrelation times directly, but note that there is no evidence of a difference in an integrated autocorrelation time between the EOFA and RHMC ensembles in the time evolution plots of Section \ref{appendix:more_repro_plots}.
\begin{table}[!h]
\centering
\resizebox{\linewidth}{!}{
\begin{tabular}{c|cc|cc|cc}
\hline
\hline
\rule{0cm}{0.4cm}& \multicolumn{2}{|c|}{\ul{16I}} & \multicolumn{2}{|c|}{\ul{16I-G}} & \multicolumn{2}{|c}{\ul{16ID-G}} \\
Observable & RHMC & EOFA & RHMC & EOFA & RHMC & EOFA \\
\hline
\rule{0cm}{0.4cm}$\langle P \rangle$ & 0.588087(22) & 0.588106(26) & 0.588033(24) & 0.588039(16) & 0.514251(43) & 0.514200(48) \\
$\langle \overline{\psi}_{l} \psi_{l} \rangle$ & 0.001697(5) & 0.001698(11) & 0.0017151(72) & 0.0017130(52) & 0.005543(11) & 0.005563(8) \\
$\langle \overline{\psi}_{s} \psi_{s} \rangle$ & 0.0037450(31) & 0.0037435(74) & 0.0037541(51) & 0.0037529(34) & 0.0085729(82) & 0.0085895(69) \\
$\langle \overline{\psi}_{l} \gamma_{5} \psi_{l} \rangle$ & -0.000015(14) & -0.000012(19) & -0.000003(15) & -0.000006(12) & 0.000033(13) & -0.000001(11) \\
$\langle \overline{\psi}_{s} \gamma_{5} \psi_{s} \rangle$ & -0.000001(8) & -0.000007(12) & -0.0000004(92) & -0.0000034(81) & 0.000017(10) & -0.000002(8) \\
$\chi_{t}$ & $1.03(19) \times 10^{-5}$ & $1.81(42) \times 10^{-5}$ & $2.16(47) \times 10^{-5}$ & $1.53(27) \times 10^{-5}$ & --- & --- \\
\hline
\hline
\end{tabular}
}
\caption{Average plaquettes, quark condensates, and topological susceptibilities ($\chi_{t}$) computed on the 16I, 16I-G and 16ID-G lattices and their corresponding EOFA reproduction ensembles. The ensemble averages on the 16I (16I-G) lattices were computed using MD trajectories 500-1500 (500-2500) after binning over 50 successive MD time units. The ensemble averages on the 16ID-G lattices were computed using MD trajectories 500:900 for the RHMC ensemble, and MD trajectories 960:1360 for the EOFA ensemble, after binning over 25 successive MD time units. We do not compute $\chi_{t}$ on the 16ID-G ensemble since the short 400 MD time unit measurement runs are insufficient to adequately sample the topological charge, as evidenced by the time evolutions plotted in Appendix \ref{appendix:more_repro_plots}.}
\label{tab:repro_test_basic_observables}
\end{table}

\subsection{Low Energy Spectra}

In Table \ref{tab:repro_test_spectra} we list results for the pion, kaon, Omega baryon, and residual masses, computed on the 16I ensemble. These calculations were performed using a measurement package previously introduced in Ref.~\cite{Blum:2014tka}, and based on the all-mode averaging (AMA) technique of Ref.~\cite{PhysRevD.88.094503}. Five \textit{exact} light quark propagators were computed per trajectory using a deflated, mixed-precision CG solver with 600 low-mode deflation vectors and a tight stopping residual $r = 10^{-8}$, while \textit{sloppy} propagators were computed for all time slices using a reduced stopping residual $r = 10^{-4}$. Strange quark propagators were computed with the tight residual $r = 10^{-8}$ for all time slices using ordinary CG with no deflation. AMA correlation functions were then computed by time-translational averaging of the sloppy propagators, using the available exact propagators to correct for bias. The light quark propagators were computed using Coulomb gauge-fixed wall (W) sources, with either local (L) or wall sinks; the strange quark propagators were computed using Coulomb gauge-fixed wall or $Z_{3}$ box ($Z_{3}B$) sources, and in both cases local sinks.

The pion and kaon masses were extracted by fitting to the asymptotic, large Euclidean time limit of the respective two-point correlation function,
\begin{equation}
\langle 0 | \overline{\mathscr{O}}(t) \mathscr{O}(0) | 0 \rangle \stackrel{t \rightarrow \infty}{\simeq} \frac{\langle 0 | \overline{\mathscr{O}}(t) | X \rangle \langle X | \mathscr{O}(0) | 0 \rangle}{2 m_{X} V} \left( e^{-m_{X} t} \pm e^{-m_{X} \left( T - t \right)} \right),
\end{equation}
where $\mathscr{O}$ denotes the choice of interpolating operator, $X \in \{ \pi, K \}$ is the ground state to which $\mathscr{O}$ couples, and $V$ and $T$ are the spatial volume and temporal extent of the lattice, respectively. In particular, we performed simultaneous fits to the $\langle PP^{LW} \rangle$, $\langle PP^{WW} \rangle$, and $\langle AP^{LW} \rangle$ correlators, with $P(x) = \overline{\psi}(x) \gamma_{5} \psi(x)$ and $A(x) = \overline{\psi}(x) \gamma_{5} \gamma_{4} \psi(x)$, and the first (second) superscript denotes the sink (source) type. The Omega baryon mass was extracted from the two-point correlation function
\begin{equation}
C_{\Omega \Omega}^{s_{1} s_{2}}(t) = \sum_{i=1}^{3} \sum_{\vec{x}} \langle 0 | \overline{\mathscr{O}}_{\Omega}^{s_{1}}(\vec{x},t)_{i} \mathscr{O}_{\Omega}^{s_{2}}(0)_{i} | 0 \rangle
\end{equation}
with the interpolating operator
\begin{equation}
\mathscr{O}_{\Omega}(x)_{i} = \varepsilon_{abc} \left( s_{a}^{\dagger}(x) C \gamma_{i} s_{b}(x) \right) s_{c}(x),
\end{equation}
$s_{1} = L$ and $s_{2} \in \{ W, Z_{3}B \}$. The correlators were then projected onto the positive parity component
\begin{equation}
P_{+} C_{\Omega \Omega}^{s_{1} s_{2}} = \frac{1}{4} \mathrm{Tr} \left[ \frac{1}{2} \left( 1 + \gamma_{4} \right) C_{\Omega \Omega}^{s_{1} s_{2}} \right]
\end{equation}
and simultaneously fit to double exponential ans\"{a}tze with common mass terms
\begin{equation}
C_{\Omega \Omega}^{s_{1} s_{2}}(t) = (Z_{1})_{\Omega \Omega}^{s_{1} s_{2}} e^{-m_{\Omega} t} + (Z_{2})_{\Omega \Omega}^{s_{1} s_{2}} e^{-m_{\Omega}' t}.
\end{equation}
Finally, the residual mass was determined by fitting the ratio
\begin{equation}
R(t) = \frac{\langle 0 | \sum_{\vec{x}} j_{5 q}^{a} (\vec{x},t) | \pi \rangle}{\langle 0 | \sum_{\vec{x}} j_{5}^{a} (\vec{x},t) | \pi \rangle}
\end{equation}
to a constant, where $j_{5q}^{a}$ is the five-dimensional pseudoscalar density evaluated at the midpoint of the fifth dimension, and $j_{a}^{5}$ is the physical pseudoscalar density constructed from the surface fields.

\begin{table}[!h]
\centering
\begin{tabular}{c|cc}
	\hline
	\hline
	\rule{0cm}{0.4cm}& \multicolumn{2}{|c}{\ul{16I}} \\
	Observable & RHMC & EOFA \\
	\hline
	\rule{0cm}{0.4cm}$a m_{\pi}$ & 0.2424(11) & 0.2425(8) \\
	$a m_{K}$ & 0.3252(11) & 0.3253(7) \\
	$a m_{\Omega}$ & 1.003(15) & 0.994(11) \\
	$a m_{\rm res}'(m_{l})$ & 0.0030558(80) & 0.0030523(78) \\
	\hline
	\hline
\end{tabular}
\caption{Low energy spectrum on the 16I ensemble computed from 100 independent measurements beginning with MD trajectory 500 and separated by 10 MD time units. Prior to fitting the correlation functions were binned over groups of 5 measurements. Corresponding effective mass plots can be found in Appendix \ref{appendix:more_repro_plots}.}
\label{tab:repro_test_spectra}
\end{table}

In addition, we have also measured the ground state pion energy, kaon mass, and residual mass on the 16I-G and 16ID-G ensembles. While the ground state of the kaon is at rest, the ground state of the pion has nonzero momentum $\vec{p}_{100} = (\pm \pi/L, 0, 0 )$ on the 16I-G ensemble and $\vec{p}_{111} = (\pm \pi/L, \pm \pi/L, \pm \pi/L )$ on the 16ID-G ensemble due to the boundary conditions. These calculations make use of an extension of the AMA measurement package described above to $G$-parity ensembles; as discussed in Ref.~\cite{Kelly:2013ana}, this requires the inclusion of additional diagrams that are generated by the mixing of quark flavors at the lattice boundaries through the $G$-parity operation. We measure on 51 configurations of the 16I-G ensemble, beginning with trajectory 500 and with a separation of 40 MD time units, and use sloppy and exact CG stopping tolerances of $10^{-4}$ and $10^{-10}$, respectively, with a single exact solve per trajectory. We likewise measure on 21 configurations of the 16ID-G ensemble, beginning with trajectory 500 (960) for the RHMC (EOFA) ensemble and separated by 20 MD time units, and use the same AMA setup. We perform no additional binning for either ensemble since the separations between consecutive measurements are already comparable to the bin sizes used to compute the plaquette and quark condensates.

\begin{table}[!h]
\centering
\begin{tabular}{c|cc|cc}
	\hline
	\hline
	\rule{0cm}{0.4cm}& \multicolumn{2}{|c}{\ul{16I-G}} & \multicolumn{2}{|c}{\ul{16ID-G}} \\
	Observable & RHMC & EOFA & RHMC & EOFA \\
	\hline
	\rule{0cm}{0.4cm}$a E_{\pi}$ & 0.3175(43) & 0.3097(48) & 0.4457(101) & 0.4614(72) \\
	\rule{0cm}{0.4cm}$a E_{\pi}^{\rm pred}$ & 0.31197(4) & 0.31207(4) & --- & --- \\
	$a m_{K}$ & 0.3271(22) & 0.3272(28) & 0.4343(34) & 0.4382(24) \\
	$a m_{\rm res}'(m_{l})$ & 0.003140(90) & 0.003054(86) & 0.00919(14) & 0.00952(13) \\
	\hline
	\hline
\end{tabular}
\caption{Low energy spectra on the 16I-G and 16ID-G ensembles computed from 51 and 21 measurements, respectively. On the 16I-G ensemble we also predict the ground state pion energy using the fitted $a m_{\pi}$ on the 16I ensemble and the continuum dispersion relation $a E_{\pi}^{\rm pred} = \sqrt{ \left( a m_{\pi} \right)^{2} + \left( a \pi / L \right)^{2} }$. Corresponding effective mass plots can be found in Appendix \ref{appendix:more_repro_plots}.}
\label{tab:repro_test_spectra_gparity}
\end{table}

\subsection{Pseudofermion Forces on the 16I Ensemble}

TWQCD has observed that the average EOFA pseudofermion force is roughly half the size of the corresponding average RHMC pseudofermion force for a particular dynamical $N_{f} = 1$ QCD simulation with domain wall quarks performed in Ref.~\cite{Chen:2014hyy}. Following this observation, we examine the forces on the RHMC and EOFA variants of the 16I ensemble. We define a norm on the space of $\mathfrak{su}(3)$-valued pseudofermion force matrices $F_{\mu}^{a}(x) \equiv \partial_{x,\mu}^{a} S(U)$ by
\begin{equation}
\big\Vert F_{\mu}(x) \big\Vert \, \equiv \left[ \sum_{a} F_{\mu}^{a}(x) F_{\mu}^{a}(x) \right]^{1/2},
\end{equation}
and consider two measures of the force associated with a given configuration of the gauge field: the first is the RMS force
\begin{equation}
\label{eqn:L2_force}
F_{\rm RMS} \equiv \frac{1}{4 V} \left[ \sum_{x,\mu} \big\Vert F_{\mu}(x) \big\Vert^{2} \right]^{1/2}
\end{equation}
 and the second is the maximum force
\begin{equation}
\label{eqn:Linf_force}
F_{\rm max} \equiv \max_{x,\mu} \big\Vert F_{\mu}(x) \big\Vert,
\end{equation}
in both cases taken over all lattice sites and link directions. While we expect Equation \eqref{eqn:Linf_force} to be a more pertinent definition in the context of HMC simulations --- we have empirically found that acceptance is controlled by the size of $F_{\rm max}$ --- both $F_{\rm RMS}$ and $F_{\rm max}$ are, \textit{a priori}, reasonable measures of the pseudofermion force. 

\begin{figure}[!h]
\centering
\subfloat{\includegraphics[width=0.48\linewidth]{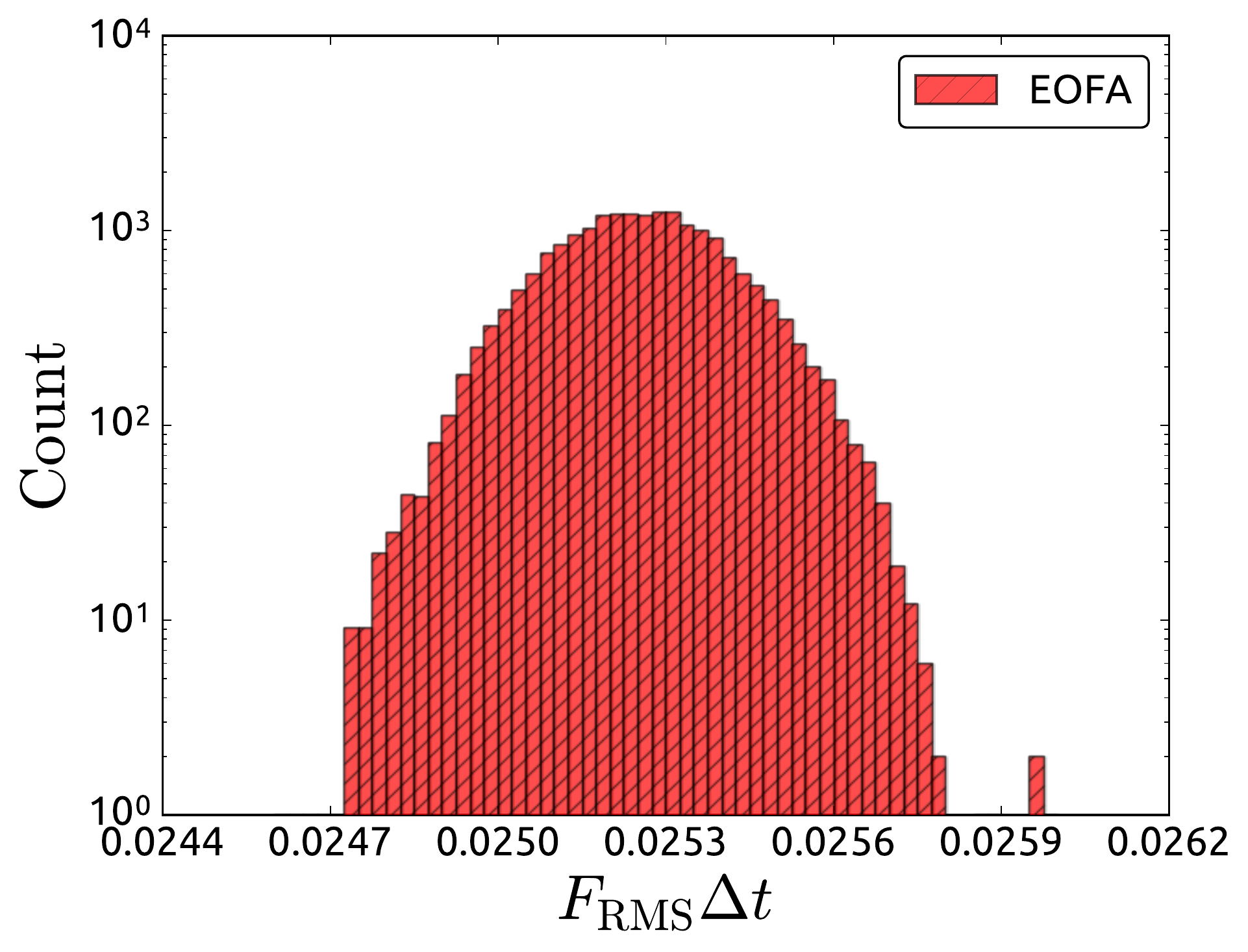}} \quad
\subfloat{\includegraphics[width=0.48\linewidth]{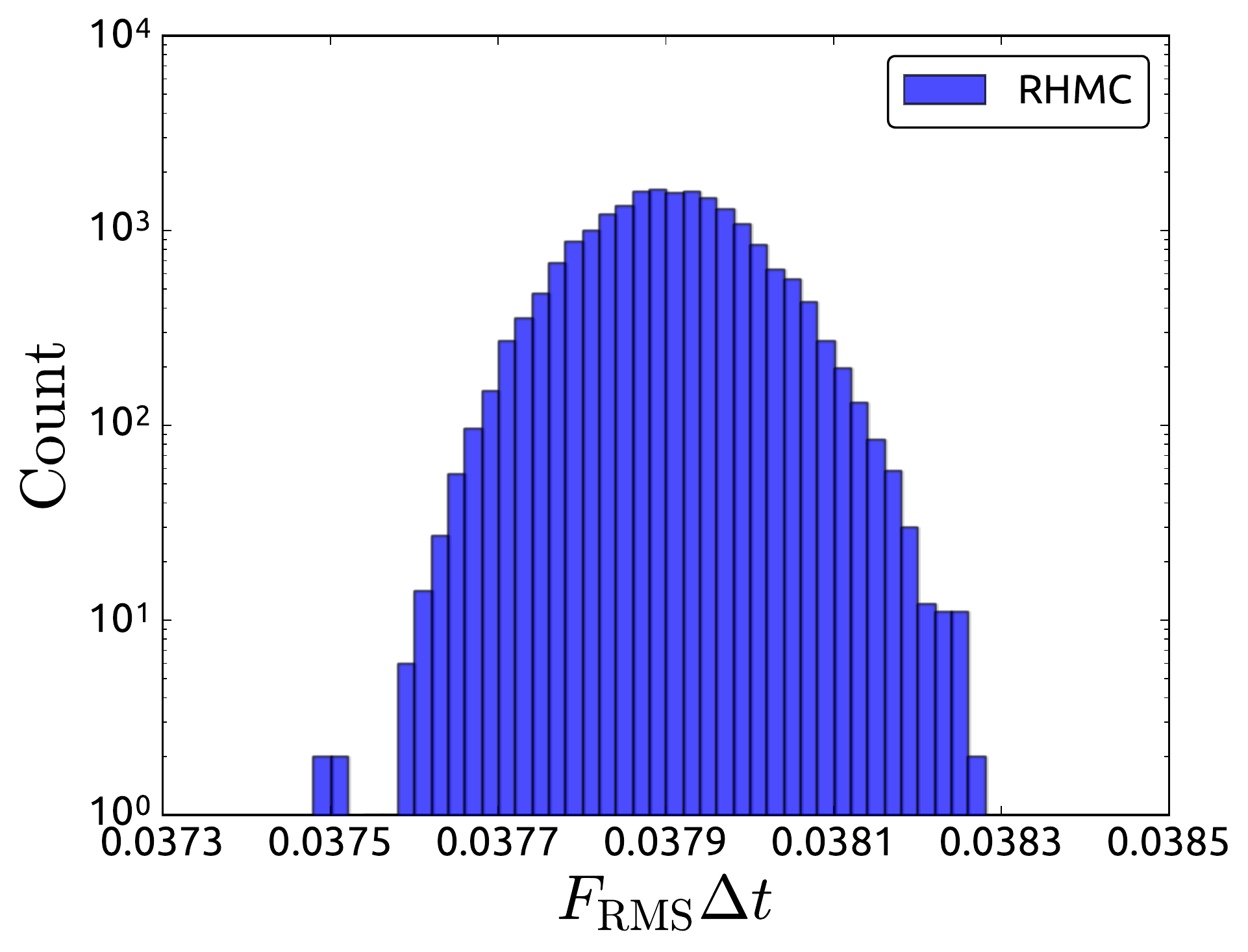}} \\
\subfloat{\includegraphics[width=0.5\linewidth]{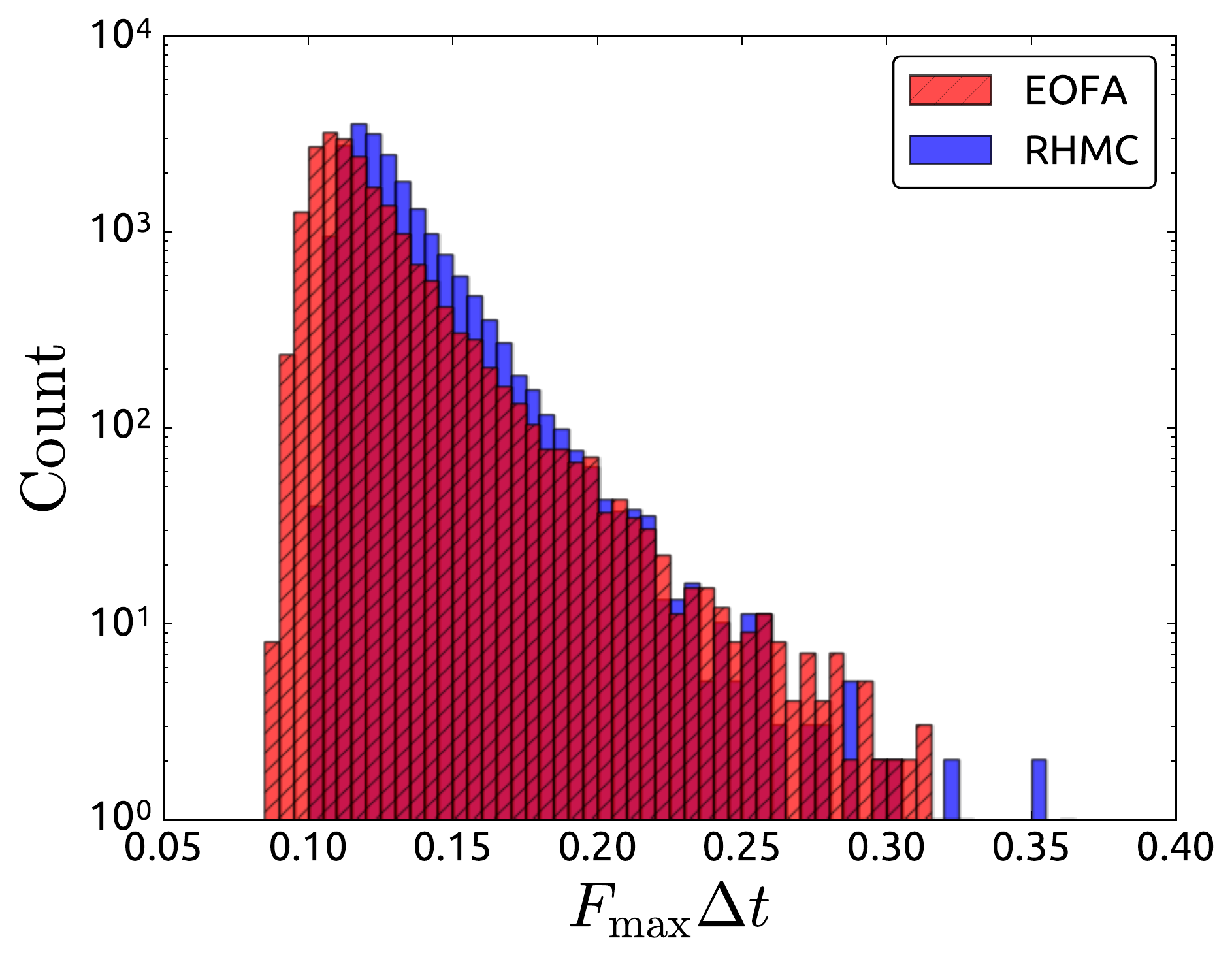}}
\caption{Histograms of the RMS and maximum pseudofermion forces associated with force evaluations falling between trajectories 500 and 1500 of the 16I HMC evolutions. $F_{\rm RMS}$ and $F_{\rm max}$ are defined by Equations \eqref{eqn:L2_force} and \eqref{eqn:Linf_force}, respectively. $\Delta t$ is the step size used to integrate the pseudofermion force contributions to the HMC evolution.}
\label{fig:16I_force_eofa_rhmc}
\end{figure}

In Figure \ref{fig:16I_force_eofa_rhmc} we compare histograms of $F_{\rm RMS}$ and $F_{\rm max}$ between the RHMC and EOFA 16I ensembles. Each data point corresponds to a single evaluation of the pseudofermion force falling between MD trajectories 500 and 1500. We find that comparing the relative sizes of the RHMC and EOFA forces is highly dependent on whether one chooses $F_{\rm RMS}$ or $F_{\rm max}$; the mean EOFA $F_{\rm RMS}$ is roughly 30\% smaller than the mean RHMC $F_{\rm RMS}$, but the distributions of $F_{\rm max}$ are nearly indistinguishable. This observation suggests that while the EOFA force distribution may have a smaller mean than the RHMC force distribution, the EOFA distribution also likely has longer tails, such that the largest forces have similar magnitudes. Since we expect the magnitude of the largest forces to correlate more strongly with the efficiency of the integrator than the magnitude of the average forces, as we have argued above, we interpret these results as suggesting that the optimal step size for an EOFA evolution should be similar to that of an RHMC simulation with the same mass parameters, even if the average force is somewhat smaller.

\begin{figure}[!h]
\centering
\subfloat{\includegraphics[width=0.48\linewidth]{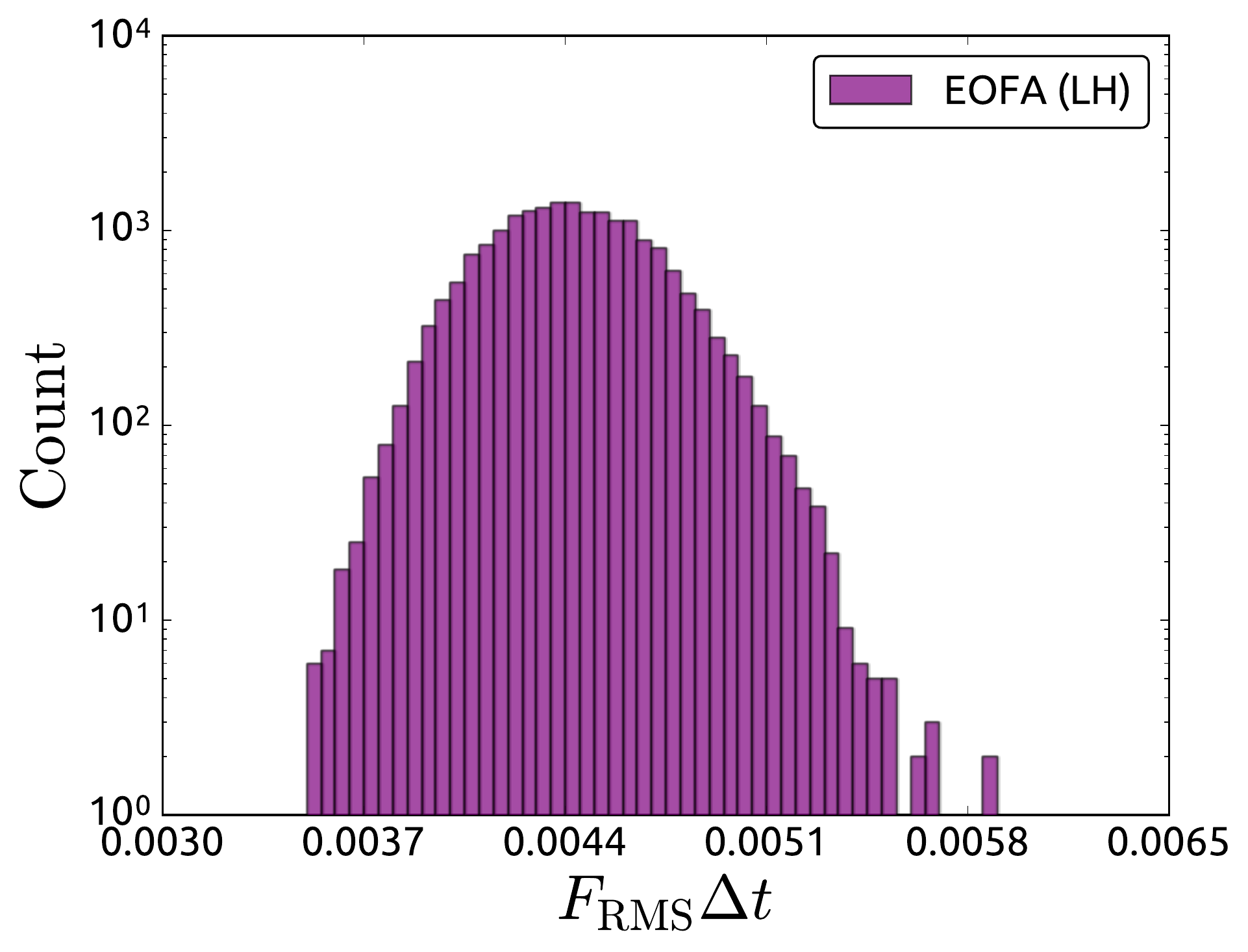}} \quad
\subfloat{\includegraphics[width=0.48\linewidth]{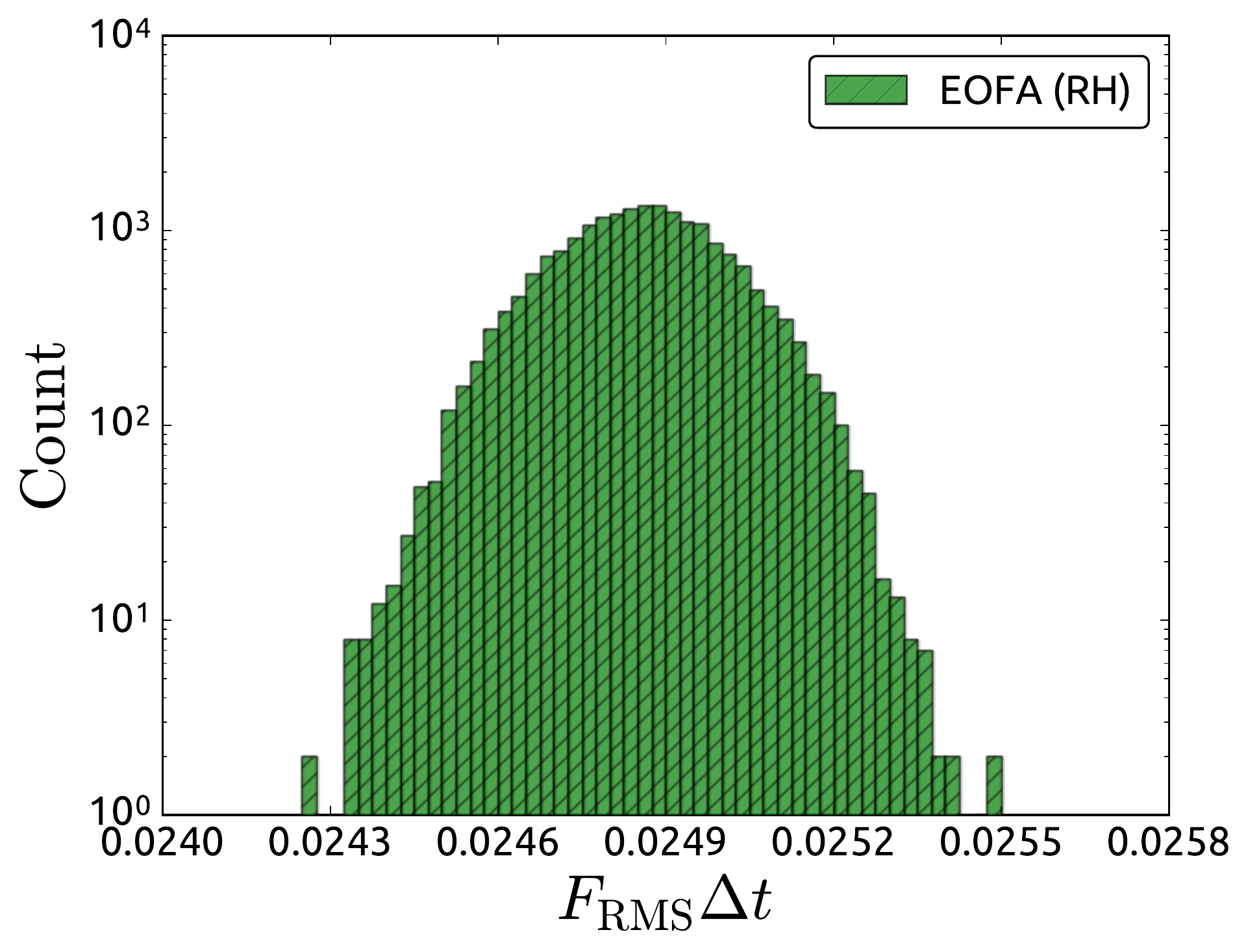}} \\
\subfloat{\includegraphics[width=0.5\linewidth]{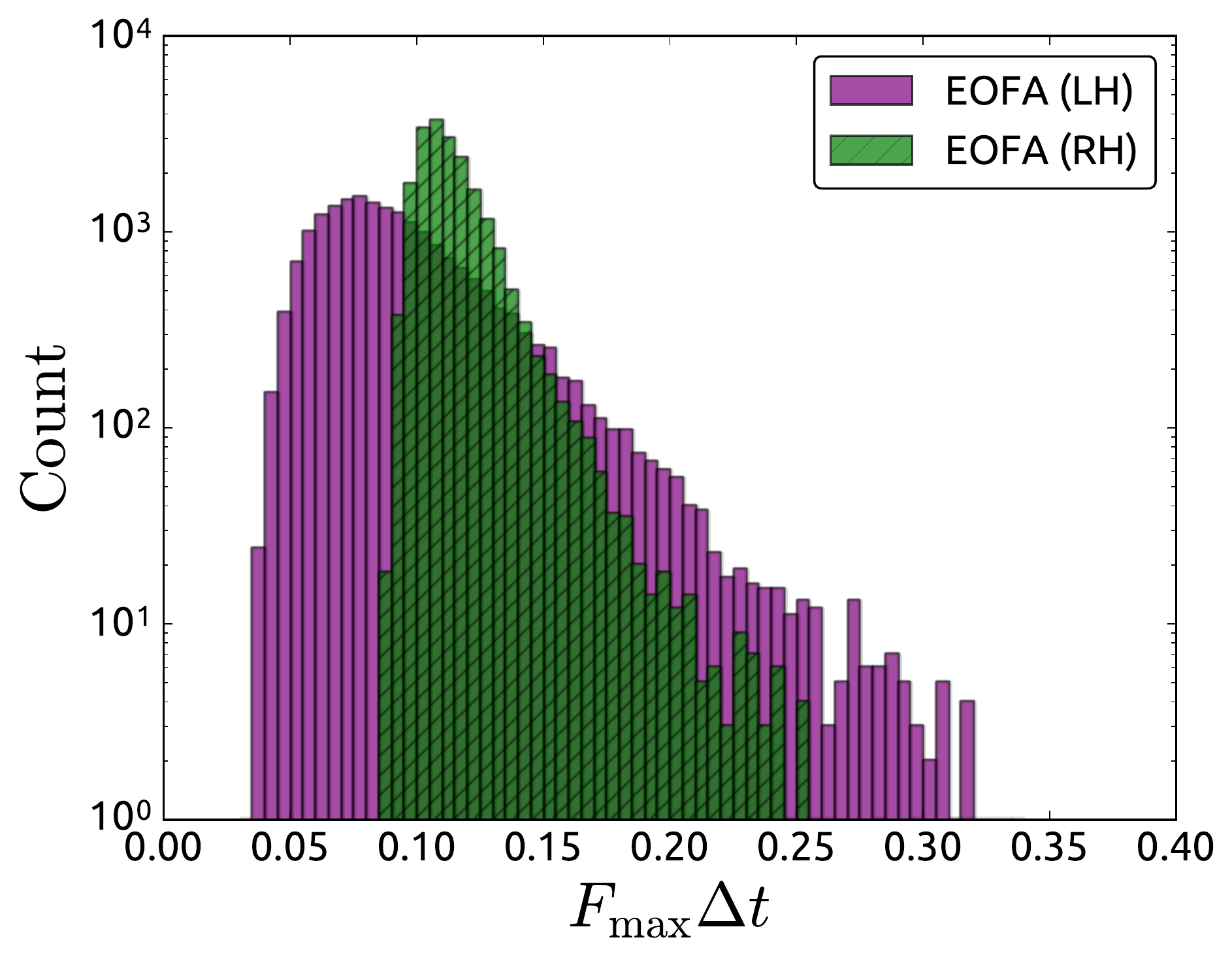}}
\caption{Histograms of the RMS and maximum pseudofermion forces associated with the left-handed and right-handed components of the pseudofermion field in Eqn.~\eqref{eqn:eofa_force}.}
\label{fig:16I_force_LH_RH}
\end{figure}

TWQCD has also observed a large hierarchy of scales in the pseudofermion forces associated with each of the two terms in Eqn.~\eqref{eqn:eofa_force}; in Ref.~\cite{Chen:2014bbc} they find that the average force associated with the first term --- involving the left-handed component of the pseudofermion field --- is more than an order of magnitude smaller than the average force associated with the second term --- involving the right-handed component --- for two different dynamical QCD simulations. They exploit this observation with a Sexton-Weingarten integration scheme, integrating the first term with a larger time step than the second, and find increased efficiency in their simulations. In Figure \ref{fig:16I_force_LH_RH} we compare histograms of the RMS and maximum left-handed and right-handed forces from 1000 thermalized configurations of the 16I EOFA ensemble. Our conclusions are analogous to the comparison between the EOFA and RHMC forces: if one considers $F_{\rm RMS}$ the left-handed force contribution is indeed substantially smaller than the right-handed force contribution, but if one instead considers $F_{\rm max}$ the force distributions are very similar in both magnitude and shape. Based on the latter observation we leave both terms in Equation \eqref{eqn:eofa_force} on the same time step in our large-scale EOFA simulations.

We also note that in these small volume test runs we have not considered applying the Hasenbusch mass preconditioning technique \cite{Hasenbusch:2002ai} to the EOFA formalism. Introducing a set of Hasenbusch masses $\{m'_{i}\}_{i=1}^{N}$, with $m_{1} < m'_{i} < m_{2}$, we can write the fermion determinant as
\begin{equation}
\label{eqn:hasenbusch_precond}
\det \left( \frac{\mathscr{D}(m_{1})}{\mathscr{D}(m_{2})} \right) = \det \left( \frac{\mathscr{D}(m_{1})}{\mathscr{D}(m'_{1})} \right) \left[ \prod_{i=1}^{N-1} \det \left( \frac{\mathscr{D}(m'_{i})}{\mathscr{D}(m'_{i+1})} \right) \right] \det \left( \frac{\mathscr{D}(m'_{N})}{\mathscr{D}(m_{2})} \right).
\end{equation}
While the left-hand side can be simulated using a single pseudofermion field, the associated forces can be large if $m_{1} \ll m_{2}$, requiring a small step size to maintain reasonable acceptance. The right-hand side, in contrast, involves $N+1$ independent pseudofermion fields, but with possibly substantially reduced forces, allowing larger step sizes to be used. For light $m_{1}$ one typically observes that the gain from increasing the step size offsets the cost of simulating extra heavy flavors, leading to a more efficient simulation. In Section \ref{sec:24ID_32ID-G} we demonstrate that Hasenbusch preconditioning allows for a substantial speed-up in the context of the 32ID-G ensemble. We also note that in addition to reducing the size of the pseudofermion forces, the Hasenbusch technique preconditions the EOFA force in the sense that the size hierarchy between the left-handed and right-handed force contributions to a single determinant disappears in the limit $m_{i}' \to m_{i+1}'$. In practice, we find that the mass preconditioned simulation has comparable left-handed and right-handed force contributions even in the RMS sense.
\FloatBarrier

\section{Optimization and Tuning}
\label{sec:optimizations}
In this section we discuss preconditioning and algorithmic techniques which reduce the cost of EOFA simulations. In some cases these are extensions of well-known lattice techniques to the EOFA formalism, while in other cases they are specific to EOFA. We illustrate these techniques using bechmark tests computed with the physical quark mass, M\"{o}bius DWF 24ID ensemble, and report timing results for code written in the Columbia Physics System (CPS) and running on 256-node or 512-node Blue Gene/Q partitions.

\subsection{Inversions of $\mathscr{D}_{\rm EOFA}$}
\label{subsec:inversion_tuning}

Since the majority of the computational effort in an HMC simulation is associated with repeatedly inverting the Dirac operator, techniques to more efficiently apply the Dirac operator or to otherwise accelerate these inversions can have a dramatic impact on the overall efficiency of the integrator. To address the former, we make use of the BAGEL assembler generation library \cite{Boyle20092739} to produce highly optimized kernels and fermion solvers for the Blue Gene/Q hardware. To address the latter, we make use of multiple preconditioning techniques, as well as a mixed precision defect correction CG solver.

The first preconditioning technique we apply --- ``even-odd'' or ``red-black'' preconditioning --- is well-known in the lattice QCD community. Lattice sites are labeled as even if $\left( x + y + z + t \right) \equiv 0 \pmod 2$, or odd if $\left( x + y + z + t \right) \equiv 1 \pmod 2$, inducing a $2 \times 2$ block structure on fermion operators
\begin{equation}
M = \left( \begin{array}{cc} M_{ee} & M_{eo} \\ M_{oe} & M_{oo} \end{array} \right).
\end{equation}
Standard tricks can then be used to relate the linear system $M \psi = \phi$ to a better conditioned linear system involving only the odd sub-lattice; this preconditioned system is substantially cheaper to invert since the size of the problem has been halved. After inverting on the odd sub-lattice, the even component of $\psi$ can also be recovered at modest cost, without ever needing to explicitly invert on the even sub-lattice. The details of this construction, and its extension to EOFA, are described in Appendix \ref{appendix:eo_preconditioning}. 

The second preconditioning technique we apply --- Cayley-form preconditioning --- is unique to EOFA, and was introduced in Ref.~\cite{Murphy:2016ywx}. The generic linear system one needs to solve in the context of EOFA has the form
\begin{equation}
\label{eqn:generic_eofa_system}
\Big( H(m_{1}) + \beta \Delta_{\pm}(m_{2},m_{3}) P_{\pm} \Big) \psi = \phi,
\end{equation}
where $H = \gamma_{5} R_{5} \mathscr{D}_{\rm EOFA}$. For M\"{o}bius domain wall fermions $\mathscr{D}_{\rm EOFA}$ is dense in $ss'$, and thus considerably more expensive to invert than $\mathscr{D}_{\rm DWF}$, which has a tridiagonal $ss'$ stencil, in terms of wall clock time. However, Eqn.~\eqref{eqn:eofa_dop} suggests that Eqn.~\eqref{eqn:generic_eofa_system} can be related to an equivalent system in terms of $\mathscr{D}_{\rm DWF}$ by using $\widetilde{\mathscr{D}}^{-1}$ as a preconditioner. We elaborate on the mathematical details in Appendix \ref{appendix:preconditioned_mobius}, and, in particular, demonstrate that $\Delta_{\pm} \widetilde{\mathscr{D}}$ has a relatively simple, rank-one form, allowing for substantially more efficient EOFA inversions --- even when $\beta \ne 0$ --- by working with the preconditioned system. This technique also has the advantage that it allows for EOFA simulations which re-use existing high-performance code for applying $\mathscr{D}_{\rm DWF}$ with little modification.

Finally, we use a restarted, mixed precision defect correction solver to perform the conjugate gradient inversions of the fully preconditioned EOFA system. For memory bandwidth-limited calculations --- such as applying the Dirac operator --- single precision computations can be performed at approximately half the cost of full double precision computations. In the defect correction approach to mixed precision CG, the following algorithm is used:
\begin{enumerate}
	\item Solve the Dirac equation in single precision arithmetic using a reduced stopping tolerance (typically $10^{-4}$ or $10^{-5}$).
	\item Compute the current residual using the (single precision) solution in full double precision arithmetic.
	\item If the desired final tolerance (typically $10^{-8}$ or smaller) has been reached, stop. Otherwise, return to step 1, using the residual vector computed in step 2 as the new CG source.
\end{enumerate}
We observe that this algorithm outperforms straight double precision CG by approximately a factor of 2 --- as one would expect if the calculation is truly memory bandwidth-limited --- provided the local lattice volume on each node is sufficiently large to avoid communications bottlenecks.

\begin{figure}[!h]
\centering
\includegraphics[width=0.95\linewidth]{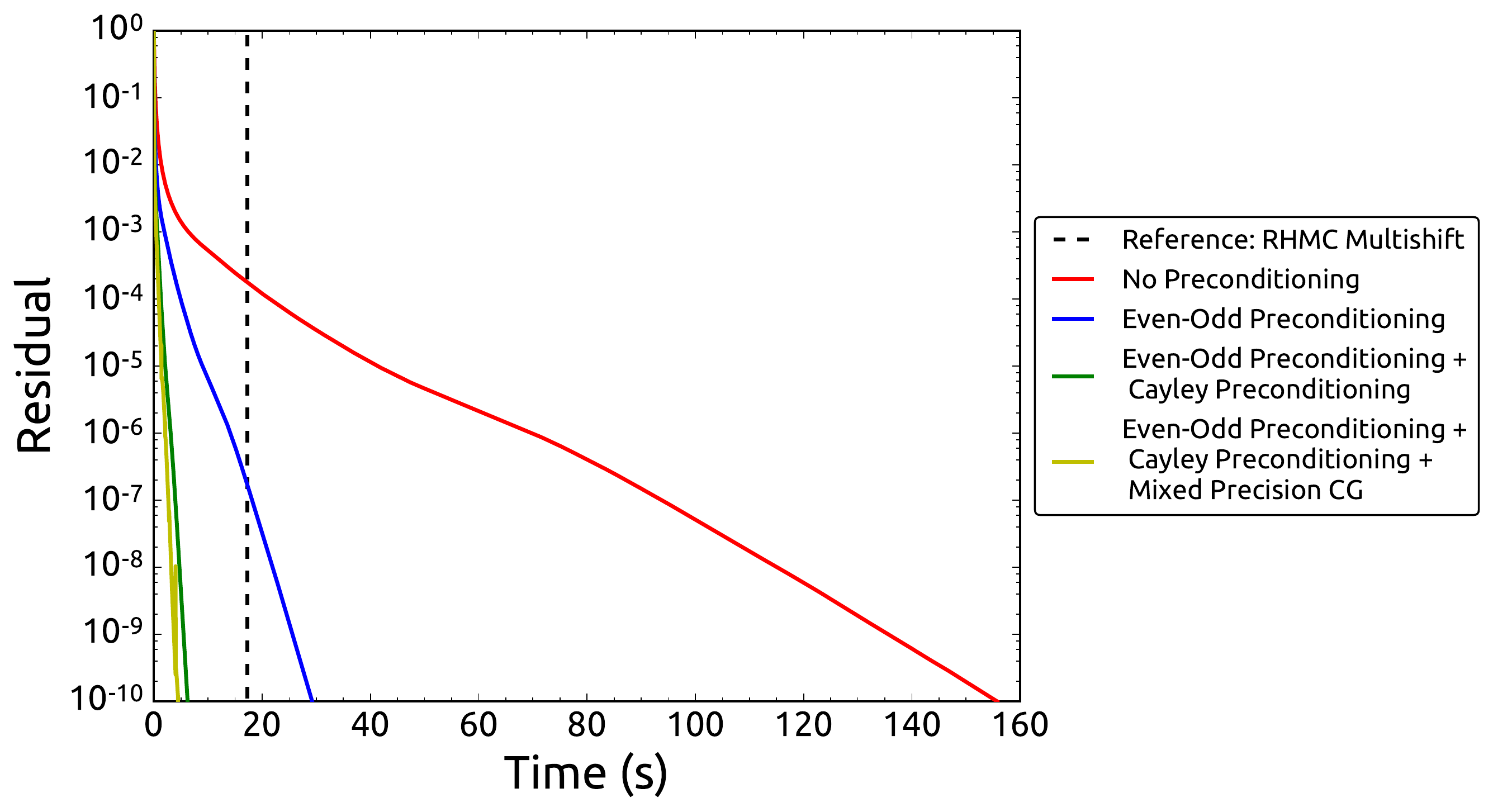}
\caption{Wall clock time required to solve Eqn.~\eqref{eqn:generic_eofa_system} to a stopping tolerance of $10^{-10}$ at the physical strange quark mass on the 24ID ensemble, as the preconditioning and algorithmic refinements discussed in the text are introduced sequentially. The dashed vertical line corresponds to the time required to apply $( \mathscr{D}_{\rm DWF}^{\dagger} \mathscr{D}_{\rm DWF})^{-1/2}$ by solving Equation \eqref{eqn:rhmc_generic_system} using the high-performance implementations of even-odd preconditioned $\mathscr{D}_{\rm DWF}$ and multishift CG in the BAGEL library.}
\label{fig:inversion_benchmark}
\end{figure}

In Figure \ref{fig:inversion_benchmark} we plot the CG residual as a function of the wall clock running time of the inverter for a series of benchmark inversions of Equation \eqref{eqn:generic_eofa_system} on the 24ID ensemble. These benchmarks show the inverter performance as we sequentially introduce even-odd preconditioning, Cayley-form preconditioning, and finally, mixed precision CG. We also plot the time required to solve the family of linear systems 
\begin{equation}
\label{eqn:rhmc_generic_system}
\left( \mathscr{D}^{\dagger}_{\rm DWF} \mathscr{D}_{\rm DWF} + \beta_{k} \right) \psi = \phi
\end{equation}
using multishift CG for the same set of poles $\{ \beta_{k} \}$ used in the rational approximation to $x^{-1/2}$ in the RHMC evolution that generated the 24ID ensemble. This allows a baseline estimate of the cost of evaluating the EOFA Hamiltonian or pseudofermion force against the cost of evaluating the RHMC Hamiltonian or pseudofermion force at the same quark mass. We observe a factor of 3.9 speed-up for fully preconditioned EOFA over the even-odd preconditioned RHMC system. In both cases the underlying operator being inverted is $\mathscr{D}_{\rm DWF}$; the slower RHMC benchmark demonstrates the overhead associated with multishift CG relative to solving a single system with standard CG, both due to the inability to fully utilize mixed precision methods and due to the additional linear algebra required at each iteration.

\subsection{Heatbath}
\label{subsec:heatbath_tuning}

\begin{figure}[!h]
\centering
\subfloat{\includegraphics[width=0.6\linewidth]{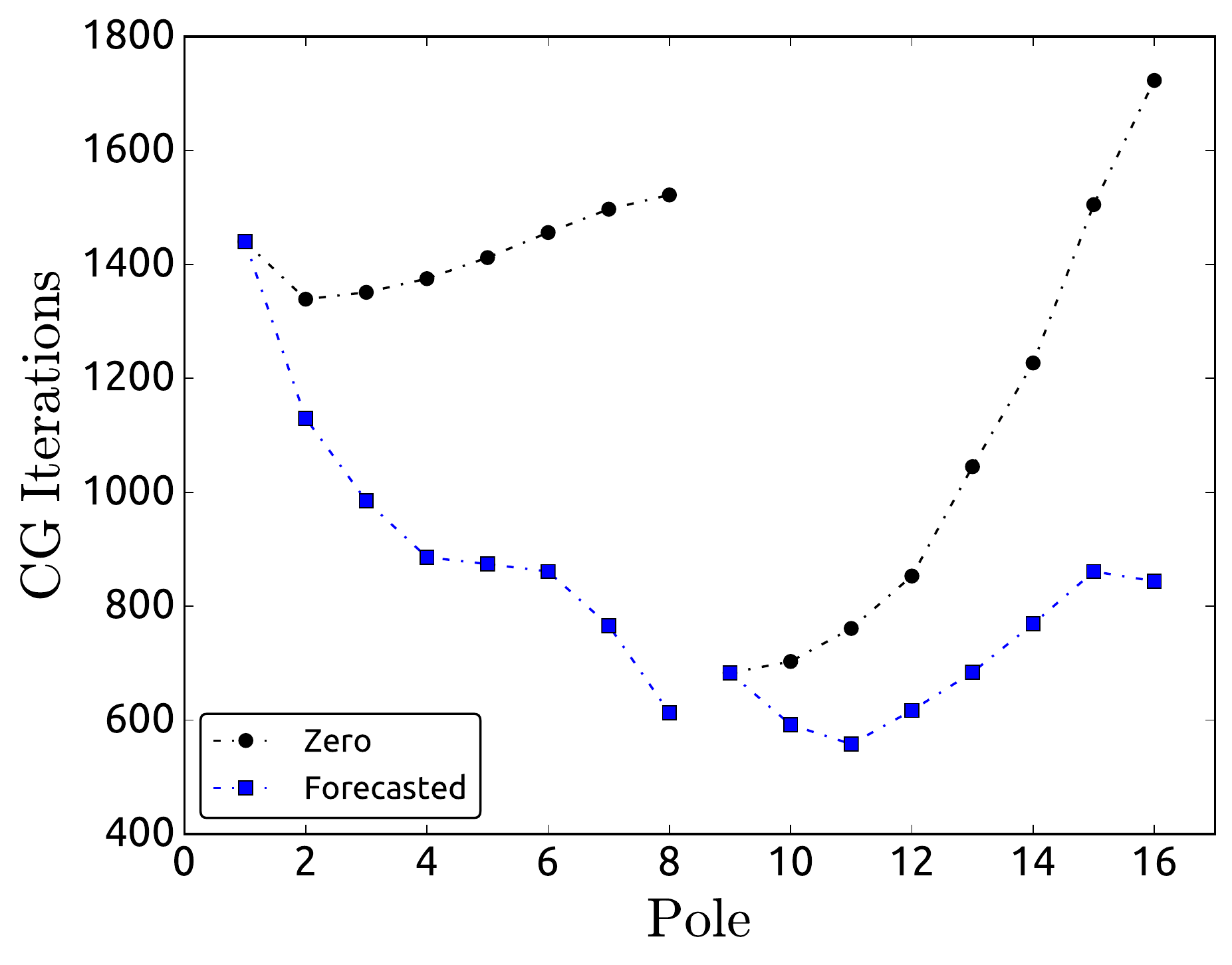}}
\caption{CG iterations required to invert Equation \eqref{eqn:generic_eofa_system} for each of the 16 values of $\beta$ entering into a rational approximation of $\mathcal{M}_{\rm EOFA}^{-1/2}$ with 8 poles on the 24ID ensemble. The first 8 poles ($\beta = -\gamma_{l}$) are associated with the first (LH) term in Equation \eqref{eqn:eofa_heatbath}, while the second 8 poles ($\beta = - \beta_{l} \gamma_{l}$) are associated with the second (RH) term. We find no improvement from using solutions to the LH system to forecast solutions to the RH system and vice-versa, since the Dirac operator being inverted in either case is evaluated with a different quark mass.}
\label{fig:heatbath_chrono_timing}
\end{figure}

Achieving the full performance improvement suggested by the inversion benchmarks in Section \ref{subsec:inversion_tuning} is complicated by the form of the EOFA heatbath, which is expected to be more expensive than the RHMC heatbath, even with efficient EOFA code. Applying $\mathcal{M}_{\rm EOFA}^{-1/2}$ (Eqn.~\eqref{eqn:eofa_heatbath}) requires two independent CG inversions per pole used in the rational approximation to $x^{-1/2}$, since multishift CG is not applicable: we use two algorithmic techniques to reduce this cost. The first is a forecasting technique initially proposed by Brower et al.~\cite{Brower:1995vx} in the context of more general HMC simulations, and later used successfully by TWQCD in the context of the EOFA heatbath \cite{Chen:2014hyy}. The idea is the following: given a set of solutions $\{ \psi_{k} \}_{k=1}^{N}$ to Equation \eqref{eqn:generic_eofa_system} for $N$ different poles $\{ \beta_{k} \}_{k=1}^{N}$, one can use the linear combination
\begin{equation}
\psi_{N+1} = \sum_{k=1}^{N} c_{k} \psi_{k}
\end{equation}
minimizing the functional
\begin{equation}
\label{eqn:cg_functional}
\Phi \left[ \psi \right] = \psi^{\dagger} \Big( H + \beta_{N+1} \Delta_{\pm} P_{\pm} \Big) \psi - \phi^{\dagger} \psi - \psi^{\dagger} \phi
\end{equation}
as the initial CG guess for the next inversion with pole $\beta_{N+1}$. The coefficients $c_{k}$ satisfy
\begin{equation}
\sum_{k=1}^{N} c_{k} \psi_{l}^{\dagger} \Big( H + \beta_{N+1} \Delta_{\pm} P_{\pm} \Big) \psi_{k} = \psi_{l}^{\dagger} \phi,
\end{equation}
and can be computed explicitly using \textit{e.g.} Gauss-Jordan elimination. Since Equation \eqref{eqn:cg_functional} is the same functional minimized by the conjugate gradient algorithm itself, accurate initial guesses can be computed for modest $N$ provided the $\{ \beta_{k} \}_{k=1}^{N+1}$ are similar in magnitude. In Figure \ref{fig:heatbath_chrono_timing} we test this forecasting technique using the 24ID ensemble and a rational approximation with 8 poles, and find that the iteration count required to solve Eqn.~\eqref{eqn:generic_eofa_system} to a tolerance of $10^{-10}$ is more than halved for the last few poles. 

The second technique we have used to accelerate the heatbath is motivated by observing that the coefficients entering into Equation \eqref{eqn:eofa_heatbath} span several orders of magnitude for a typical rational approximation to $x^{-1/2}$. We find typical values $k \alpha_{l} \gamma_{l}^{2} / \alpha_{0} \sim \mathcal{O}(10^{-3}-10^{-5})$, suggesting that the inversions can be performed with reduced stopping tolerances relative to the desired accuracy of $\mathcal{M}_{\rm EOFA}^{-1/2} \psi$, since the solution vectors are ultimately multiplied by small coefficients when the result is formed. We have explored the following simple optimization scheme to relax the stopping conditions for each pole:
\begin{enumerate}
\item Choose a desired tolerance for the heatbath, $\varepsilon_{\rm tol}$, where $\varepsilon$ is defined by Equation \eqref{eqn:heatbath_rel_err}.
\item Choose one of the inversions required to compute $\mathcal{M}_{\rm EOFA}^{-1/2}$ according to Equation \eqref{eqn:eofa_heatbath}, and relax the stopping tolerance until the overall error in the heatbath $\varepsilon$ reaches $\varepsilon_{\rm tol}$.
\item Iterate over each inversion until all stopping conditions have been tuned.
\end{enumerate}
We report results for the 24ID ensemble in Table \ref{tab:heatbath_tuning_24ID}. Using a rational approximation with 6 poles, and $\varepsilon_{\rm tol} = 10^{-10}$, we observe that the total heatbath time is more than halved while only slightly increasing the error. We have also checked that the final error and heatbath running time after tuning is insensitive to the exact order in which the stopping tolerances are tuned.

\begin{table}[!h]
\centering
\begin{tabular}{ccc}
\hline
\hline
\rule{0cm}{0.4cm}& $\varepsilon$ & Total Heatbath Time \\
Untuned & $1.52 \times 10^{-11}$ & 129.5 s \\
Tuned & $7.79 \times 10^{-11}$ & 68.9 s \\
\hline
\hline
\end{tabular}
\caption{The relative error ($\varepsilon$) and total running time for the EOFA heatbath on the 24ID ensemble before and after applying the tuning algorithm discussed in the text.}
\label{tab:heatbath_tuning_24ID} 
\end{table}  
\FloatBarrier

\section{Large-Scale EOFA Calculations}
\label{sec:24ID_32ID-G}
In this section we turn to two ongoing ensemble generation calculations currently being performed by the RBC/UKQCD collaboration. The first is a strong-coupling $N_{f} = 2+1$ $24^{3} \times 64 \times 24$ Iwasaki+DSDR lattice (24ID) intended for exploratory studies and calculations requiring high statistics \cite{coarse_ensembles}. The second (32ID) has been used for a first-principles calculation of the ratio of Standard Model $CP$-violation parameters $\epsilon'/\epsilon$ from $\Delta I = 1/2$ $K \rightarrow \pi \pi$ decays in Ref.~\cite{Bai:2015nea}. RBC/UKQCD is currently generating more gauge field configurations to reduce the statistical errors in the $\Delta I = 1/2$ decay amplitudes. Both ensembles have physical quark masses and large volumes, allowing for tests of the performance of EOFA in the context of state-of-the-art domain wall fermion calculations.

Tables \ref{tab:production_hmc_details} and \ref{tab:production_integrator_layouts} summarize the details of the integrator parameters and nesting for these evolutions. The ensembles labeled RHMC correspond to the evolutions of Ref.~\cite{coarse_ensembles} (24ID) and Ref.~\cite{Bai:2015nea} (32ID-G). For the ensembles marked EOFA, we have changed the strange quark (light quark) action to EOFA for the 24ID (32ID-G) ensemble and retuned the details of the integrator as described in the remainder of the section. For the 32ID-G ensemble --- where, due to the $G$-parity flavor doubling, the EOFA action naturally describes the degenerate light quark pair --- we have also switched from an Omelyan integrator to a force gradient integrator, and inserted additional Hasenbusch preconditioning determinants. 

\begin{table}[!h]
\centering
\begin{tabular}{cccccc}
\hline
\hline
\rule{0cm}{0.4cm}Ensemble & Integrator & $\delta \tau$ & $r_{\rm FG}$ & $r_{\rm MD}$ & $r_{\rm MC}$ \\
\hline
\rule{0cm}{0.4cm}24ID (RHMC) & Force Gradient QPQPQ & 0.0833 & $10^{-5}$ & $10^{-7}$ & $10^{-10}$ \\
24ID (EOFA) & Force Gradient QPQPQ & 0.0833 & $10^{-5}$ & $10^{-7}$ & $10^{-10}$ \\
\hline
\rule{0cm}{0.4cm}32ID-G (RHMC) & Omelyan ($\lambda = 0.22$) & 0.0625 & --- & $10^{-7}$ & $10^{-10}$ \\
32ID-G (EOFA) & Force Gradient QPQPQ & 0.1667 & $10^{-5}$ & $10^{-7}$ & $10^{-10}$ \\
\hline
\hline
\end{tabular}
\caption{Basic integrator and HMC details for the generation of the 24ID and 32ID-G ensembles. We denote the coarsest time step used to evolve the outermost level by $\delta \tau$, and the CG stopping tolerances used for the force gradient forecasting, molecular dynamics, and Monte Carlo steps by $r_{\rm FG}$, $r_{\rm MD}$, and $r_{\rm MC}$, respectively. We elaborate on the details of the integrator nesting in Table \ref{tab:production_integrator_layouts}.}
\label{tab:production_hmc_details}
\end{table}

\begin{table}[!h]
\centering
\resizebox{\linewidth}{!}{
\begin{tabular}{cccc}
\hline
\hline
\rule{0cm}{0.4cm}Ensemble & Level & Action & Update \\
\hline
\rule{0cm}{0.4cm}\multirow{4}{*}{24ID (RHMC)} & 1 & $\mathrm{RatQuo}_{1/2}(0.085,1.0)$ & 1:1 \\ 
& \multirow{2}{*}{2} & $\mathrm{Quo}(0.00107,0.00789)$ + $\mathrm{Quo}(0.00789,0.0291)$ + $\mathrm{Quo}(0.0291,0.095)$ + & \multirow{2}{*}{1:1} \\ 
& & $\mathrm{Quo}(0.095,0.3)$ + $\mathrm{Quo}(0.3,0.548)$ + $\mathrm{Quo}(0.548,1.0)$ & \\
& 3 & Gauge + DSDR & 1:1 \\ 
\hline
\rule{0cm}{0.4cm}\multirow{4}{*}{24ID (EOFA)} & 1 & $\mathrm{EOFA}(0.085,1.0)$ & 1:1 \\ 
& \multirow{2}{*}{2} & $\mathrm{Quo}(0.00107,0.00789)$ + $\mathrm{Quo}(0.00789,0.0291)$ + $\mathrm{Quo}(0.0291,0.095)$ + & \multirow{2}{*}{1:1} \\ 
& & $\mathrm{Quo}(0.095,0.3)$ + $\mathrm{Quo}(0.3,0.548)$ + $\mathrm{Quo}(0.548,1.0)$ & \\
& 3 & Gauge + DSDR & 1:1 \\
\hline
\rule{0cm}{0.4cm}\multirow{4}{*}{32ID-G (RHMC)} & 1 & $\mathrm{RatQuo}_{1/2}(0.0001,0.007)$ & 1:1 \\ 
& 2 & $\mathrm{RatQuo}_{1/2}(0.007,1.0)$ + $\mathrm{RatQuo}_{1/4}(0.045,1.0)$ & 1:2 \\ 
& 3 & DSDR & 1:2 \\
& 4 & Gauge & 1:1 \\ 
\hline
\rule{0cm}{0.4cm}\multirow{5}{*}{32ID-G (EOFA)} & \multirow{3}{*}{1} & $\mathrm{EOFA}(0.0001,0.0058)$ + $\mathrm{EOFA}(0.0058,0.0149)$ + $\mathrm{EOFA}(0.0149,0.059)$ + & \multirow{3}{*}{5:1} \\
& & $\mathrm{EOFA}(0.059,0.177)$ + $\mathrm{EOFA}(0.177,0.45)$ + & \\
& & $\mathrm{EOFA}(0.45,1.0)$ + $\mathrm{RatQuo}_{1/4}(0.045,1.0)$ & \\
& 2 & DSDR & 1:2 \\
& 3 & Gauge & 1:1 \\ 
\hline
\hline
\end{tabular}
}
\caption{Integrator layouts for the 24ID and 32ID-G ensembles. The notation A:B for the update scheme denotes the number of steps of the next innermost integrator level (A) per step of the current level (B).}
\label{tab:production_integrator_layouts}
\end{table}

\subsection{24ID Ensemble}

We use the 24ID ensemble as a straightforward benchmark of RHMC against an equivalent EOFA simulation to describe a physical heavy quark flavor. Here this is the strange quark, but $N_{f} = 2+1+1$ simulations with dynamical strange and charm quarks are another obvious target of EOFA. We make no serious attempt to retune the integrator after switching to EOFA beyond tuning the heatbath step with the following procedure:
\begin{enumerate}
\item Compute the largest and smallest eigenvalues of $\mathcal{M}_{\rm EOFA}$ (Eqn.~\eqref{eqn:eofa_action}) for a few thermalized configurations of the gauge field, and use these measurements to inform the bounds of the rational approximations to $x^{-1/2}$ constructed via the Remez algorithm.
\item Add poles to the rational approximation, with all CG stopping tolerances set to $r_{\rm MC}$, until $\varepsilon < r_{\rm MC}$ (Eqn.~\eqref{eqn:heatbath_rel_err}) is reached. 
\item With the rational approximation now fixed from step 2, tune the CG stopping tolerances corresponding to each pole, following the procedure described in Section \ref{subsec:heatbath_tuning}, and keeping $\varepsilon < r_{\rm MC}$.
\end{enumerate} 
After tuning the heatbath, we then ran a single trajectory of the RHMC evolution and the EOFA evolution on a 256-node Blue Gene/Q partition. For the EOFA ensemble, we compare two schemes. The first (``dense'') is a straightforward implementation of M\"{o}bius DWF as proposed in Ref.~\cite{Chen:2014hyy}: we invert Equation \eqref{eqn:generic_eofa_system} directly, where $H = \gamma_{5} R_{5} \mathscr{D}_{\rm EOFA}$ and the other dense 5D operators appearing in the EOFA action are listed explicitly in Appendix \ref{appendix:eofa_operators_mobius}. We also do not apply the final step in our heatbath tuning procedure, leaving all CG stopping tolerances in the heatbath fixed at $r_{\rm MC} = 10^{-10}$. In the second EOFA scheme (``preconditioned'') we fully tune the heatbath step and apply the Cayley-form preconditioning detailed in Appendix \ref{appendix:preconditioned_mobius} to inversions of Equation \eqref{eqn:generic_eofa_system}. Timing breakdowns for the strange quark part of the evolution are reported in Table \ref{tab:24ID_timing_breakdown}. 

\begin{table}[!ht]
\centering
\resizebox{\linewidth}{!}{
\begin{tabular}{c|cc|cc|cc}
\hline
\hline
& \multicolumn{2}{|c|}{\textbf{RHMC}} & \multicolumn{2}{|c|}{\textbf{EOFA} (Dense)} & \multicolumn{2}{|c}{\textbf{EOFA} (Preconditioned)} \\
Step & Time (s) & \% & Time (s) & \% & Time (s) & \% \\
\hline
\rule{0cm}{0.4cm}Heatbath & 42.6 & 2.7 & 340.6 & 15.1 & 68.9 & 15.5 \\
Force gradient integration (total) & 1485.6 & 94.8 & 1840.6 & 81.8 & 355.9 & 80.1 \\
Final Hamiltonian evaluation & 39.4 & 2.5 & 68.8 & 3.1 & 19.8 & 4.4 \\
\hline
\rule{0cm}{0.4cm}Total & 1567.6 & --- & 2250.0 & --- & 444.6 & --- \\
(Total RHMC) / Total & 1.0 & --- & 0.7 & --- & 3.5 & --- \\
\hline
\hline
\end{tabular}
}
\caption{Strange quark timings for a single MD trajectory of the 24ID ensemble on a 256-node Blue Gene/Q partition. We compare RHMC to EOFA with (``preconditioned'') and without (``dense'') Cayley-form preconditioning.}
\label{tab:24ID_timing_breakdown}
\end{table}

We observe that the dense EOFA formalism is actually somewhat slower than RHMC: the additional complexity of the EOFA heatbath, together with the more expensive inversions of the dense 5D operator $\mathscr{D}_{\rm EOFA}$, negate the expected performance gains from the simpler forms of the Hamiltonian and force evaluations. We emphasize, however, that we have made no attempt to retune the integrator details to optimize for EOFA. After introducing Cayley-form preconditioning --- so that we are inverting the tridiagonal operator $\mathscr{D}_{\rm DWF}$ rather than $\mathscr{D}_{\rm EOFA}$ when we solve Equation \eqref{eqn:generic_eofa_system} --- we find that EOFA outperforms RHMC by a significant factor of 3.5.

\subsection{32ID-G Ensemble}

One particularly promising feature of EOFA in the context of $G$-parity ensembles is the potential for aggressive Hasenbusch mass preconditioning of the light quark determinant; this makes the 32ID-G ensemble a particularly interesting case study since the EOFA formalism is used to describe a physical mass light quark pair. In Ref.~\cite{Bai:2015nea} the RBC/UKQCD collaboration observed that mass preconditioning is not particularly effective for the RHMC light quark determinant, since each molecular dynamics step requires one multishift inversion of $\mathscr{D}^{\dagger} \mathscr{D}$ evaluated at the numerator quark mass and two multishift inversions of $\mathscr{D}^{\dagger} \mathscr{D}$ evaluated at the denominator quark mass. The latter two solves become prohibitively expensive if many intermediate masses are introduced, negating the expected gain from integrating the preconditioned pseudofermion forces with larger step sizes. The EOFA force, on the other hand, is no more expensive to evaluate than the force associated with the standard quotient action (Eqn.~\eqref{eqn:quo_action}), so it is natural to expect better performance from Hasenbusch preconditioning. 

In Table \ref{tab:32ID-G_tuning_runs} we list details of the tuning runs we have used to explore potential schemes for evolving the 32ID-G ensemble with EOFA light quarks. We started by switching from an Omelyan integrator, for which the leading errors are $\mathcal{O}(\delta \tau^{2})$, to a force gradient integrator, for which the leading errors are $\mathcal{O}(\delta \tau^{4})$, and studied the effects of inserting mass preconditioning determinants one at a time (runs 1-7). We then identified two promising mass preconditioning schemes --- one with four intermediate masses (runs 8-10), and the other with five intermediate masses (runs 11-14) --- and continued tuning the step size, CG stopping conditions, and heatbath, to optimize the job time per trajectory and Monte Carlo acceptance. The initial RHMC scheme used in Ref.~\cite{Bai:2015nea} corresponds to run 1, and the final EOFA scheme we have adopted for our continuing ensemble generation corresponds to run 12. 

\begin{table}[!h]
\centering
\resizebox{0.95\linewidth}{!}{
	\begin{tabular}{cccccccc}
		\hline
		\hline
		\rule{0cm}{0.4cm}Run & Integrator Type & Light Hasenbusch Masses & $\Delta \tau$ & $r_{\rm MD}$ & $N_{\rm traj}$ & Acceptance & Efficiency \\
		\hline
		\rule{0cm}{0.4cm}\textbf{1} & \textbf{O} & \textbf{0.007} & \textbf{0.0625} & $\bm{10^{-8}}$ & \textbf{850} & \textbf{88\%} & \textbf{---} \\
		2 & O & --- & 0.0625 & $10^{-8}$ & 10 & 40\% & 1.2 \\
		3 & FG & 0.043 & 0.0625 & $10^{-8}$ & 10 & 100\% & 2.0 \\
		4 & FG & 0.018, 0.12 & 0.0625 & $10^{-8}$ & 10 & 100\% & 1.8 \\
		5 & FG & 0.0118, 0.0412, 0.23 & 0.0625 & $10^{-8}$ & 10 & 100\% & 1.7 \\
		6 & FG & 0.0075, 0.023, 0.11, 0.4 & 0.0625 & $10^{-8}$ & 10 & 100\% & 1.7 \\
		7 & FG & 0.0058, 0.0149, 0.059, 0.177, 0.45 & 0.0625 & $10^{-8}$ & 10 & 100\% & 1.5 \\
		\hline
		\rule{0cm}{0.4cm}8 & FG & 0.0103, 0.029, 0.12, 0.41 & 0.1000 & $10^{-6}$ & 15 & 67\% & 4.0 \\
		9 & FG & 0.0103, 0.029, 0.12, 0.41 & 0.1000 & $10^{-7}$ & 20 & 95\% & 3.0 \\
		10 & FG & 0.0103, 0.029, 0.12, 0.41 & 0.1667 & $10^{-7}$ & 20 & 75\% & 4.5 \\
		\hline
		\rule{0cm}{0.4cm}11 & FG & 0.0058, 0.0149, 0.059, 0.177, 0.45 & 0.1000 & $10^{-6}$ & 40 & 80\% & 3.0 \\
		\textbf{12} & \textbf{FG} & \textbf{0.0058, 0.0149, 0.059, 0.177, 0.45} & \textbf{0.1667} & $\mathbf{10^{-7}}$ & \textbf{850} & \textbf{93\%} & \textbf{4.2} \\
		13 & FG & 0.0058, 0.0149, 0.059, 0.177, 0.45 & 0.2000 & $10^{-7}$ & 60 & 65\% & 4.5 \\
		14 & FG & 0.0058, 0.0149, 0.059, 0.177, 0.45 & 0.2000 & $10^{-8}$ & 25 & 72\% & 3.9 \\
		\hline
		\hline
	\end{tabular}
}
\caption{HMC details for the production ensemble generation run (1) of Ref.~\cite{Bai:2015nea}, as well as 13 tuning runs after switching to EOFA light quarks (2-14). We use the following notation: ``O'' denotes the Omelyan integrator, ``FG'' denotes the force gradient integrator, ``$N_{\rm traj}$'' is the number of trajectories generated for the timing run, ``acceptance'' is the fraction of gauge field configurations which were accepted in the final Monte Carlo step, and ``efficiency'' is the ratio of the total job time per trajectory for the specified integration scheme to the total job time per trajectory of the scheme used in run 1. Entries in bold correspond to the original RHMC scheme (1) and the final, fully tuned EOFA scheme (12).}
\label{tab:32ID-G_tuning_runs}
\end{table}

We find, in practice, that Hasenbusch mass preconditioning is extremely effective for the EOFA light quark determinant. In addition to reducing the size of the pseudofermion force, we also observe that the largest eigenvalue of the EOFA action, Equation \eqref{eqn:eofa_action}, decreases rapidly as $m_{2} \rightarrow m_{1}$. As a consequence, the heatbath is also less expensive with Hasenbusch preconditioning, since, as we increase the number of intermediate masses, we can simultaneously decrease the range and number of poles entering into the rational approximation used for each determinant. Table \ref{tab:32ID-G_heatbath_vs_nlhsb} summarizes the measured spectral range, the heatbath error, and the total heatbath cost for each of the runs 2-7. For this ensemble the first Hasenbusch mass reduces the cost of the heatbath by more than a factor of two, and subsequent Hasenbusch masses essentially leave the cost fixed.

\begin{table}[!h]
\centering
\begin{tabular}{ccccccc}
	\hline
	\hline
	\rule{0cm}{0.4cm}$N_{\rm LHSB}$ & Mass Ratio & $\lambda_{\rm min}$ & $\lambda_{\rm max}$ & $N_{\rm poles}$ & $\varepsilon$ & $\Delta t_{\rm HB}$ (s) \\
	\hline
	\rule{0cm}{0.4cm}0 & 0.0001/1.0 & 1.0 & 1150 & 11 & $6.91 \times 10^{-11}$ & 5263.2 \\
	\hline
	\rule{0cm}{0.4cm}\multirow{2}{*}{1} & 0.0001/0.043 & 1.0 & 33.3 & 7 & $3.50 \times 10^{-11}$ & \multirow{2}{*}{2226.6} \\
	& 0.043/1.0 & 1.0 & 22.8 & 7 & $6.82 \times 10^{-12}$ & \\
	\hline
	\rule{0cm}{0.4cm}\multirow{3}{*}{2} & 0.0001/0.018 & 1.0 & 13.5 & 6 & $2.13 \times 10^{-11}$ & \multirow{3}{*}{2043.8} \\
	& 0.018/0.12 & 1.0 & 6.4 & 5 & $1.11 \times 10^{-11}$ & \\
	& 0.12/1.0 & 1.0 & 8.3 & 6 & $6.18 \times 10^{-12}$ & \\
	\hline
	\rule{0cm}{0.4cm}\multirow{4}{*}{3} & 0.0001/0.0118 & 1.0 & 8.9 & 6 & $6.08 \times 10^{-12}$ & \multirow{4}{*}{2307.8} \\
	& 0.0118/0.0412 & 1.0 & 3.3 & 4 & $4.09 \times 10^{-11}$ & \\
	& 0.0412/0.23 & 1.0 & 5.5 & 5 & $1.29 \times 10^{-11}$ & \\
	& 0.23/1.0 & 1.0 & 4.3 & 5 & $1.11 \times 10^{-11}$ & \\
	\hline
	\rule{0cm}{0.4cm}\multirow{5}{*}{4} & 0.0001/0.0075 & 1.0 & 5.9 & 5 & $9.63 \times 10^{-12}$ & \multirow{5}{*}{2080.7} \\
	& 0.0075/0.023 & 1.0 & 2.8 & 4 & $3.55 \times 10^{-11}$ & \\
	& 0.023/0.11 & 1.0 & 4.6 & 5 & $1.00 \times 10^{-11}$ & \\
	& 0.11/0.4 & 1.0 & 3.6 & 4 & $1.98 \times 10^{-11}$ & \\
	& 0.4/1.0 & 1.0 & 2.5 & 4 & $2.34 \times 10^{-11}$ & \\
	\hline
	\rule{0cm}{0.4cm}\multirow{6}{*}{5} & 0.0001/0.0058 & 1.0 & 4.7 & 5 & $1.11 \times 10^{-11}$ & \multirow{6}{*}{2289.0} \\
	& 0.0058/0.0149 & 1.0 & 2.3 & 4 & $1.64 \times 10^{-11}$ & \\
	& 0.0149/0.059 & 1.0 & 3.7 & 4 & $9.65 \times 10^{-11}$ & \\
	& 0.059/0.177 & 1.0 & 3.0 & 4 & $4.14 \times 10^{-11}$ & \\
	& 0.177/0.45 & 1.0 & 2.5 & 4 & $2.71 \times 10^{-11}$ & \\
	& 0.45/1.0 & 1.0 & 2.2 & 4 & $1.64 \times 10^{-11}$ & \\
	\hline
	\hline
\end{tabular}
\caption{Measured spectral range of $\mathcal{M}_{\rm EOFA}$, heatbath relative error ($\varepsilon$), and total time for the heatbath step ($\Delta t_{\rm HB}$), using $N_{\rm LHSB}$ intermediate mass preconditioning steps and an order $N_{\rm poles}$ rational approximation to $x^{-1/2}$, with all CG stopping tolerances set to $r_{\rm MC} = 10^{-10}$. Timings are reported for a 512-node Blue Gene/Q partition.}
\label{tab:32ID-G_heatbath_vs_nlhsb}
\end{table}

\begin{figure}[!h]
\centering
\subfloat[RHMC Ensemble]{\includegraphics[width=0.85\linewidth]{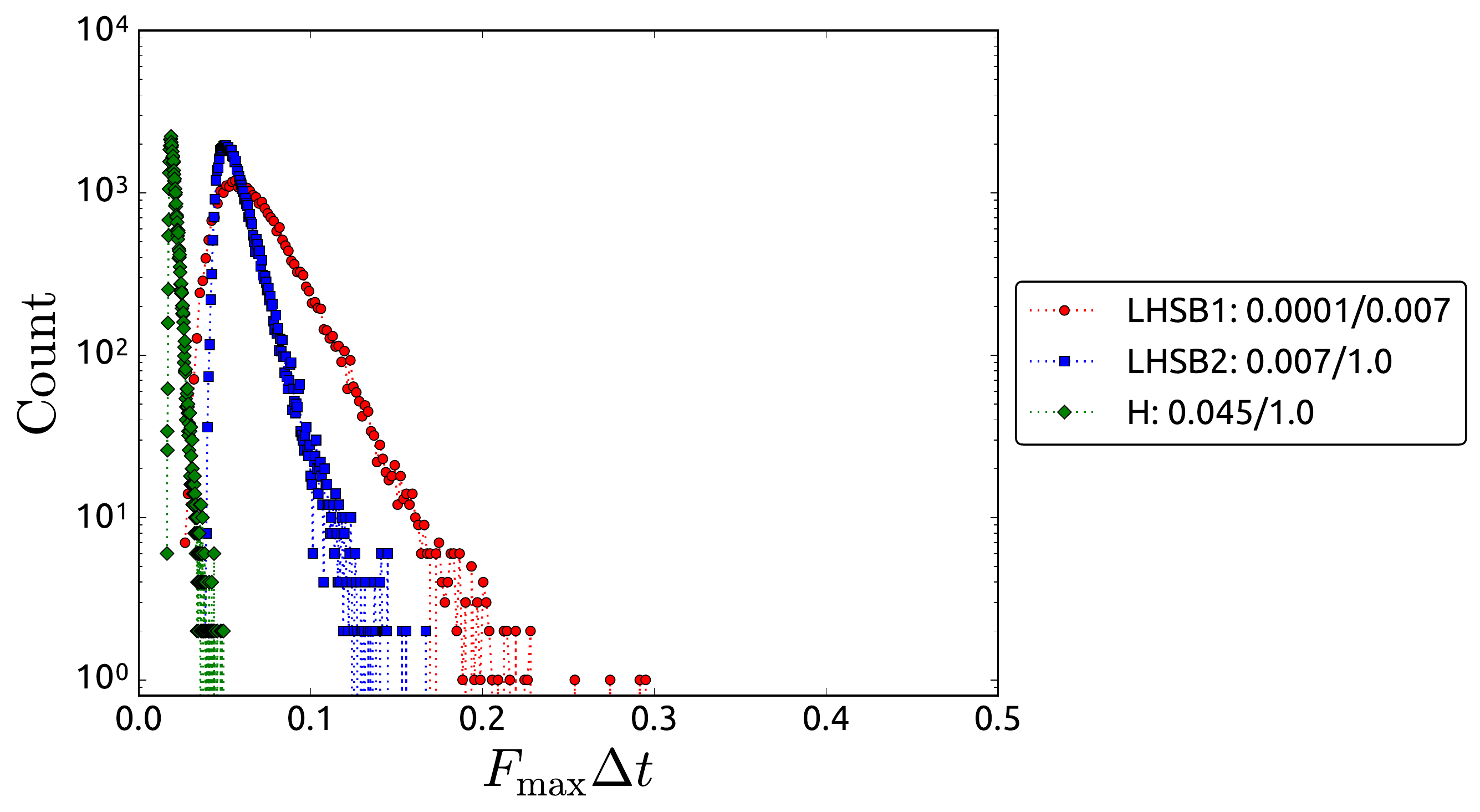}} \\
\subfloat[EOFA Ensemble]{\includegraphics[width=0.85\linewidth]{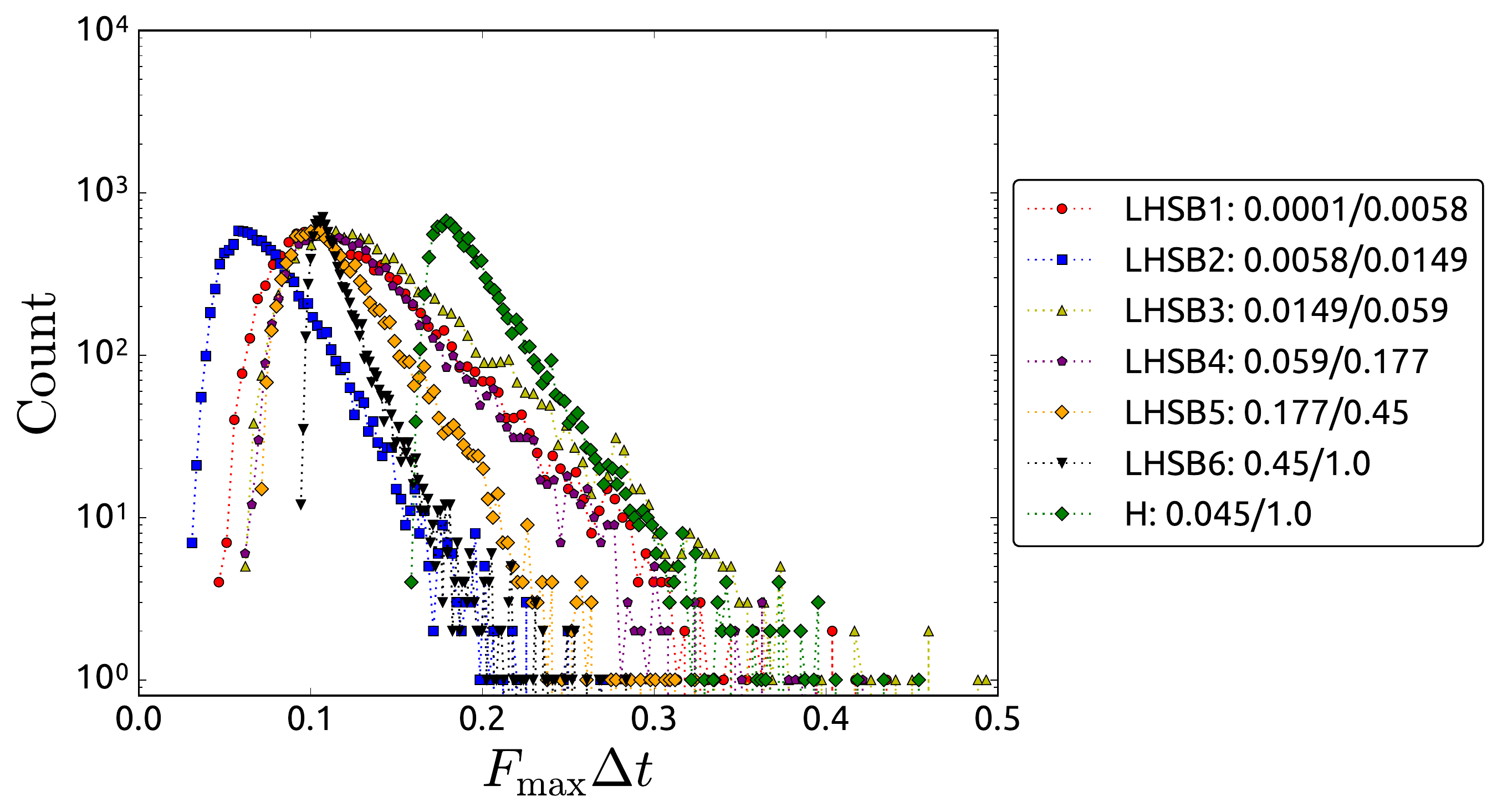}}
\caption{Histograms of the maximum force, defined by Equation \eqref{eqn:Linf_force}, measured between trajectories 500 and 1350 on the 32ID-G RHMC ensemble and measured between trajectories 1350 and 2200 on the 32ID-G EOFA ensemble. We use the abbreviation ``LHSB'' in the legends to denote the various mass ratios entering into our mass preconditioning scheme for the light quark determinant, and ``H'' to denote the strange quark determinant.}
\label{fig:32ID-G_force_hists}
\end{figure}

For each of the runs 2-7 we generated ten trajectories, beginning from the same seed configuration, and analyzed the resulting distributions of $F_{\rm RMS}$ and $F_{\rm max}$. In panel (a) of Figure \ref{fig:32ID-G_force_hists} we plot distributions of $F_{\rm max}$ from 850 trajectories of the production RHMC ensemble generation calculation (run 1). Since we are using exactly the same RHMC action for the strange quark on the RHMC and EOFA ensembles, we tune by adjusting the number and magnitude of the intermediate light Hasenbusch masses such that the forces associated with each of the light quark determinants are comparable to the strange quark force. This allows us to simplify the integrator layout to a three-level scheme, with the light and strange quark determinants updated on the same level. We find that four intermediate Hasenbusch masses are sufficient to ensure that the strange quark force is dominant in the sense of $F_{\rm RMS}$, and that five intermediate Hasenbusch masses are sufficient in the sense of $F_{\rm max}$. Panel (b) shows an analogous force distribution for the latter mass preconditioning scheme. 

In runs 8-10 we explore further tuning of a scheme with four light Hasenbusch masses, and in runs 11-14 we explore further tuning of a scheme with five light Hasenbusch masses. We note that the Monte Carlo acceptance is relatively poor in runs 8-10 --- as we argued in Section \ref{sec:16_cubed_reproduction_tests}, this is consistent with the view that the acceptance should be controlled by the largest integration errors accrued during the trajectory, which are proportional to $F_{\rm max}$ rather than $F_{\rm RMS}$ --- and thus have abandoned this mass preconditioning scheme in favor of the scheme used in runs 11-14. We have then tuned the step size of the outermost integrator level ($\delta \tau$) and the CG stopping tolerance used in the molecular dynamics evolution ($r_{\rm MD}$) to minimize the mean time required to generate an accepted gauge field configuration, resulting in the scheme of run 12. In addition, we have applied the heatbath tuning procedure described in Section \ref{subsec:heatbath_tuning} in all of the runs 8-14, allowing us to relax CG stopping tolerances for the individual solves in the heatbath, while keeping the overall error bounded by $r_{\rm MC} = 10^{-10}$. For the final scheme (12) this optimization further reduced the cost of the light quark heatbath from approximately 2300 s, as reported in Table \ref{tab:32ID-G_heatbath_vs_nlhsb}, to approximately 850 s after tuning.

Comparing the fully tuned EOFA scheme (12) to the original RHMC scheme (1) in Table \ref{tab:32ID-G_tuning_runs}, we find that we are able to generate EOFA trajectories a factor of 4.2 times faster than RHMC trajectories, while maintaining a slightly higher acceptance rate of 93\%. We emphasize, however, that this improved performance is only partially attributable to the simpler form of the EOFA Hamiltonian and force evaluations: we have also switched from an Omelyan integrator to a force gradient integrator, retuned the step sizes and integrator layout, and, in some cases, applied optimizations to the EOFA simulation that are not applicable to RHMC simulations (\textit{e.g.} mixed precision CG). Figure \ref{tab:tricks} briefly summarizes the respective techniques used in the RHMC and EOFA evolution schemes. We have now adopted the EOFA scheme tested in run 12 for ensemble generation in our ongoing $\Delta I = 1/2$ $K \rightarrow \pi \pi$ calculation \cite{Kelly:2016}. We expect the resulting performance gain to enable up to four times as many measurements in our current production run as we would have been able to generate using the initial RHMC evolution scheme, enabling a significantly more precise first-principles determination of the Standard Model ratio $\epsilon'/\epsilon$.
 \begin{figure}[!h]
 	\begin{mdframed}[innertopmargin=10pt, innerbottommargin=5pt]
 		\begin{minipage}{0.48\linewidth}
 			\begin{center} \textbf{\underline{RHMC}} \end{center}
 			\begin{itemize}
 				\item Omelyan integrator ($\delta \tau = 0.0625$)
 				\item One light quark Hasenbusch mass
 				\item Multishift CG with single precision $\slashed D$ but accumulating solution and search vectors in double precision, coupled with reliable update to correct residual
 				\item Even-odd preconditioning
 			\end{itemize}
 		\end{minipage}
 		\begin{minipage}{0.48\linewidth}
 			\begin{center} \textbf{\underline{EOFA}} \end{center}
 			\begin{itemize}
 				\item Force gradient integrator ($\delta \tau = 0.1667$)
 				\item Five light quark Hasenbusch masses
 				\item Mixed precision defect correction CG
 				\item Even-odd preconditioning
 				\item Cayley-form preconditioning
 				\item Force gradient forecasting \cite{Yin:2011np}
 				\item Heatbath forecasting
 				\item Heatbath stopping tolerance tuning
 			\end{itemize}
 		\end{minipage}
 		\vspace{0.1cm}
 	\end{mdframed}
 	\caption{Comparison of optimizations used in the RHMC 32ID-G simulation to the optimizations used in the EOFA 32ID-G simulation.}
 	\label{tab:tricks}
 \end{figure}
\FloatBarrier

%

\section{Conclusion}
\label{sec:conclusion}
In this work we have explored the viability of the exact one flavor algorithm (EOFA) as an alternative to the rational Hybrid Monte Carlo (RHMC) algorithm in molecular dynamics simulations of lattice QCD with domain wall fermions and periodic or $G$-parity boundary conditions. We have verified the formal equivalence of EOFA to RHMC through statistical tests of the EOFA action (Section \ref{sec:HMC_with_EOFA}), and checked, using a series of inexpensive, small volume ensembles with heavy pions, that physical observables such as the plaquette, quark condensates, topological susceptibility, and low energy spectrum are consistent between ensembles generated using EOFA and ensembles generated using RHMC (Section \ref{sec:16_cubed_reproduction_tests}). We have then discussed preconditioning and tuning techniques for EOFA simulations (Section \ref{sec:optimizations} and Appendix \ref{appendix:eo_preconditioning}), and finally, demonstrated that EOFA can substantially outperform RHMC for state-of-the-art lattice QCD simulations with large volumes and physical quark masses (Section \ref{sec:24ID_32ID-G}). In particular, we find that we are able to generate gauge field configurations for the ongoing RBC/UKQCD calculation of the $\Delta I = 1/2$ $K \rightarrow \pi \pi$ decay amplitudes a factor of 4.2 times faster with EOFA. The keys to this dramatic speed-up are a novel preconditioning technique which relates inversions of the EOFA Dirac operator ($\mathscr{D}_{\rm EOFA}$) to cheaper inversions of the standard domain wall fermion Dirac operator ($\mathscr{D}_{\rm DWF}$), and the ability to apply mixed precision defect correction solvers and extensive Hasenbusch mass preconditioning in the context of EOFA.

Future work will explore further physics applications of EOFA. We intend to generate variants of the 24ID ensemble with non-degenerate up and down quark masses in the near future. These ensembles will enable exploratory studies of isospin breaking effects in the meson and baryon spectra, as well as in other precision lattice calculations such as the extraction of the CKM matrix element $V_{us}$ from semileptonic kaon decays \cite{Boyle:2015hfa}. Other potential applications include domain wall QCD simulations with dynamical charm quarks in the sea, and simulations with light, $SU(3)$-symmetric quarks. The latter simulations could be used, for example, to better constrain the strange quark dependence of our $SU(3)$ chiral perturbation theory studies \cite{Mawhinney:2015sfj}, or to probe the location of the critical point separating the crossover and first-order phase transition regions in three-flavor domain wall QCD at finite temperature. 
\FloatBarrier

\section*{Acknowledgments}
The authors are grateful to Peter Boyle, Ting-Wai Chiu, and Norman Christ for helpful discussions in support of this work. Calculations were performed using the Blue Gene/Q computers of the RIKEN-BNL Research Center and Brookhaven National Lab. The software used includes the CPS QCD code (\url{https://github.com/RBC-UKQCD/CPS}) \cite{cps}, supported in part by the USDOE SciDAC program, the BAGEL (\url{http://www2.ph.ed.ac.uk/~paboyle/bagel/Bagel.html}) assembler kernel generator for high-performance optimized kernels and fermion solvers \cite{Boyle20092739}, and the Grid data parallel C++ QCD library (\url{https://github.com/paboyle/Grid}) \cite{Boyle:2015tjk}. R.D.M.~and D.J.M.~are supported in part by U.S.~DOE grant \#DE-SC0011941. C.J.~is supported in part by U.S.~DOE Contract \#AC-02-98CH10886 (BNL). C.K.~is supported by the Intel Corporation.

\bibliographystyle{apsrev4-1}
\bibliography{EOFA_paper}

\appendix

\section{EOFA Operators for Shamir and M\"{o}bius DWF}
\label{appendix:eofa_operators}
In this appendix we list the operators which enter into $\mathscr{D}_{\rm EOFA}$ (Eqn.~\eqref{eqn:eofa_dop}) and the EOFA action (Eqn.~\eqref{eqn:eofa_action}). The more general case of DWF with Zolotarev-type domain wall fermions is constructed implicitly in Ref.~\cite{Chen:2014hyy}; we explicitly list these operators for the more restrictive cases of Shamir and M\"{o}bius DWF used in our simulations. We use $\Theta_{s}$ to denote the discrete Heaviside theta function
\begin{equation}
\Theta_{s} = 
\begin{dcases}
0, \,\,\, s < 0 \\
1, \,\,\, s \geq 0
\end{dcases}
\end{equation}
and assume even $L_{s}$. The operators $\Omega_{\pm}$ and $\Delta_{\pm}$ are related by the identity
\begin{equation}
\Delta_{\pm} = k \Omega_{\pm} \Omega_{\pm}^{\dagger},
\end{equation}
and the M\"{o}bius operators reduce to the corresponding Shamir operators in the limit $c=d=1/2$. We note that the dense M\"{o}bius expressions listed here are not used inside the inverter; we instead invert the preconditioned system discussed in Appedix \ref{appendix:preconditioned_mobius}.

\subsection{Shamir Kernel}
\label{appendix:eofa_operators_shamir}

\begin{equation}
k = m_{2} - m_{1}
\end{equation}

\begin{equation}
\left[ \Omega_{+} \right]_{ss'} = \delta_{s,L_{s}-1} \, \delta_{s',0}
\end{equation}

\begin{equation}
\left[ \Omega_{-} \right]_{ss'} = \delta_{s,0} \, \delta_{s',0}
\end{equation}

\begin{equation}
\left[ \Delta_{+}(m_{1},m_{2}) \right]_{ss'} = \left( m_{2} - m_{1} \right) \delta_{s,L_{s}-1} \, \delta_{s',L_{s}-1}
\end{equation}

\begin{equation}
\left[ \Delta_{-}(m_{1},m_{2}) \right]_{ss'} = \left( m_{2} - m_{1} \right) \delta_{s,0} \, \delta_{s',0}
\end{equation}

\begin{equation}
\left[ M_{+}(m) \right]_{ss'} = \delta_{ss'} - \delta_{s,s'+1} + m \delta_{s,L_{s}-1} \, \delta_{s',0}
\end{equation}

\begin{equation}
\left[ M_{-}(m) \right]_{ss'} = \delta_{ss'} - \delta_{s,s'-1} + m \delta_{s,0} \, \delta_{s',L_{s}-1}
\end{equation}

\begin{equation}
\big[ \widetilde{\mathscr{D}}(m) \big]_{ss'} = \delta_{ss'}
\end{equation}

\begin{equation}
\big[ \widetilde{\mathscr{D}}(m)^{-1} \big]_{ss'} = \delta_{ss'}
\end{equation}

\subsection{M\"{o}bius Kernel}
\label{appendix:eofa_operators_mobius}

\begin{equation}
k = \frac{2 c \left( m_{2} - m_{1} \right) \left( c + d \right)^{2 L_{s}}}{\left[ \left( c + d \right)^{L_{s}} + m_{1} \left( c - d \right)^{{Ls}} \right] \left[ \left( c + d \right)^{L_{s}} + m_{2} \left( c - d \right)^{{Ls}} \right]}
\end{equation}

\begin{equation}
\left( \Omega_{+} \right)_{ss'} = (-1)^{s+1} \frac{\left( c - d \right)^{L_{s}-s-1}}{\left( c + d \right)^{L_{s}-s}} \, \delta_{s',0}
\end{equation}

\begin{equation}
\left( \Omega_{-} \right)_{ss'} = (-1)^{s} \frac{\left( c - d \right)^{s}}{\left( c + d \right)^{s+1}} \, \delta_{s',0}
\end{equation}

\begin{equation}
\label{eqn:mobius_Dp}
\left[ \Delta_{+}(m_{1},m_{2}) \right]_{ss'} = \frac{ (-1)^{s+s'} 2 c \left( m_{2} - m_{1} \right) \left( c + d \right)^{s+s'} \left( c - d \right)^{2 \left( L_{s} - 1 \right) - s - s'} }{ \left[ \left( c + d \right)^{L_{s}} + m_{1} \left( c - d \right)^{{Ls}} \right] \left[ \left( c + d \right)^{L_{s}} + m_{2} \left( c - d \right)^{{Ls}} \right] }
\end{equation}

\begin{equation}
\label{eqn:mobius_Dm}
\left[ \Delta_{-}(m_{1},m_{2}) \right]_{ss'} = \frac{ (-1)^{s+s'} 2 c \left( m_{2} - m_{1} \right) \left( c + d \right)^{2 \left( L_{s} - 1 \right) - s - s'} \left( c - d \right)^{s + s'} }{ \left[ \left( c + d \right)^{L_{s}} + m_{1} \left( c - d \right)^{{Ls}} \right] \left[ \left( c + d \right)^{L_{s}} + m_{2} \left( c - d \right)^{{Ls}} \right] }
\end{equation}

\begin{multline}
\label{eqn:mobius_Mp}
\left[ M_{+}(m) \right]_{ss'} = 
\frac{\left( -1 \right)^{s-s'} 2 c \left( c + d \right)^{Ls-s+s'-1} \left( c - d \right)^{s-s'-1}}{\left( c + d \right)^{Ls} + m \left( c - d \right)^{Ls}} \, \Theta_{s-s'-1} \\
+ \frac{\left( c + d \right)^{Ls-1} - m \left( c - d \right)^{Ls-1}}{\left( c + d \right)^{Ls} + m \left( c - d \right)^{Ls}} \, \delta_{ss'} \\
+ \frac{\left( -1 \right)^{s-s'+1} 2 c m \left( c + d \right)^{s'-s-1} \left( c - d \right)^{Ls+s-s'-1}}{\left( c + d \right)^{Ls} + m \left( c - d \right)^{Ls}} \, \Theta_{s'-s-1}
\end{multline}

\begin{multline}
\label{eqn:mobius_Mm}
\left[ M_{-}(m) \right]_{ss'} = 
\frac{\left( -1 \right)^{s'-s+1} 2 c m \left( c + d \right)^{s-s'-1} \left( c - d \right)^{L_{s}-s+s'-1}}{\left( c + d \right)^{Ls} + m \left( c - d \right)^{Ls}} \, \Theta_{s-s'-1} \\
+ \frac{\left( c + d \right)^{Ls-1} - m \left( c - d \right)^{Ls-1}}{\left( c + d \right)^{Ls} + m \left( c - d \right)^{Ls}} \, \delta_{ss'} \\
+ \frac{\left( -1 \right)^{s-s'} 2 c \left( c + d \right)^{L_{s}+s-s'-1} \left( c - d \right)^{s'-s-1}}{\left( c + d \right)^{Ls} + m \left( c - d \right)^{Ls}} \, \Theta_{s'-s-1} 
\end{multline}

\begin{multline}
\big[ \widetilde{\mathscr{D}}(m) \big]_{ss'} = \left( c + d \right) \delta_{ss'} + \left( c - d \right) P_{+} \delta_{s,s'+1} + \left( c - d \right) P_{-} \delta_{s,s'-1} \\
- m \left( c - d \right) P_{+} \delta_{s,0} \, \delta_{s',L_{s}-1} - m \left( c - d \right) P_{-} \delta_{s,L_{s}-1} \, \delta_{s',0}
\end{multline}

\begin{multline}
\big[ \widetilde{\mathscr{D}}(m)^{-1} \big]_{ss'} = \left[ \frac{m \left( -1 \right)^{s-s'+1} \left( c + d \right)^{s'-s-1} \left( c - d \right)^{L_{s}+s-s'}}{\left( c + d \right)^{L_{s}} + m \left( c - d \right)^{L_{s}}} + \frac{\left( -1 \right)^{s-s'} \left( c - d \right)^{s-s'}}{\left( c + d \right)^{s-s'+1}} \, \Theta_{s-s'} \right] P_{+} \\
+
\left[ \frac{m \left( -1 \right)^{s'-s+1} \left( c + d \right)^{s-s'-1} \left( c - d \right)^{L_{s}+s'-s}}{\left( c + d \right)^{L_{s}} + m \left( c - d \right)^{L_{s}}} + \frac{\left( -1 \right)^{s'-s} \left( c - d \right)^{s'-s}}{\left( c + d \right)^{s'-s+1}} \, \Theta_{s'-s} \right] P_{-}
\end{multline}
\FloatBarrier

\section{Four-Dimensional Even-Odd Preconditioning}
\label{appendix:eo_preconditioning}
The inversions required to compute the exact one flavor Hamiltonian can be accelerated using a standard checkerboarding technique: we label lattice sites as ``even'' if $\left( x + y + z + t \right) \equiv 0 \pmod 2$ or ``odd'' if $\left( x + y + z + t \right) \equiv 1 \pmod 2$. This naturally induces a block structure in the Dirac operator $\mathscr{D}$, which can be $LDU$ decomposed as
\begin{equation}
\underbrace{\left( \begin{array}{cc} \mathscr{D}_{ee} & \mathscr{D}_{eo} \\ \mathscr{D}_{oe} & \mathscr{D}_{oo} \end{array} \right)}_{\mathscr{D}} = \underbrace{\left( \begin{array}{cc} 1 & 0 \\ \mathscr{D}_{oe} \mathscr{D}_{ee}^{-1} & 1 \end{array} \right)}_{L} \underbrace{\left( \begin{array}{cc} \mathscr{D}_{ee} & 0 \\ 0 & \mathscr{D}_{oo} - \mathscr{D}_{oe} \mathscr{D}_{ee}^{-1} \mathscr{D}_{eo} \end{array} \right)}_{D} \underbrace{\left( \begin{array}{cc} 1 & \mathscr{D}_{ee}^{-1} \mathscr{D}_{eo} \\ 0 & 1 \end{array} \right)}_{U}.
\end{equation}
Left-multiplying the linear system $\mathscr{D} \psi = \phi$ by 
\begin{equation}
L^{-1} = \left( \begin{array}{cc} 1 & 0 \\ - \mathscr{D}_{oe} \mathscr{D}_{ee}^{-1} & 1 \end{array} \right)
\end{equation}
results in the equivalent system
\begin{equation}
\left( \begin{array}{c} \mathscr{D}_{ee} \psi_{e} + \mathscr{D}_{eo} \psi_{o} \\ \left( \mathscr{D}_{oo} - \mathscr{D}_{oe} \mathscr{D}_{ee}^{-1} \mathscr{D}_{eo} \right) \psi_{o} \end{array} \right) = \left( \begin{array}{c} \phi_{e} \\ \phi_{o} - \mathscr{D}_{oe} \mathscr{D}_{ee}^{-1} \phi_{e} \end{array} \right),
\end{equation}
leading to the following trick: assuming $\mathscr{D}_{ee}^{-1}$ is available in an explicit form, it suffices to invert
\begin{equation}
\left( \mathscr{D}_{oo} - \mathscr{D}_{oe} \mathscr{D}_{ee}^{-1} \mathscr{D}_{eo} \right) \psi_{o} = \widetilde{\phi}_{o},
\end{equation}
with $\widetilde{\phi}_{o} \equiv \phi_{o} - \mathscr{D}_{oe} \mathscr{D}_{ee}^{-1} \phi_{e}$. This system only involves the odd sublattice, and is thus substantially cheaper to invert than $\mathscr{D}$ using an iterative algorithm like CG. The solution on the even sublattice can then be reconstructed for a trivial additional cost as 
\begin{equation}
\psi_{e} = \mathscr{D}_{ee}^{-1} \left( \phi_{e} - \mathscr{D}_{eo} \psi_{o} \right).
\end{equation} 
This technique is already well understood in the context of RHMC with Shamir or M\"{o}bius DWF; in this appendix we describe how to generalize the method to the exact one flavor algorithm.

In the context of EOFA, the generic linear system one needs to invert takes the form 
\begin{equation}
\label{eqn:hermit_indef_sys}
\Big( H(m_{1}) + \beta \Delta_{\pm}(m_{2},m_{3}) P_{\pm} \Big) \psi = \phi.
\end{equation}
We choose to multiply by an overall factor of $\gamma_{5} R_{5}$, rewriting the system as
\begin{equation}
\label{eqn:non_hermit_sys}
\Big( \mathscr{D}_{\rm EOFA}(m_{1}) + \beta \gamma_{5} R_{5} \Delta_{\pm}(m_{2},m_{3}) P_{\pm} \Big) \psi = \gamma_{5} R_{5} \phi,
\end{equation}
for the following reasons: first, we wish to re-use the existing high-performance implementation of the Wilson $\slashed{D}$ kernel in the BAGEL library without modification, and second, overall factors of $\gamma_{5} R_{5}$ will cancel inside the inverter since we use CG applied to the normal equations and $(\gamma_{5} R_{5})^{\dagger} (\gamma_{5} R_{5}) = 1$. Since $\left( \mathscr{D}_{\rm EOFA} \right)_{eo} = \left( \mathscr{D}_{\rm DWF} \right)_{eo}$ and $\Delta_{\pm} \propto \delta_{xx'}$ in the 4D bulk, only the operators coupling sites of the same parity need to be modified to implement even-odd preconditioned EOFA. We take somewhat different approaches for the Shamir and M\"{o}bius cases.

\subsection{Shamir Kernel}

Recall that for the Shamir kernel $\mathscr{D}_{\rm DWF} = \mathscr{D}_{\rm EOFA}$, so the extension of an inverter for the even-odd preconditioned $\mathscr{D}_{\rm DWF}$ operator to instead solve Eqn.~\eqref{eqn:non_hermit_sys} is straightforward. With $\mathscr{D} = \mathscr{D}_{\rm DWF}$, the same parity fermion matrix has the tridiagonal block structure
\begin{multline}
\label{eqn:Ddwf_ooee}
\left( \mathscr{D}_{\rm DWF} \right)_{ee} = \left( \mathscr{D}_{\rm DWF} \right)_{oo} = \delta_{xx'} \, \Big\{ \left( 5 - M_{5} \right) \delta_{ss'} - P_{+} \delta_{s,s'+1} - P_{-} \delta_{s,s'-1} \\
+ m_{1} P_{+} \delta_{s,0} \delta_{s',L_{s}-1} + m_{1} P_{-} \delta_{s,L_{s}-1} \delta_{s',0} \Big\}.
\end{multline}
One can check by explicit calculation that the shift operators have the form
\begin{equation}
\begin{dcases}
\beta \gamma_{5} R_{5} \Delta_{+}(m_{2},m_{3}) P_{+} = \beta \left( m_{3} - m_{2} \right) P_{+} \delta_{xx'} \delta_{s,0} \delta_{s',L_{s}-1} \\
\beta \gamma_{5} R_{5} \Delta_{-}(m_{2},m_{3}) P_{-} = - \beta \left( m_{3} - m_{2} \right) P_{-} \delta_{xx'} \delta_{s,L_{s}-1} \delta_{s',0}
\end{dcases},
\end{equation}
so one can consider the operator appearing in Eqn.~\eqref{eqn:non_hermit_sys} as a slight generalization of Eqn.~\eqref{eqn:Ddwf_ooee} to
\begin{multline}
\mathscr{D}_{ee} = \mathscr{D}_{oo} = \delta_{xx'} \, \Big\{ \left( 5 - M_{5} \right) \delta_{ss'} - P_{+} \delta_{s,s'+1} - P_{-} \delta_{s,s'-1} \\
+ d_{+} P_{+} \delta_{s,0} \delta_{s',L_{s}-1} + d_{-} P_{-} \delta_{s,L_{s}-1} \delta_{s',0} \Big\},
\end{multline}
with
\begin{equation}
d_{-} = m_{1} - \beta \left( m_{3} - m_{2} \right) \delta_{i,-}
\end{equation}
and
\begin{equation}
d_{+} = m_{1} + \beta \left( m_{3} - m_{2} \right) \delta_{i,+},
\end{equation}
where the index $i$ denotes the chirality of the shift operator. $\mathscr{D}_{ee}^{-1}$ can be efficiently applied using the $LDU$ decomposition of $\mathscr{D}_{ee}$, again as a slight generalization of the standard Shamir DWF case.

\subsection{M\"{o}bius Kernel and Cayley-Form Preconditioning}
\label{appendix:preconditioned_mobius}

Using Eqn.~\eqref{eqn:eofa_dop_def} we can write $\mathscr{D}_{\rm EOFA}$ in the form
\begin{equation}
\left( \mathscr{D}_{\rm EOFA} \right)_{xx',ss'} = \left( D_{W} \right)_{xx'} \delta_{ss'} + \delta_{xx'} \left( \mathscr{D}^{\perp} \right)_{ss'}.
\end{equation}
The action of the operator appearing in Eqn.~\eqref{eqn:non_hermit_sys} on lattice sites of the same parity, then, is given by
\begin{multline}
\mathscr{D}_{ee} = \mathscr{D}_{oo} = \delta_{xx'} \Big\{ \left( 4 - M_{5} \right) \delta_{ss'} + \left( M_{+}(m_{1}) \right)_{ss'} P_{+} + \left( M_{-}(m_{1}) \right)_{ss'} P_{-} \\
+ \beta \gamma_{5} R_{5} \left( \Delta_{\pm}(m_{2},m_{3}) \right)_{ss'} P_{\pm} \Big\},
\end{multline}
with $M_{+}$, $M_{-}$, and $\Delta_{\pm}$ as defined in equations \eqref{eqn:mobius_Dp}-\eqref{eqn:mobius_Mm}. The matrix elements of $\mathscr{D}_{ee}^{-1} = \mathscr{D}_{oo}^{-1}$ can be found by explicit numerical inversion as part of the setup cost; this is a trivial overhead since it suffices to invert only the $ss'$ subblock of $\mathscr{D}_{ee}$. In this form, the exact factorization of the fermion determinant in Eqn.~\eqref{eqn:EOFA_factorization} comes at the cost of dense $L_{s} \times L_{s}$ matrix operations. We argue that it is possible to do significantly better by introducing an additional preconditioning step.

We note that the system defined by Eqn.~\eqref{eqn:non_hermit_sys} can be more efficiently inverted for the case of M\"{o}bius DWF by using the operator $\widetilde{\mathscr{D}}^{-1}$ as a right preconditioner, resulting in an equivalent system in terms of $\mathscr{D}_{\rm DWF}$. For the special case $\beta = 0$ this is straightforward: observing that the relationship between $\mathscr{D}_{\rm EOFA}$ and $\mathscr{D}_{\rm DWF}$ (Eqn.~\eqref{eqn:eofa_dop}) can be used to manipulate
\begin{equation}
\mathscr{D}_{\rm EOFA} \psi = \mathscr{D}_{\rm EOFA} \cdot \widetilde{\mathscr{D}} \cdot \underbrace{\widetilde{\mathscr{D}}^{-1} \psi}_{\equiv \psi'} = \mathscr{D}_{\rm DWF} \psi',
\end{equation}
it suffices to solve $\mathscr{D}_{\rm DWF} \psi' = \gamma_{5} R_{5} \phi$, from which $\psi = \widetilde{\mathscr{D}} \psi'$ can be recovered at the cost of a single additional matrix multiplication. While we observe that $\mathscr{D}_{\rm DWF}^{\dagger} \mathscr{D}_{\rm DWF}$ has a slightly larger condition number than $\mathscr{D}_{\rm EOFA}^{\dagger} \mathscr{D}_{\rm EOFA}$, leading to a modest increase in the total number of CG iterations required to invert the system, $\mathscr{D}_{\rm DWF}$ also has a tridiagonal stencil in the fifth dimension, and can thus be applied in $\mathcal{O}(L_{s})$ operations --- unlike the $\mathcal{O}(L_{s}^{2})$ operations required for the dense $\mathscr{D}_{\rm EOFA}$ --- leading to a substantial reduction in wall clock time for the inversion.

The $\beta \neq 0$ case is more involved, but can be treated in a similar manner. Right preconditioning Eqn.~\eqref{eqn:non_hermit_sys} with $\widetilde{\mathscr{D}}^{-1}$ leads to
\begin{equation}
\label{eqn:precond_mobius_eofa_system}
\left( \mathscr{D}_{\rm DWF}(m_{1}) \pm \beta R_{5} \Delta_{\pm}(m_{2},m_{3}) \widetilde{\mathscr{D}} P_{\pm} \right) \psi' = \gamma_{5} R_{5} \phi,
\end{equation}
where we have used $\gamma_{5} P_{\pm} = \pm P_{\pm}$. We define a new, preconditioned, shift operator $\widetilde{\Delta}_{\pm}$ by
\begin{equation}
\widetilde{\Delta}_{\pm}(m_{1},m_{2}) \equiv R_{5} \Delta_{\pm}(m_{1},m_{2}) \widetilde{\mathscr{D}} P_{\pm},
\end{equation}
and note that since $(\widetilde{\Delta})_{eo} = (\widetilde{\Delta})_{oe} = 0$, Eqn.~\eqref{eqn:precond_mobius_eofa_system} can be inverted efficiently even with $\beta \neq 0$ provided we can apply the operator $(\mathscr{D}_{\rm DWF})_{ee} \pm \beta \widetilde{\Delta}_{\pm}$ and its inverse in $\mathcal{O}(L_{s})$ operations. This turns out to be possible after observing that $\widetilde{\Delta}_{\pm}$ is rank-one, \textit{i.e.}~it can be written as a vector outer product
\begin{equation}
\label{eqn:delta_rank_one}
\widetilde{\Delta}_{\pm} = u_{\pm} \otimes v_{\pm}.
\end{equation} To see this, we start by decomposing $\widetilde{\mathscr{D}}$ into  its chiral components --- $\widetilde{\mathscr{D}} = \widetilde{\mathscr{D}}_{+} P_{+} + \widetilde{\mathscr{D}}_{-} P_{-}$ --- in terms of which we can also decompose
\begin{equation}
\widetilde{\Delta}_{\pm} = R_{5} \Delta_{\pm} \widetilde{\mathscr{D}}_{\pm} P_{\pm}.
\end{equation}
The 5D structure of these operators can be worked out by direct calculation, leading to Eqn.~\eqref{eqn:delta_rank_one}, with
\begin{equation}
\begin{dcases}
\left( u_{+} \right)_{s} = \left( -1 \right)^{s} \frac{\left( c - d \right)^{s}}{\left( c + d \right)^{L_{s}+s+1}} \left( \left( c + d \right)^{L_{s}} + m_{1} \left( c - d \right)^{L_{s}} \right) \\
\left(v_{+}\right)_{s} = k \, \delta_{s,L_{s}-1} \\
\left( u_{-} \right)_{s} = \left( -1 \right)^{s+1} \frac{\left( c - d \right)^{L_{s}-1-s}}{\left( c + d \right)^{2L_{s}-s}} \left( \left( c + d \right)^{L_{s}} + m_{1} \left( c - d \right)^{L_{s}} \right) \\
\left(v_{-}\right)_{s} = k \, \delta_{s,0} \\
\end{dcases}.
\end{equation}
Matrix-vector products involving the preconditioned shift operator and a pseudofermion field can be computed from this decomposition as
\begin{equation}
\label{eqn:Mooee_shift_p}
\begin{dcases}
\left( \widetilde{\Delta}_{+} \psi \right)_{s} = k \left( u_{+} \right)_{s} P_{+} \psi_{L_{s}-1} \\
\left( \widetilde{\Delta}_{+}^{\dagger} \psi \right)_{s}= k \, \delta_{s,L_{s}-1} \, P_{+} \left[ \sum_{s'=0}^{L_{s}-1} \left( u_{+} \right)_{s'} \psi_{s'} \right] \\
\end{dcases}
\end{equation}
and
\begin{equation}
\label{eqn:Mooee_shift_m}
\begin{dcases}
\left( \widetilde{\Delta}_{-} \psi \right)_{s} = k \left( u_{-} \right)_{s} P_{-} \psi_{0} \\
\left( \widetilde{\Delta}_{-}^{\dagger} \psi \right)_{s} = k \, \delta_{s,0} \, P_{-} \left[ \sum_{s'=0}^{L_{s}-1} \left( u_{-} \right)_{s'} \psi_{s'} \right] \\
\end{dcases}.
\end{equation}
The inverses can be applied using the Sherman-Morrison formula:
\begin{equation}
\label{eqn:sherman_morrison_mobius_inv}
\Big( (\mathscr{D}_{\rm DWF})_{ee} \pm \beta \left( u_{\pm} \otimes v_{\pm} \right) \Big)^{-1} = (\mathscr{D}_{\rm DWF})_{ee}^{-1} \mp \beta \frac{(\mathscr{D}_{\rm DWF})_{ee}^{-1} \left( u_{\pm} \otimes v_{\pm} \right) (\mathscr{D}_{\rm DWF})_{ee}^{-1}}{1 \pm \beta \langle v_{\pm}, (\mathscr{D}_{\rm DWF})_{ee}^{-1} u_{\pm} \rangle}.
\end{equation}
In terms of
\begin{equation}
x_{\pm} \equiv (\mathscr{D}_{\rm DWF})_{ee}^{-1} u_{\pm},
\end{equation}
which can be constructed numerically using the tridiagonal matrix algorithm \cite{Datta:2010:NLA:1805893}, the necessary factors can be written as
\begin{equation}
\label{eqn:Mooeeinv_shift_p}
\resizebox{\linewidth}{!}{
$\displaystyle
\begin{dcases}
1 + \beta \langle v_{+}, (\mathscr{D}_{\rm DWF})_{ee}^{-1} u_{+} \rangle = 1 + \beta k \left( x_{+} \right)_{L_{s}-1} \\
\Big( \left[ (\mathscr{D}_{\rm DWF})_{ee}^{-1} \left( u_{\pm} \otimes v_{\pm} \right) (\mathscr{D}_{\rm DWF})_{ee}^{-1} \right] \psi \Big)_{s} = \frac{k \left( x_{+} \right)_{s} }{\left(c + d \right)^{L_{s}} + m_{1} \left( c - d \right)^{L_{s}}} \, P_{+} \left[ \sum_{s'=0}^{L_{s}-1} \left( c + d \right)^{s'} \left( c - d \right)^{L_{s}-1-s'} \psi_{s'} \right] \\
\Big( \left[ (\mathscr{D}_{\rm DWF})_{ee}^{-1} \left( u_{+} \otimes v_{+} \right) (\mathscr{D}_{\rm DWF})_{ee}^{-1} \right]^{\dagger} \psi \Big)_{s} = \frac{k \left( c + d \right)^{s} \left( c - d \right)^{L_{s}-1-s} }{ \left( c + d \right)^{L_{s}} + m_{1} \left( c - d \right)^{L_{s}} } \, P_{+} \left[ \sum_{s'=0}^{L_{s}-1} \left( x_{+} \right)_{s'} \psi_{s'} \right]
\end{dcases}
$
}
\end{equation}
and
\begin{equation}
\label{eqn:Mooeeinv_shift_m}
\resizebox{\linewidth}{!}{
$\displaystyle
\begin{dcases}
1 - \beta \langle v_{-}, (\mathscr{D}_{\rm DWF})_{ee}^{-1} u_{-} \rangle = 1 - \beta k \left( x_{-} \right)_{0} \\
\Big( \left[ (\mathscr{D}_{\rm DWF})_{ee}^{-1} \left( u_{-} \otimes v_{-} \right) (\mathscr{D}_{\rm DWF})_{ee}^{-1} \right] \psi \Big)_{s} = \frac{k \left( x_{-} \right)_{s}}{ \left( c + d \right)^{L_{s}} + m_{1} \left( c - d \right)^{L_{s}} } \, P_{-} \left[ \sum_{s'=0}^{L_{s}-1} \left(c+d\right)^{L_{s}-1-s'} \left(c-d\right)^{s'} \psi_{s'} \right] \\
\Big( \left[ (\mathscr{D}_{\rm DWF})_{ee}^{-1} \left( u_{-} \otimes v_{-} \right) (\mathscr{D}_{\rm DWF})_{ee}^{-1} \right]^{\dagger} \psi \Big)_{s} = \frac{k \left(c+d\right)^{L_{s}-1-s} \left(c-d\right)^{s}}{ \left( c + d \right)^{L_{s}} + m_{1} \left( c - d \right)^{L_{s}} } \, P_{-} \left[ \sum_{s'=0}^{L_{s}-1} \left( x_{-} \right)_{s'} \psi_{s'} \right]
\end{dcases}
$
},
\end{equation}
which allow Eqn.~\eqref{eqn:sherman_morrison_mobius_inv} to be applied to a pseudofermion vector in $\mathcal{O}(L_{s})$ operations.

\begin{figure}[h]
\centering
\includegraphics[width=0.95\textwidth]{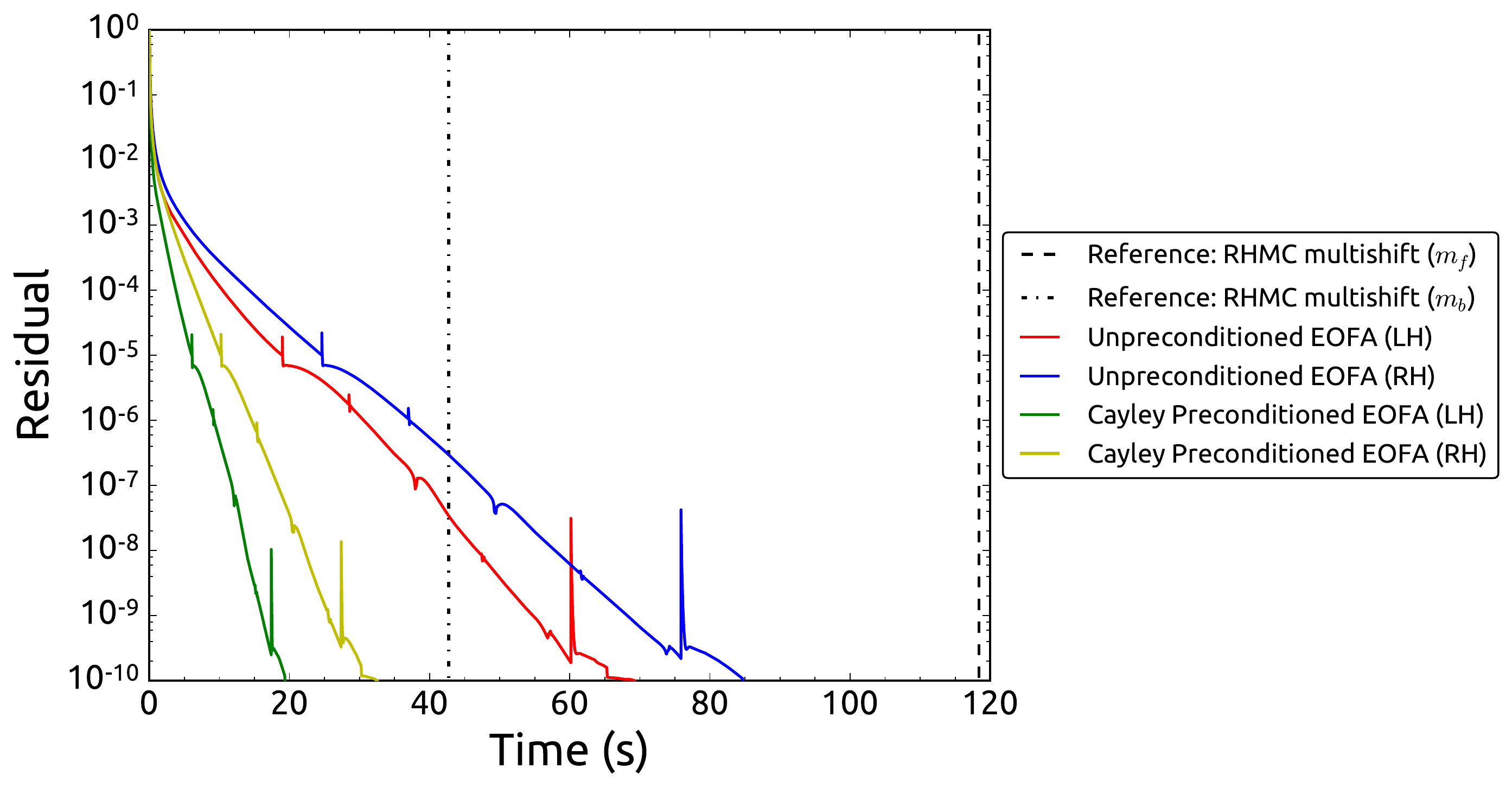}
\caption{Comparison of wall clock inversion times for the two solves required to evaluate the EOFA Hamiltonian or pseudofermion force with and without Cayley-form preconditioning for the strange quark determinant on the 24ID ensemble. The dashed vertical lines show the corresponding total cost of the multishift inversions of $\mathscr{D}_{\rm DWF}^{\dagger} \mathscr{D}_{\rm DWF}$ needed to evaluate the RHMC Hamiltonian or pseudofermion force on the same ensemble.}
\label{fig:cayley_precond_test}
\end{figure}

In Figure \ref{fig:cayley_precond_test} we benchmark representative even-odd preconditioned inversions of Eqn.~\eqref{eqn:non_hermit_sys} on the 24ID ensemble, with and without additional preconditioning by $\widetilde{\mathscr{D}}^{-1}$, at the physical strange quark mass. In addition to observing a substantial improvement in terms of wall clock time for the inversion, we note that this preconditioning scheme also has the advantage that it requires little new code --- assuming an existing high-performance implementation of $\mathscr{D}_{\rm DWF}$ --- since $\mathscr{D}_{\rm EOFA}$ is never applied directly in the preconditioned formalism.

\FloatBarrier

\clearpage
\section{Additional Plots for Small Volume Reproduction Tests}
\label{appendix:more_repro_plots}
\subsection{Evolution of the Plaquette, Quark Condensates, and Topological Charge}

\begin{figure}[!h]
\centering
\subfloat[Plaquette]{\includegraphics[width=0.49\linewidth]{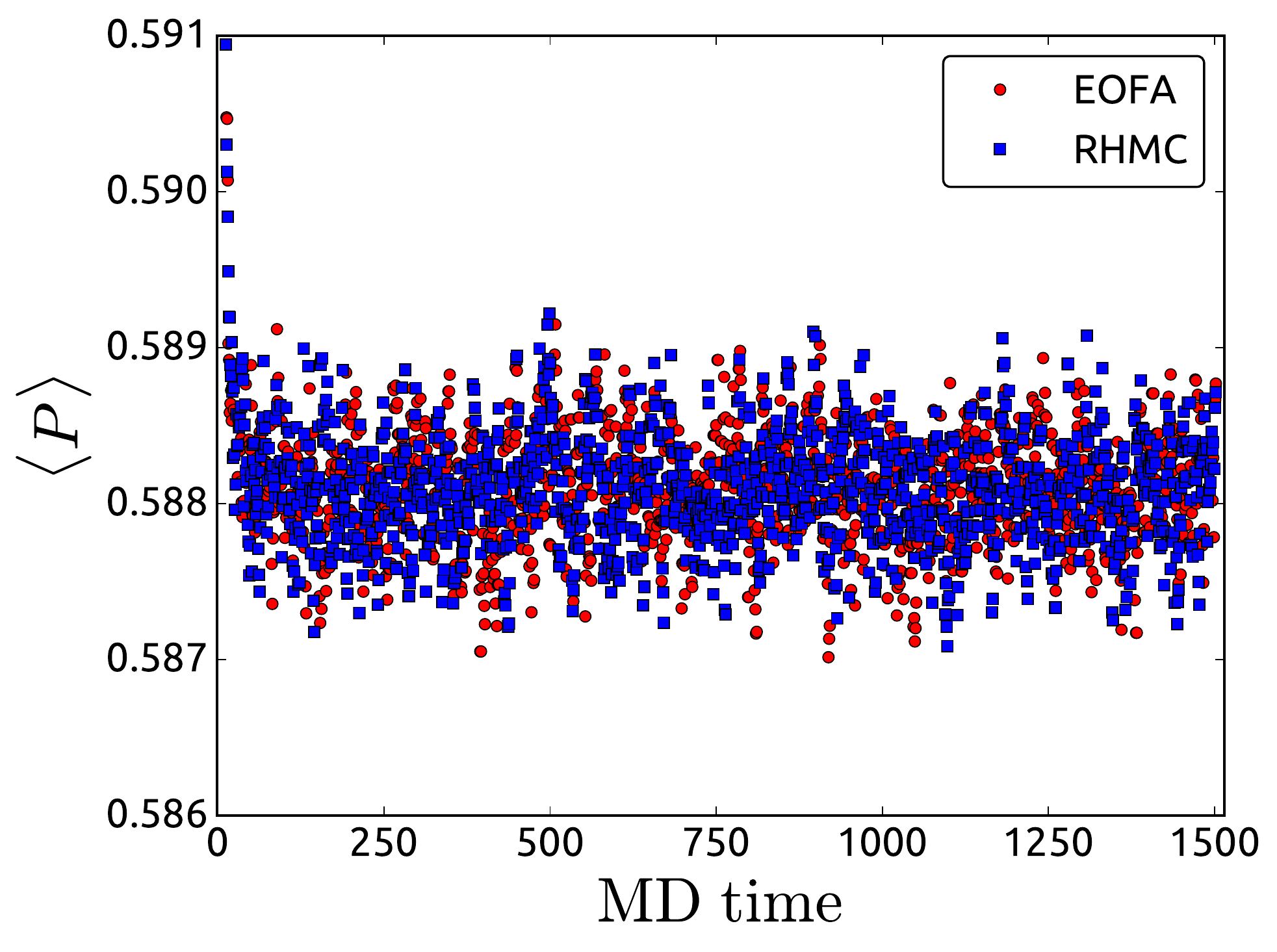}} \quad
\subfloat[Topological Charge]{\includegraphics[width=0.47\linewidth]{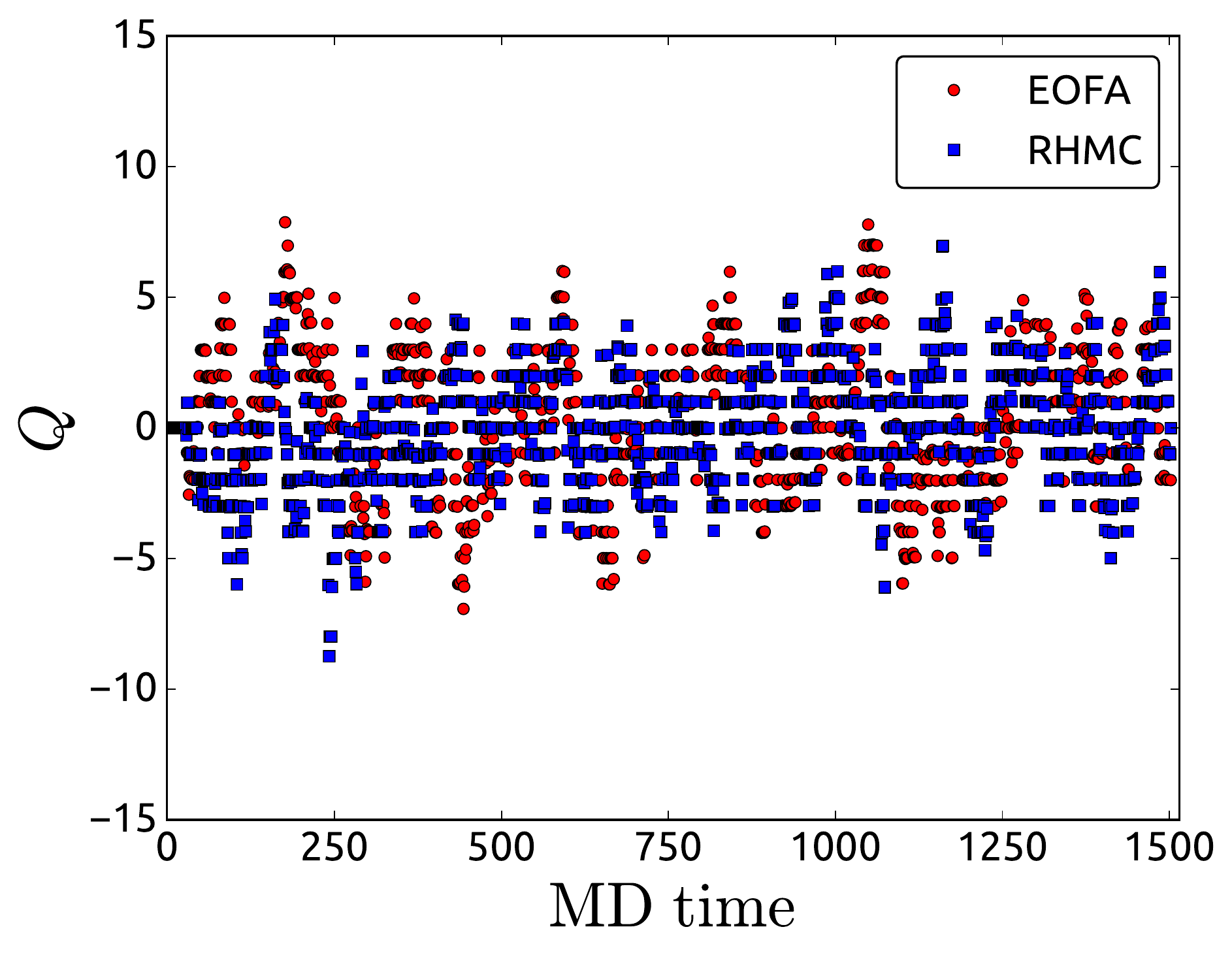}} \\
\subfloat[Chiral Condensates]{\includegraphics[width=0.48\linewidth]{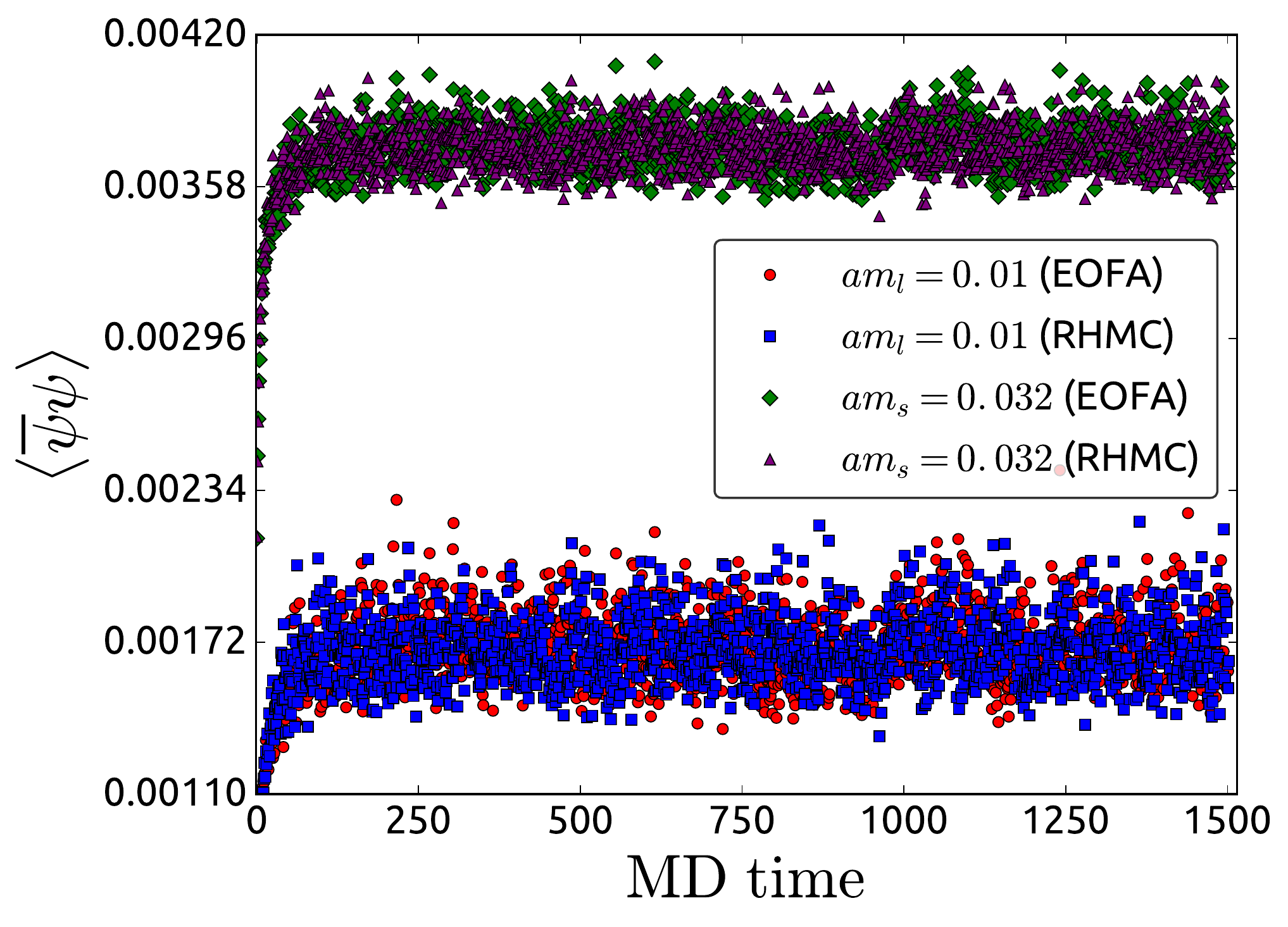}} \quad
\subfloat[Pseudoscalar Condensates]{\includegraphics[width=0.48\linewidth]{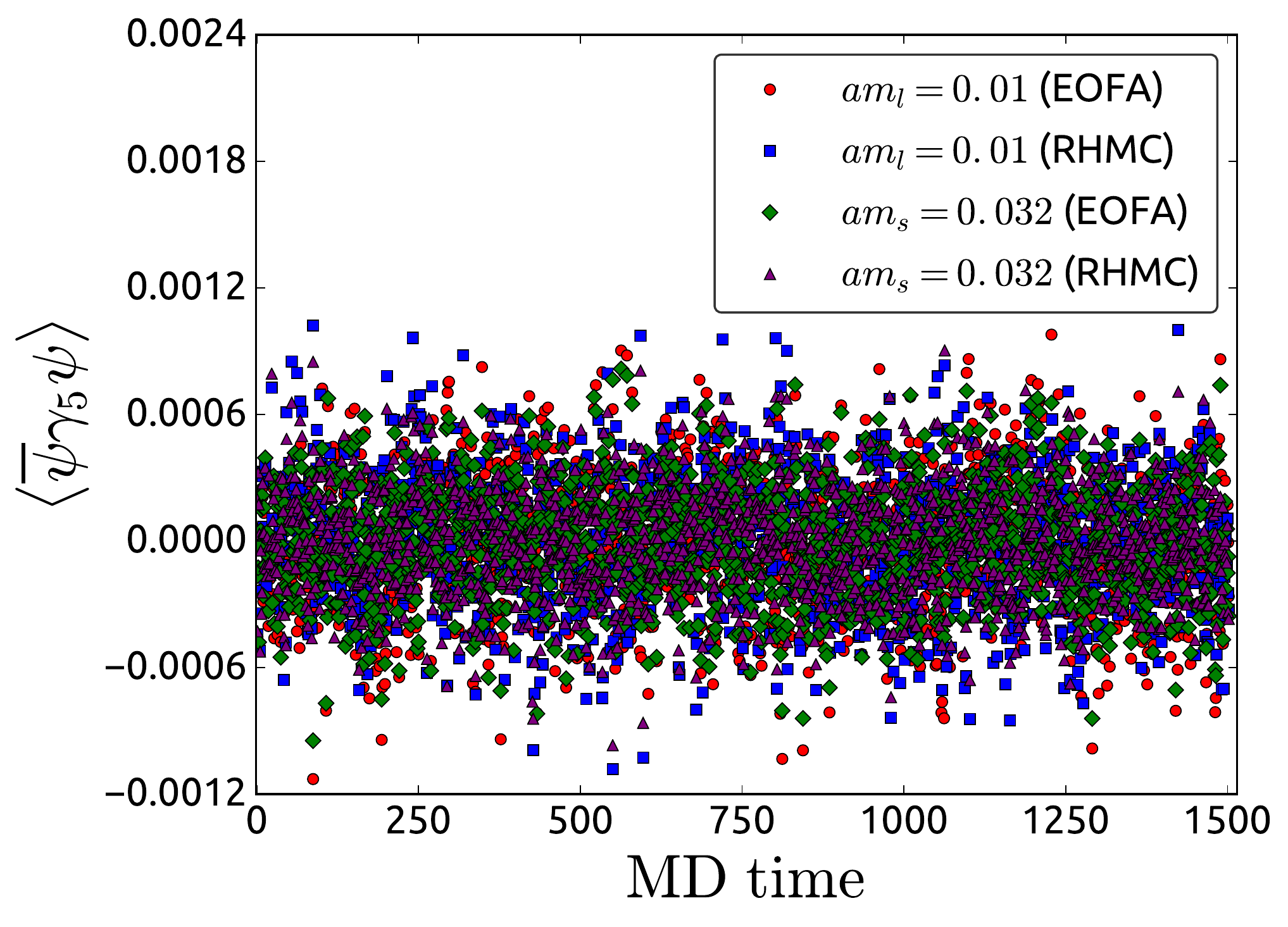}} \\
\caption{Molecular dynamics evolution of the average plaquette, topological charge, and quark condensates on the 16I ensembles.}
\end{figure}

\begin{figure}[!h]
\centering
\subfloat[Plaquette]{\includegraphics[width=0.49\linewidth]{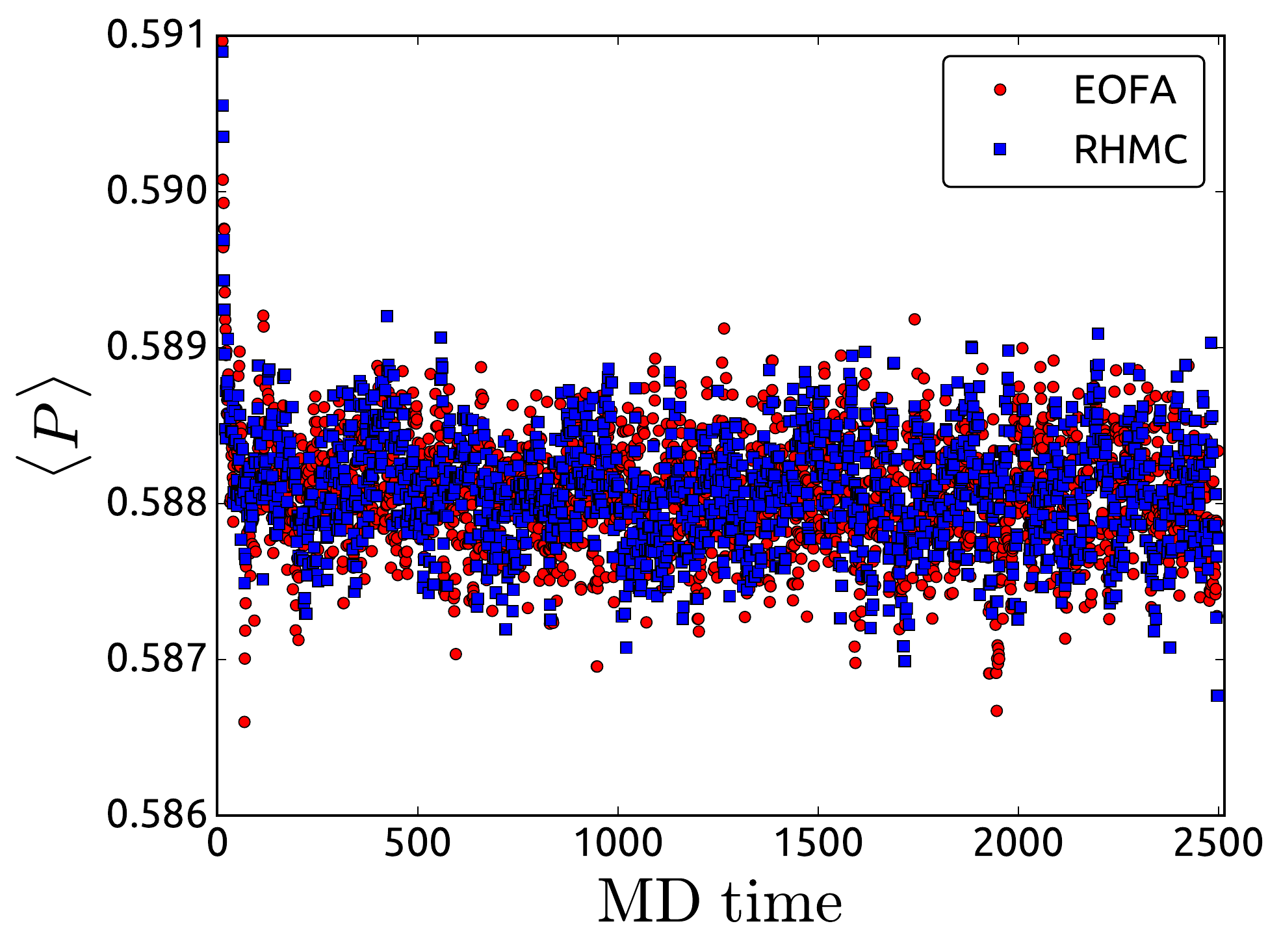}} \quad
\subfloat[Topological Charge]{\includegraphics[width=0.47\linewidth]{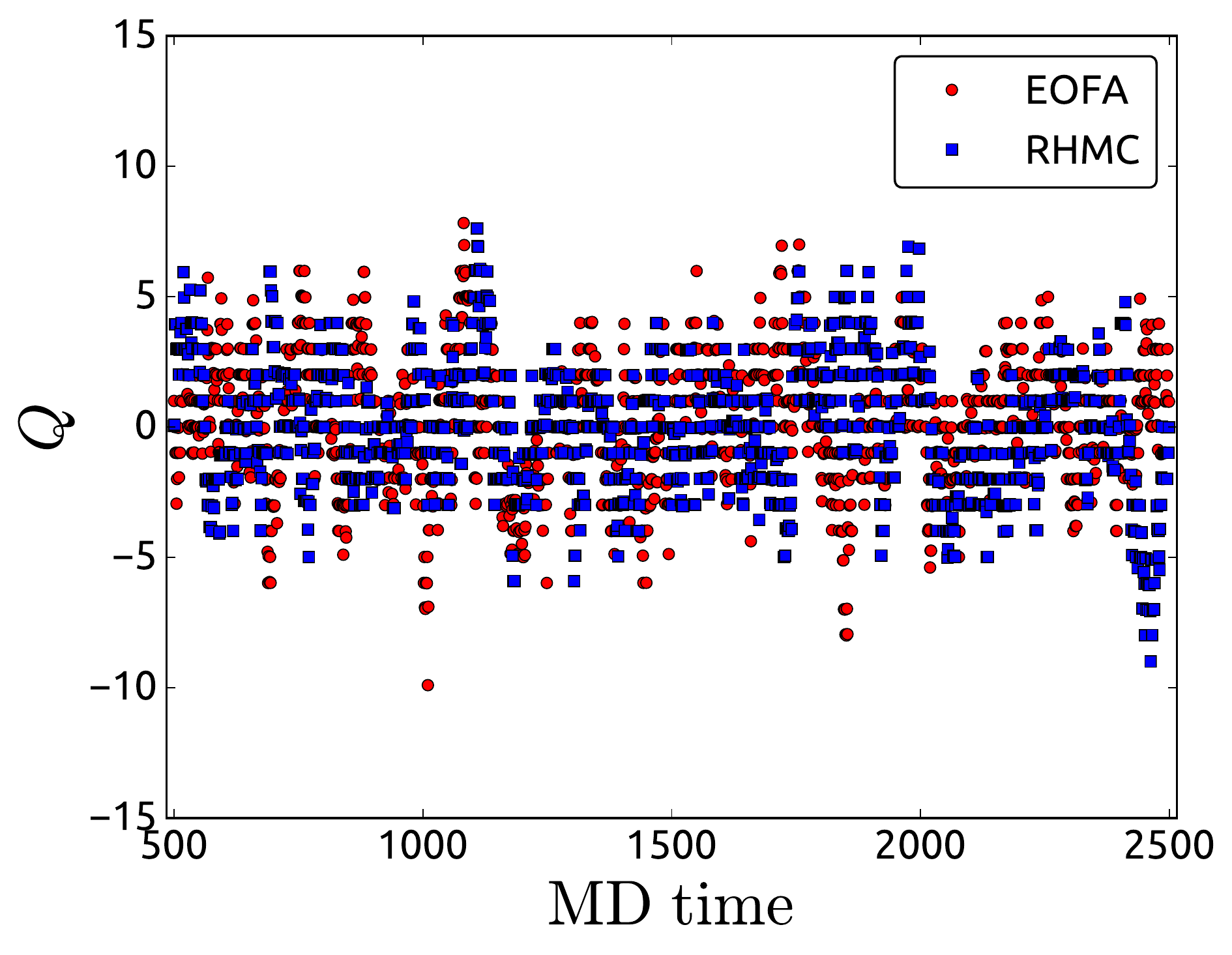}} \\
\subfloat[Chiral Condensates]{\includegraphics[width=0.48\linewidth]{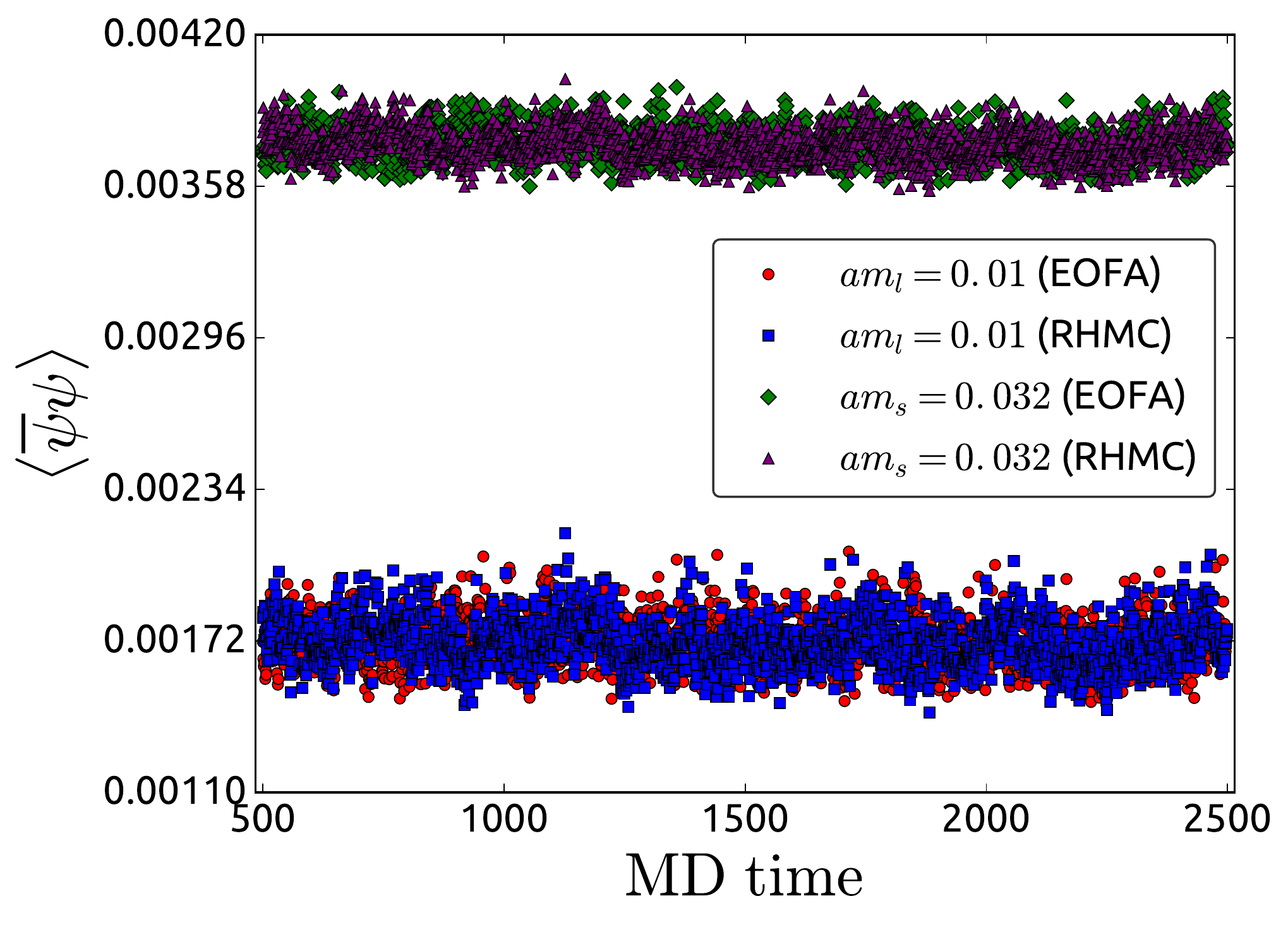}} \quad
\subfloat[Pseudoscalar Condensates]{\includegraphics[width=0.48\linewidth]{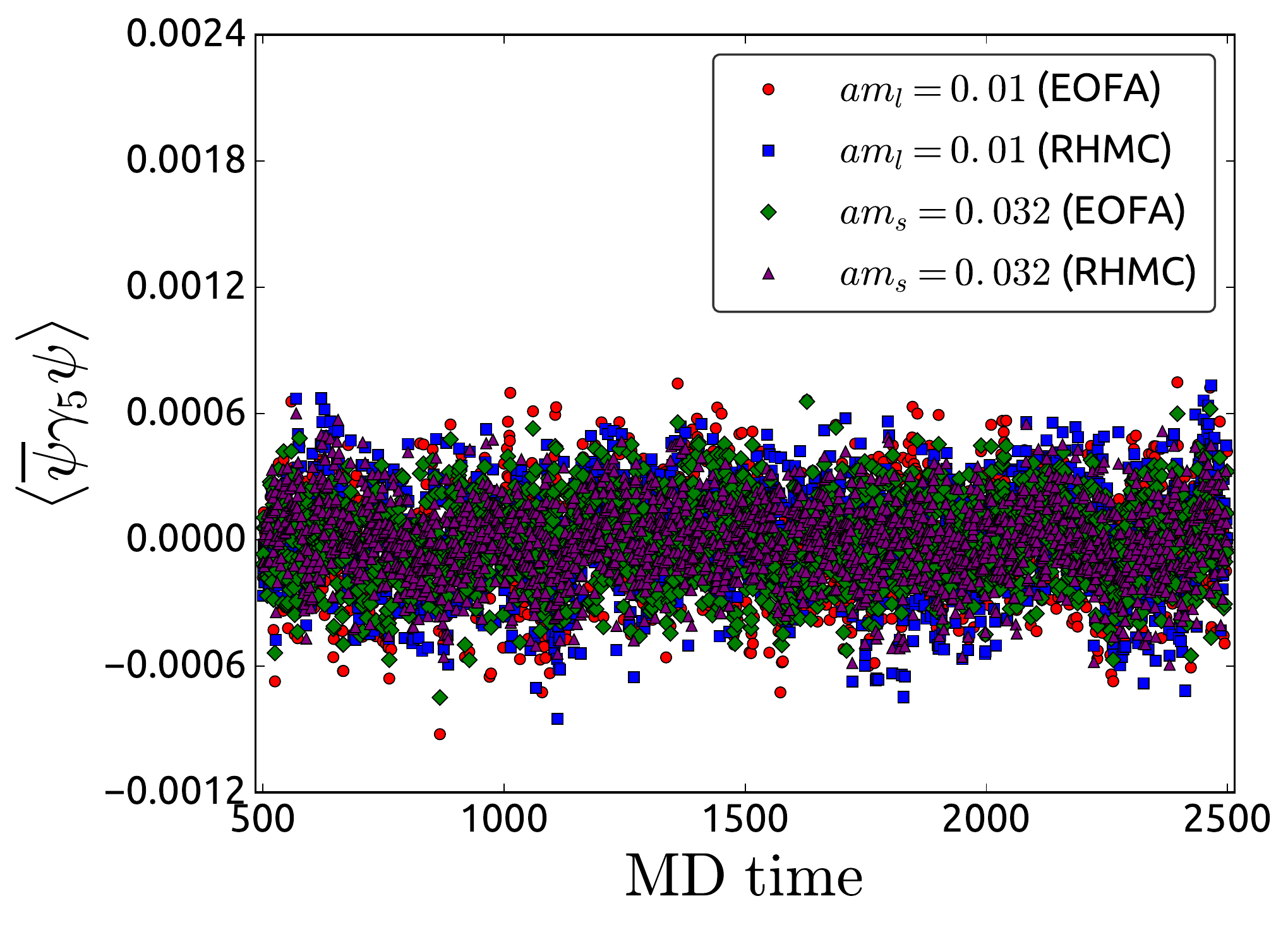}} \\
\caption{Molecular dynamics evolution of the average plaquette, topological charge, and quark condensates on the 16I-G ensembles.}
\end{figure}

\begin{figure}[!h]
\centering
\subfloat[Plaquette]{\includegraphics[width=0.49\linewidth]{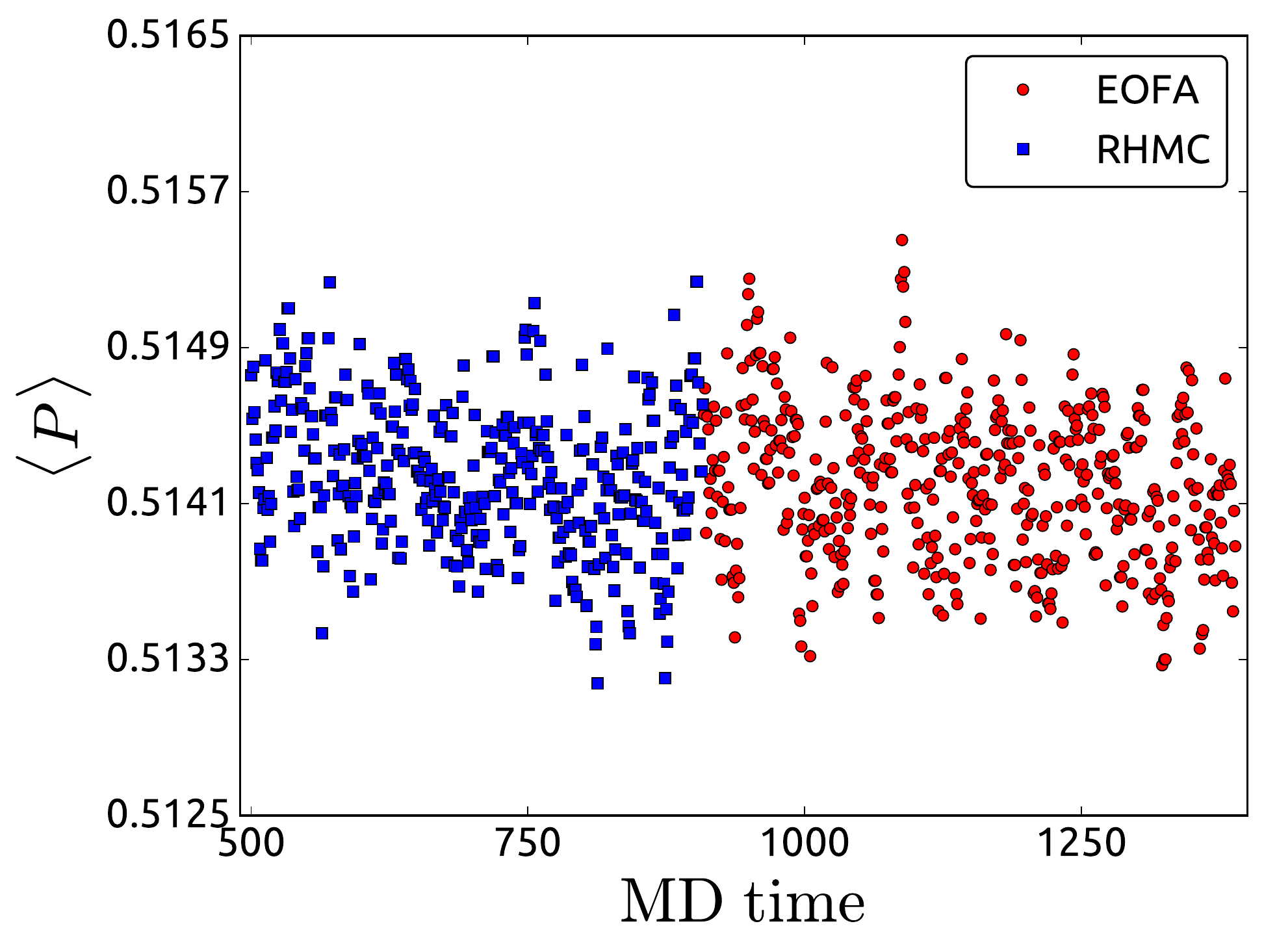}} \quad
\subfloat[Topological Charge]{\includegraphics[width=0.47\linewidth]{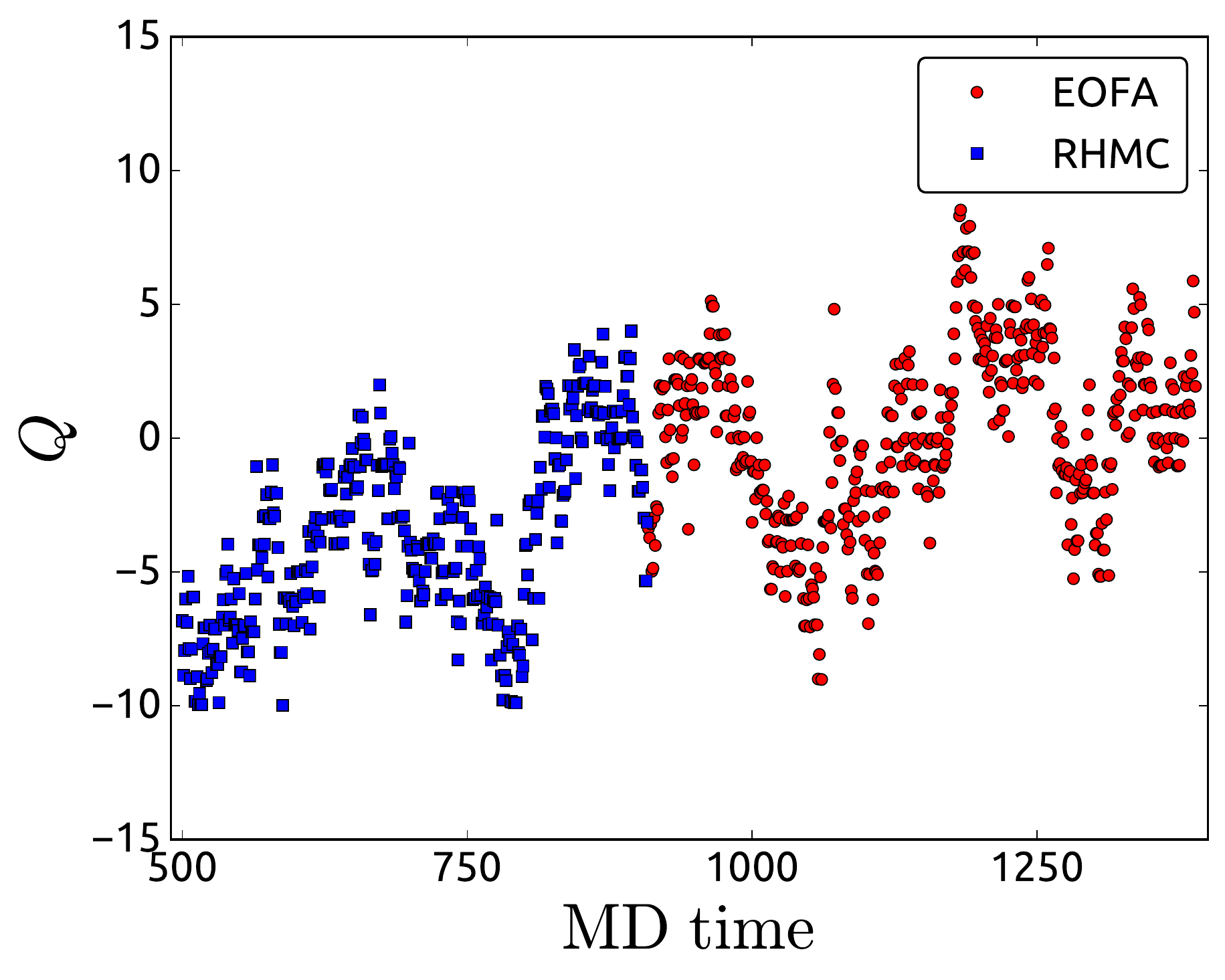}} \\
\subfloat[Chiral Condensates]{\includegraphics[width=0.48\linewidth]{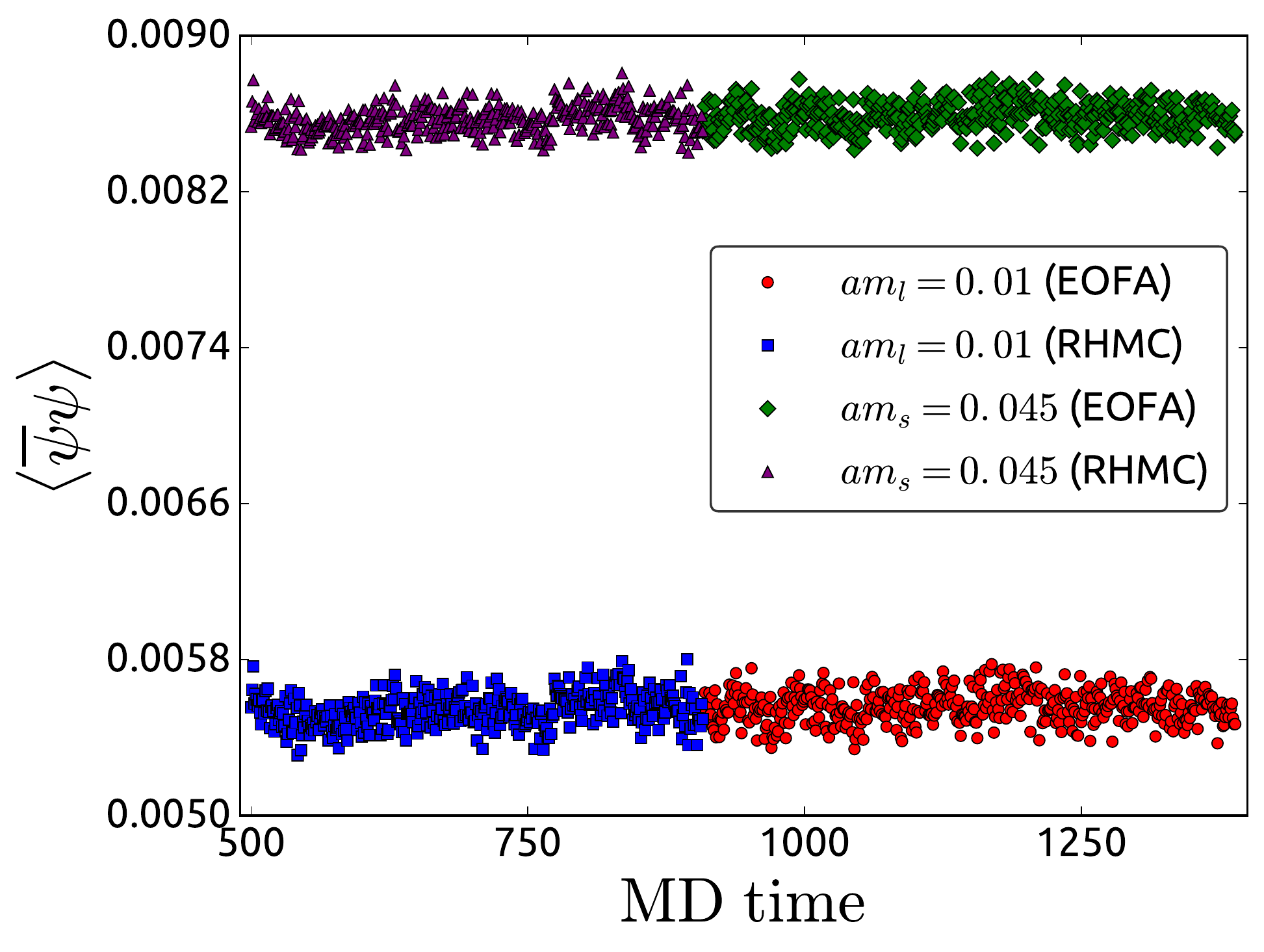}} \quad
\subfloat[Pseudoscalar Condensates]{\includegraphics[width=0.48\linewidth]{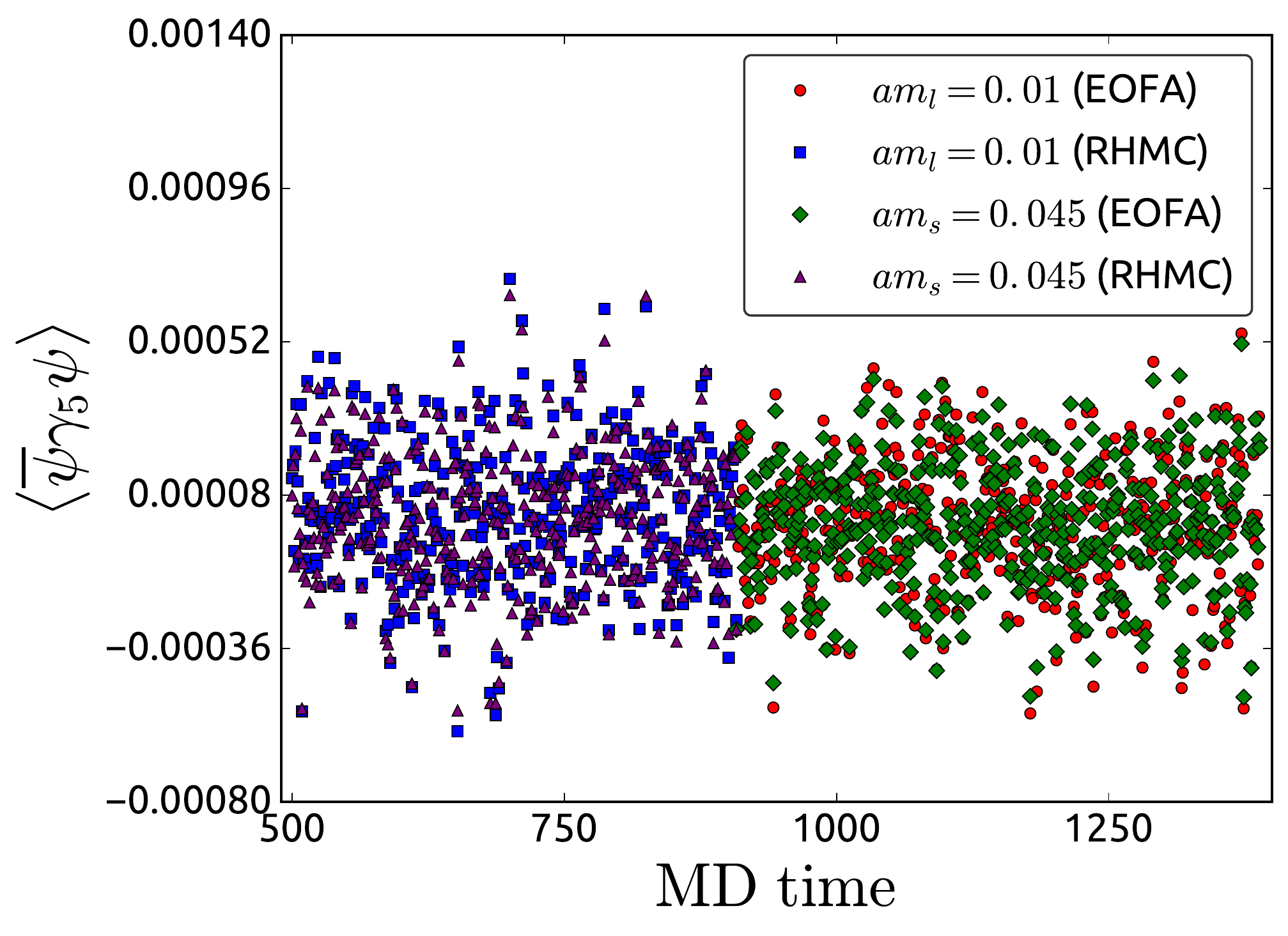}} \\
\caption{Molecular dynamics evolution of the average plaquette, topological charge, and quark condensates on the 16ID-G ensembles.}
\end{figure}
\FloatBarrier

\subsection{Effective Mass Plots}

\begin{figure}[!h]
\centering
\subfloat{\includegraphics[width=0.48\linewidth]{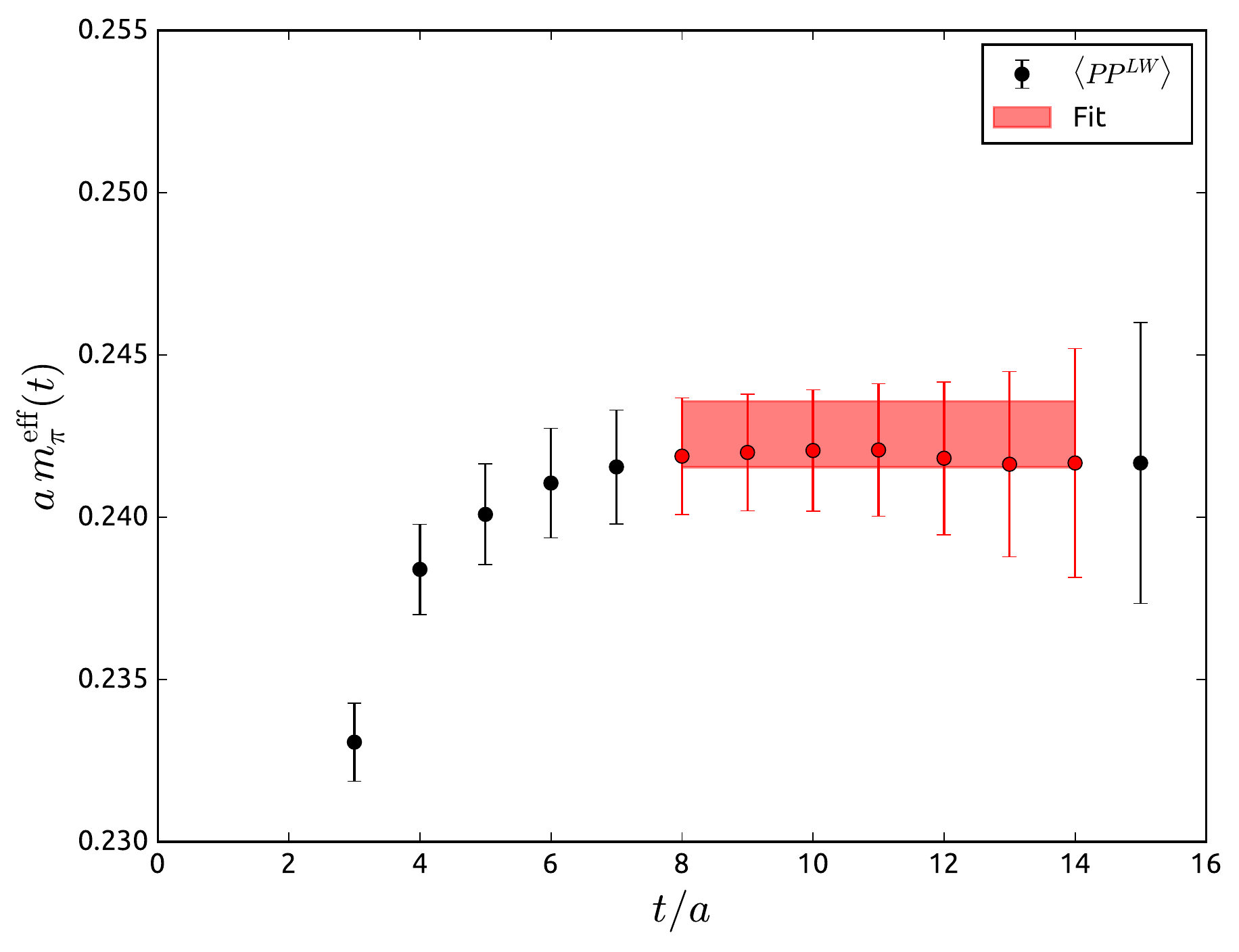}} \quad
\subfloat{\includegraphics[width=0.48\linewidth]{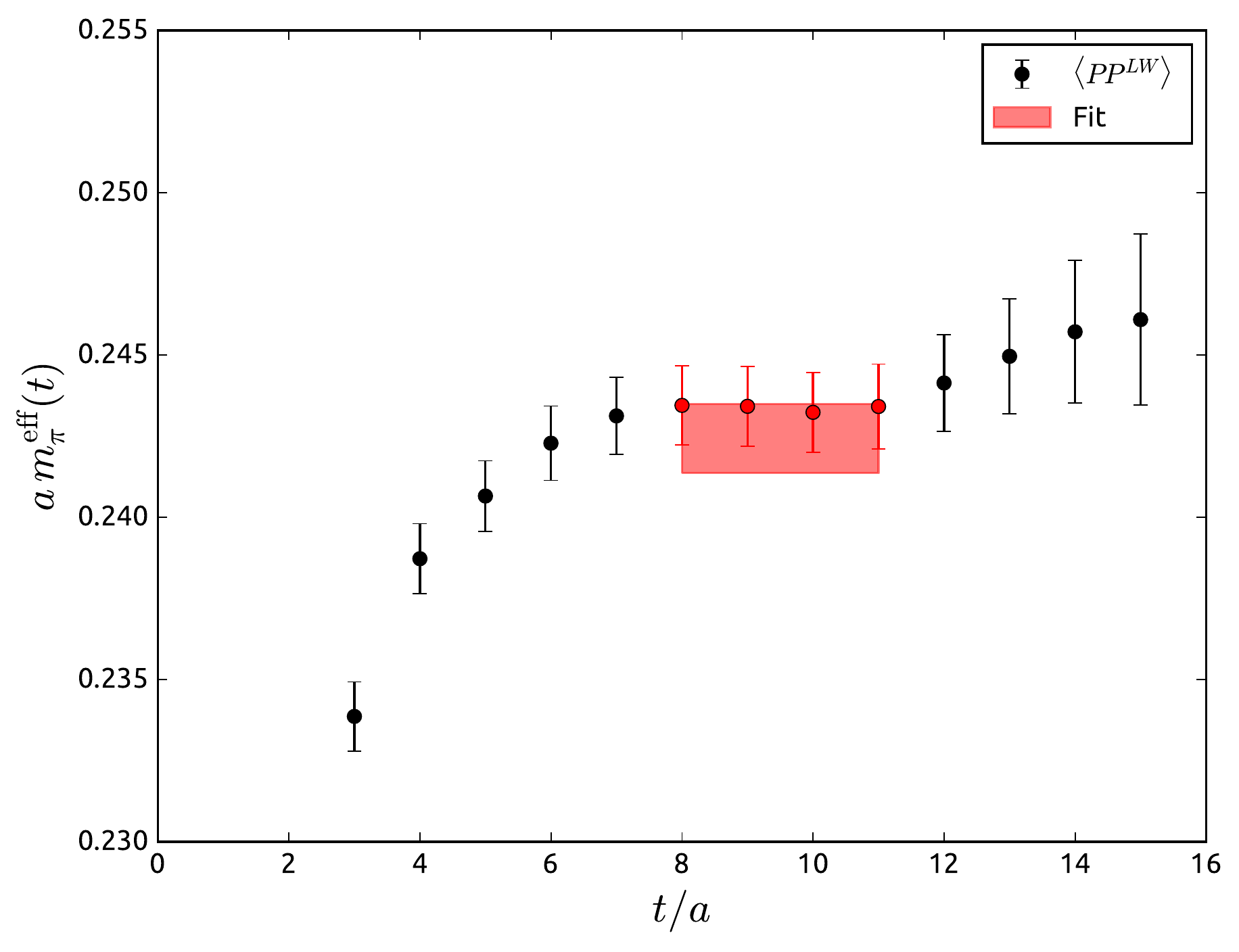}} \\
\subfloat{\includegraphics[width=0.48\linewidth]{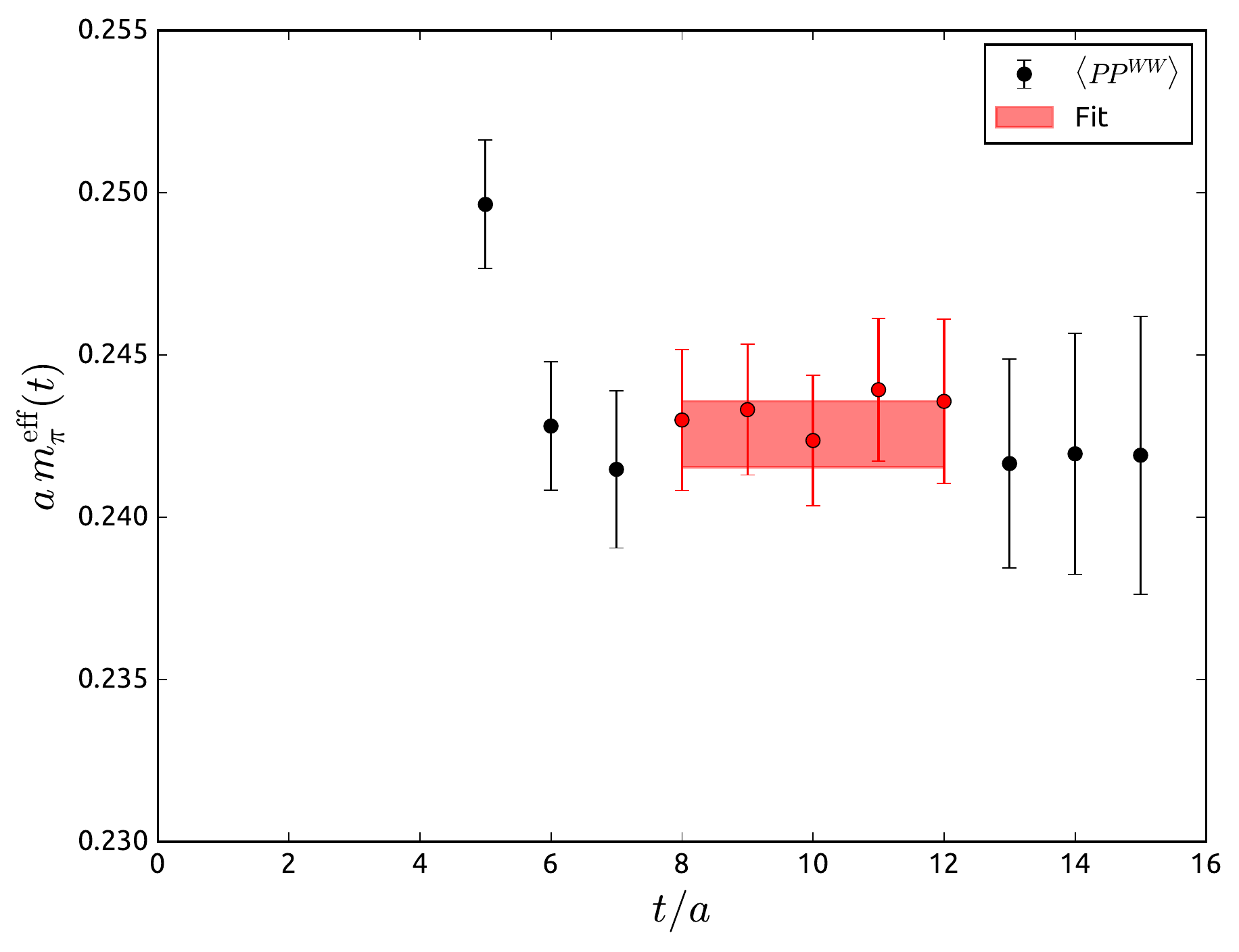}} \quad
\subfloat{\includegraphics[width=0.48\linewidth]{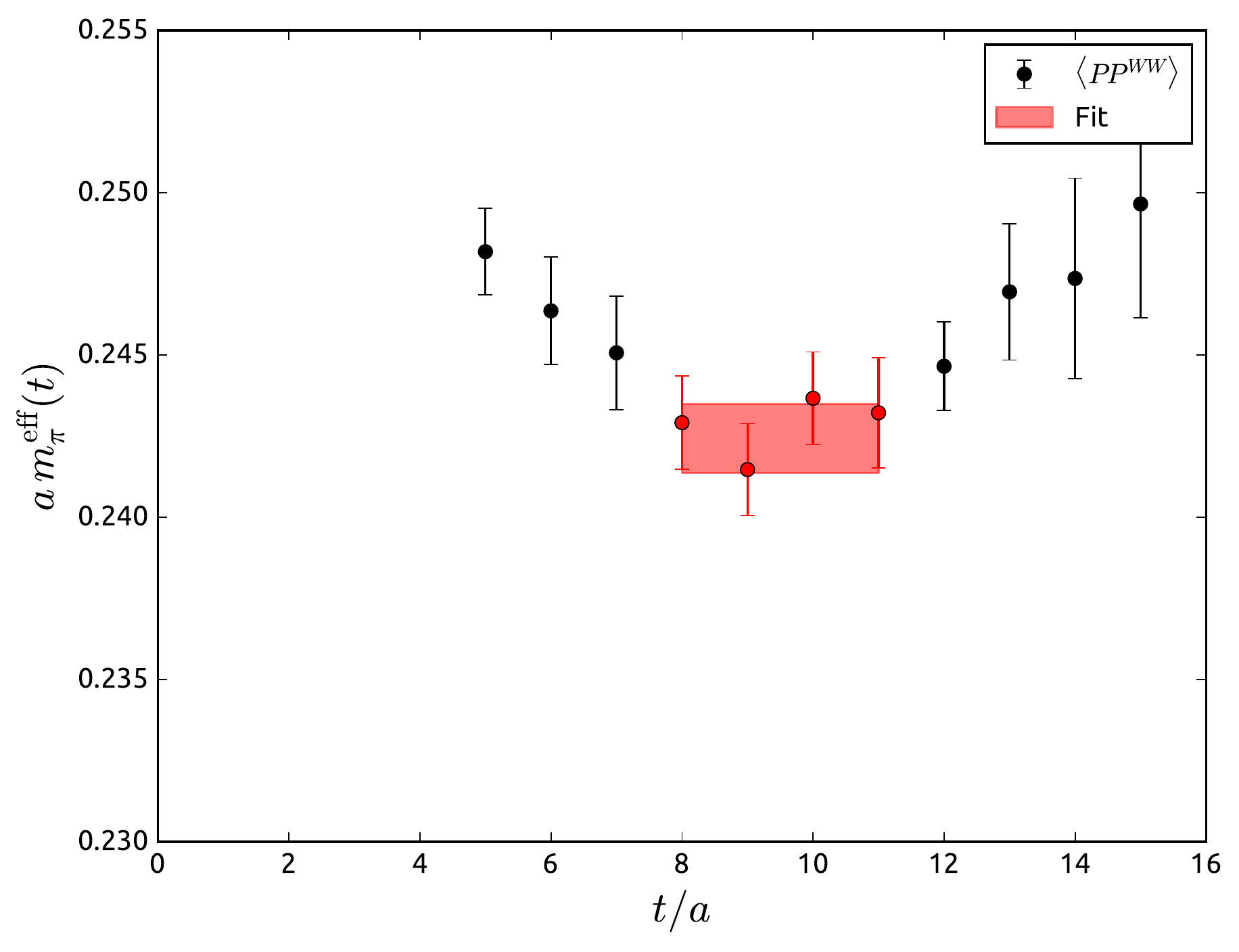}} \\
\subfloat{\includegraphics[width=0.48\linewidth]{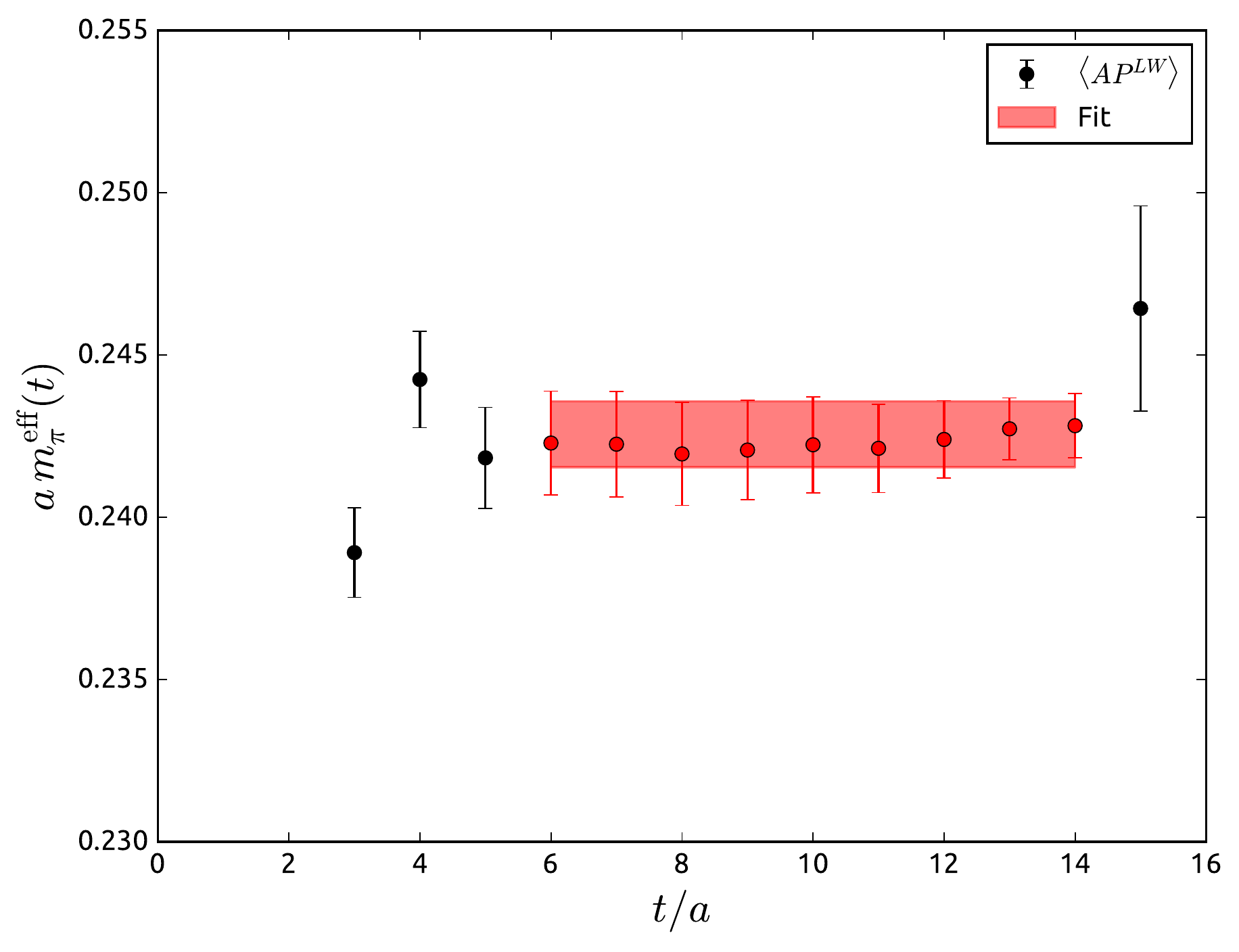}} \quad
\subfloat{\includegraphics[width=0.48\linewidth]{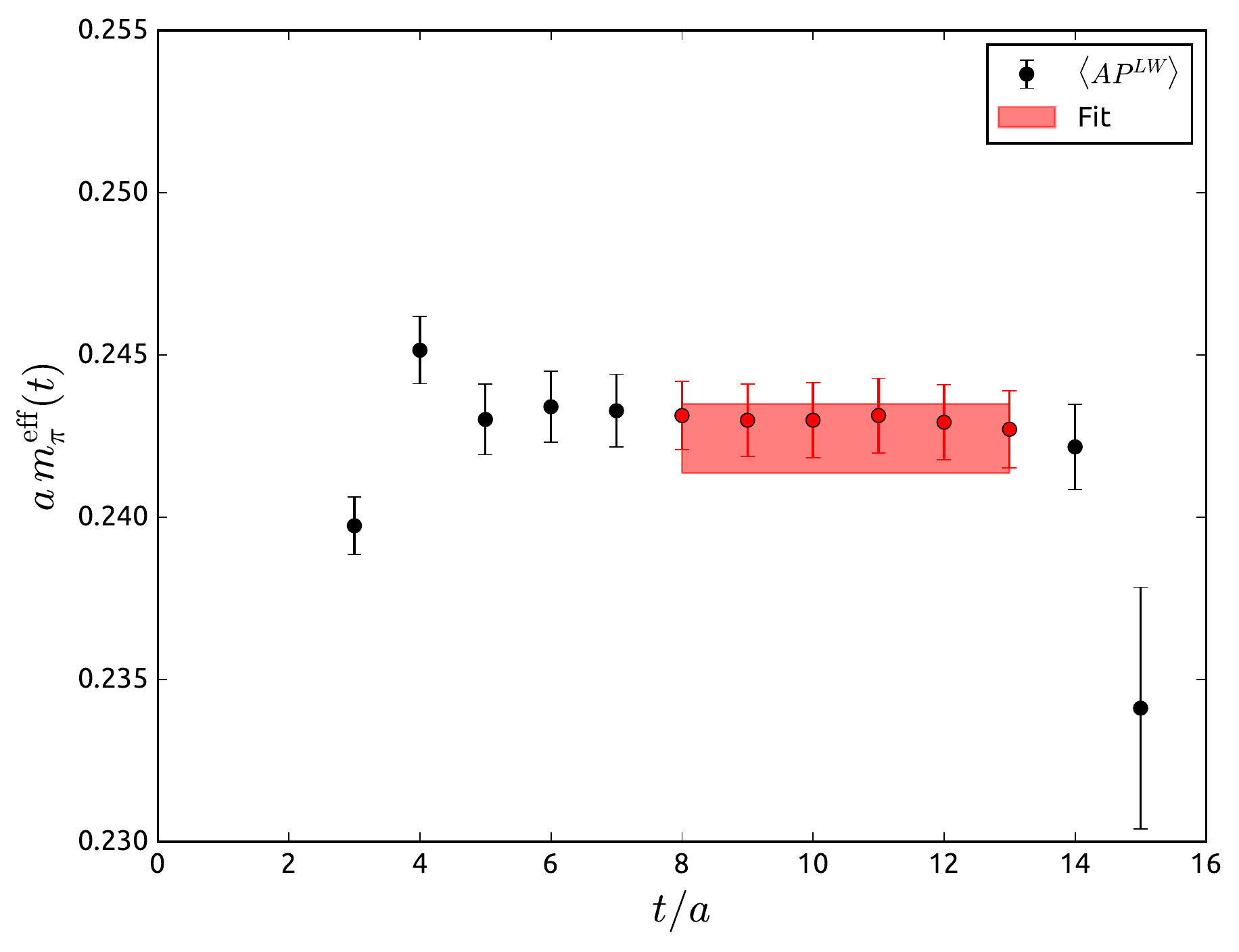}}
\caption{Effective pion mass from a simultaneous fit to the $\langle PP^{LW} \rangle$ (top), $\langle PP^{WW} \rangle$ (middle), and $\langle AP^{LW} \rangle$ (bottom) correlation functions, as measured on the EOFA (left) and RHMC (right) 16I ensembles.} 
\end{figure}

\begin{figure}[!h]
\centering
\subfloat{\includegraphics[width=0.48\linewidth]{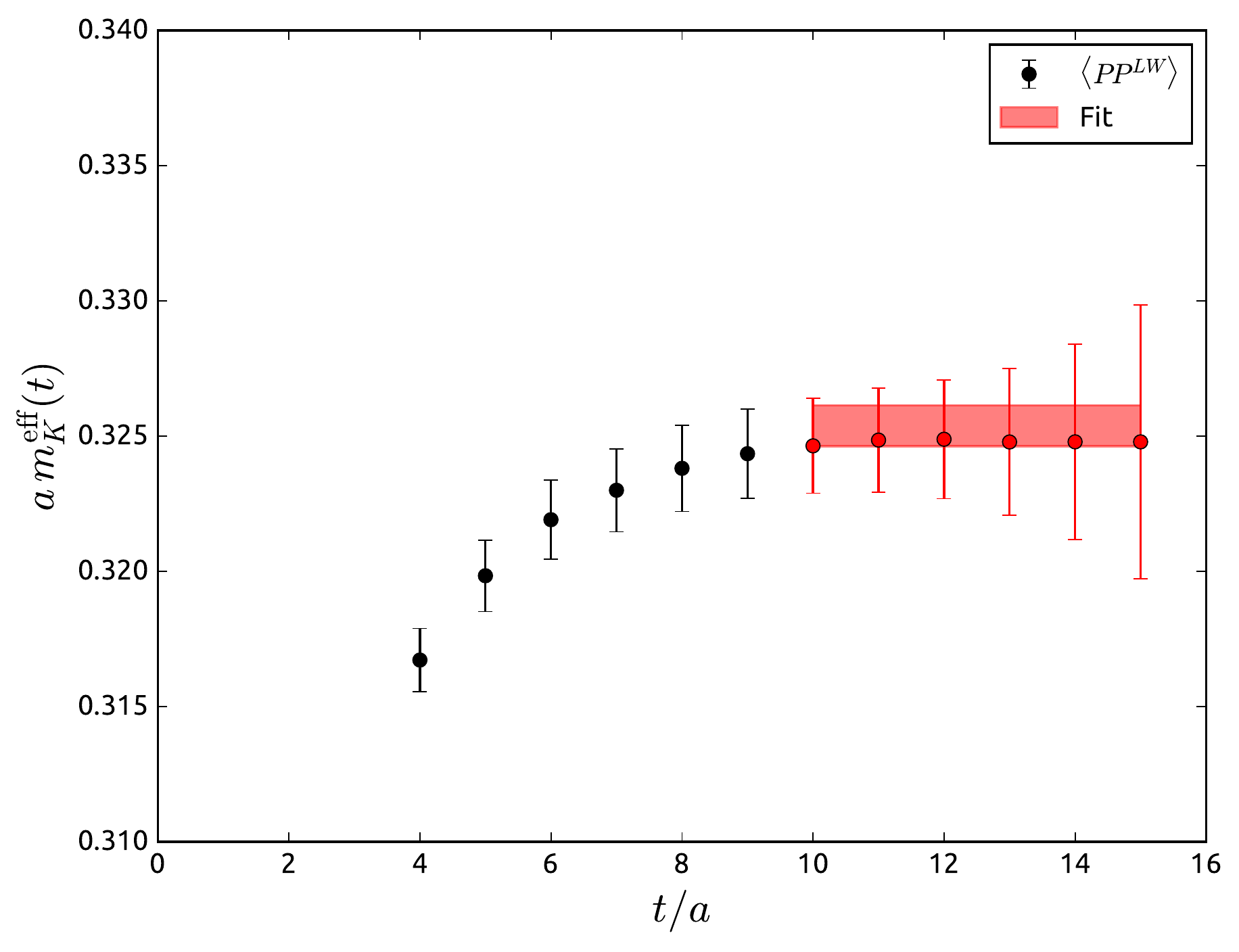}} \quad
\subfloat{\includegraphics[width=0.48\linewidth]{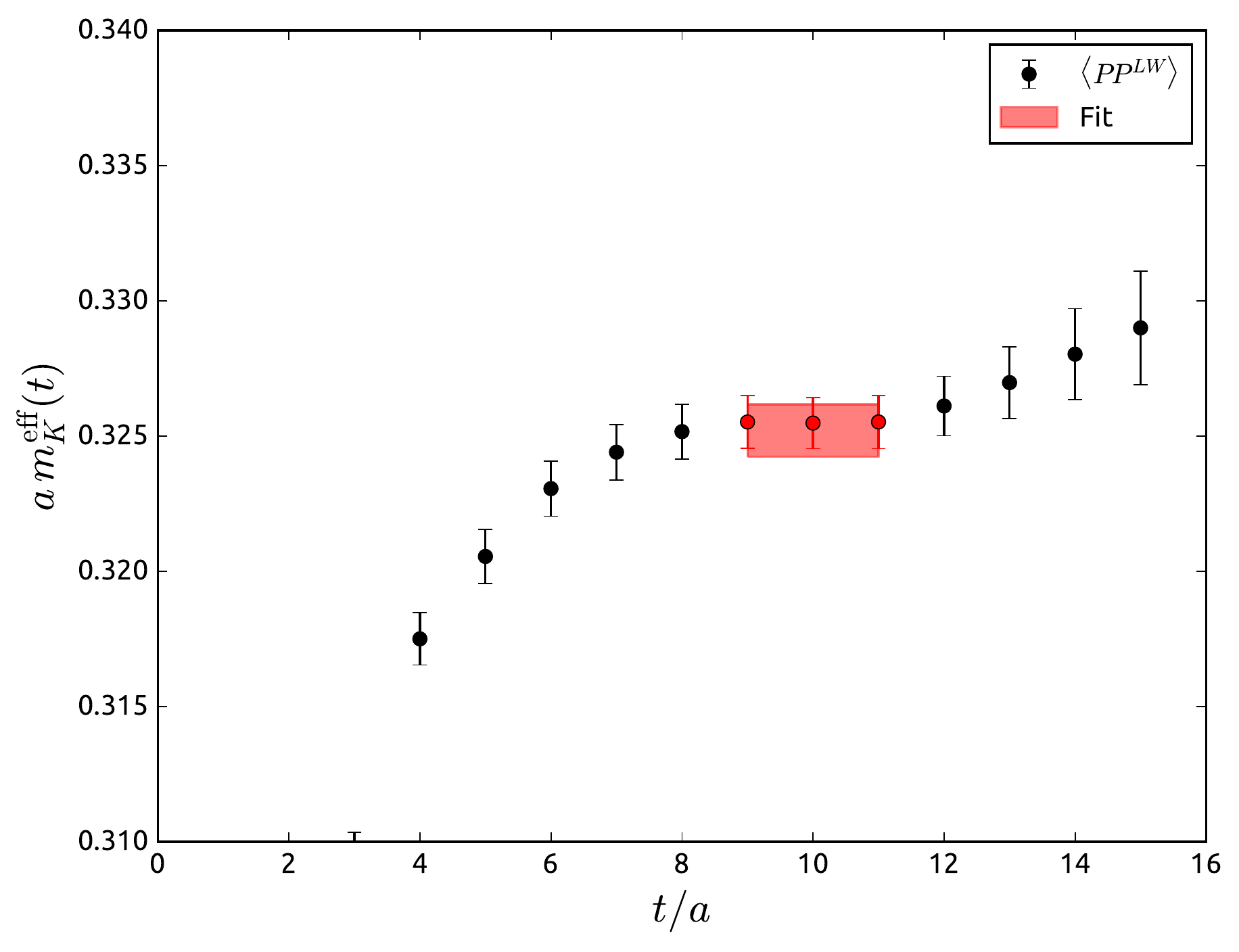}} \\
\subfloat{\includegraphics[width=0.48\linewidth]{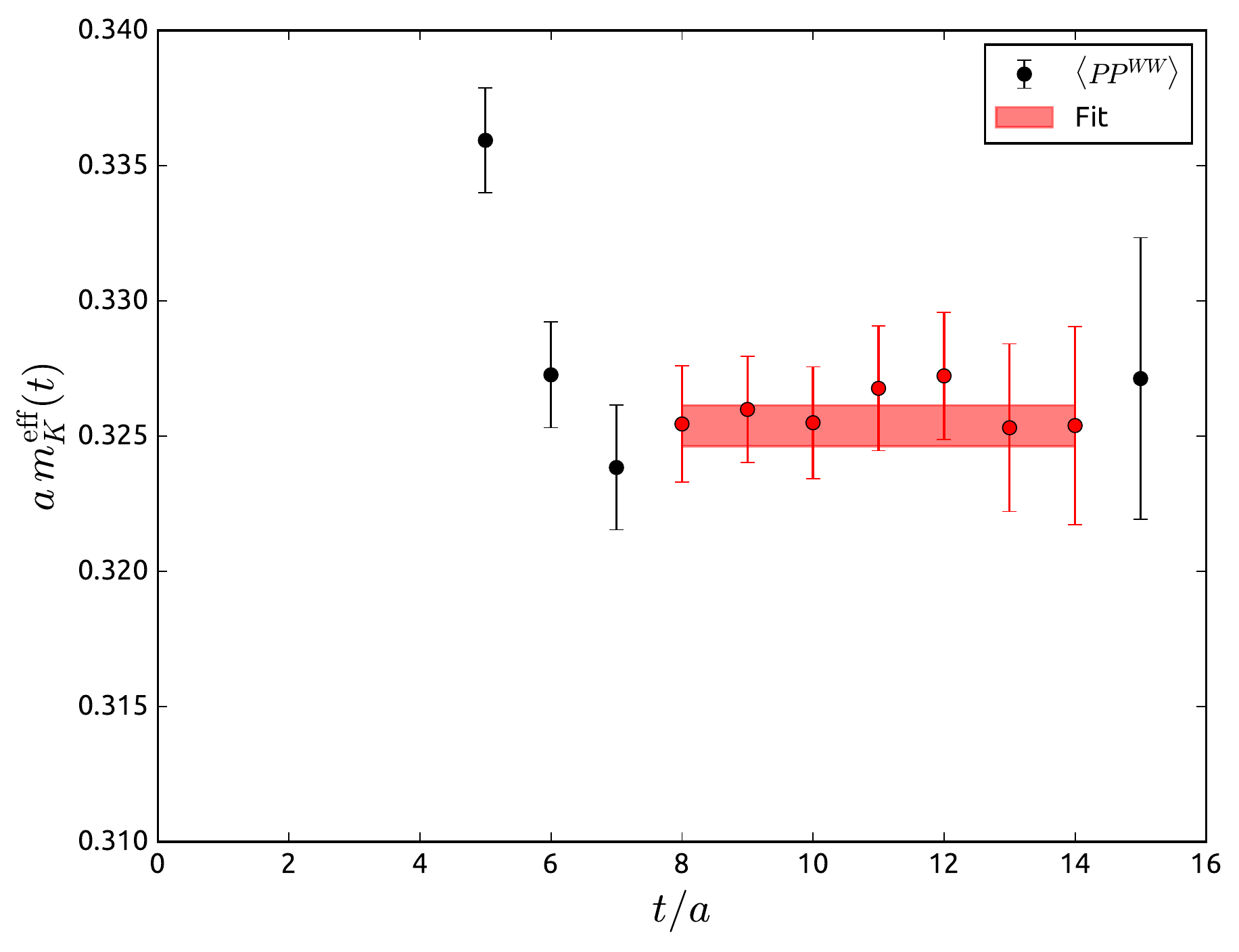}} \quad
\subfloat{\includegraphics[width=0.48\linewidth]{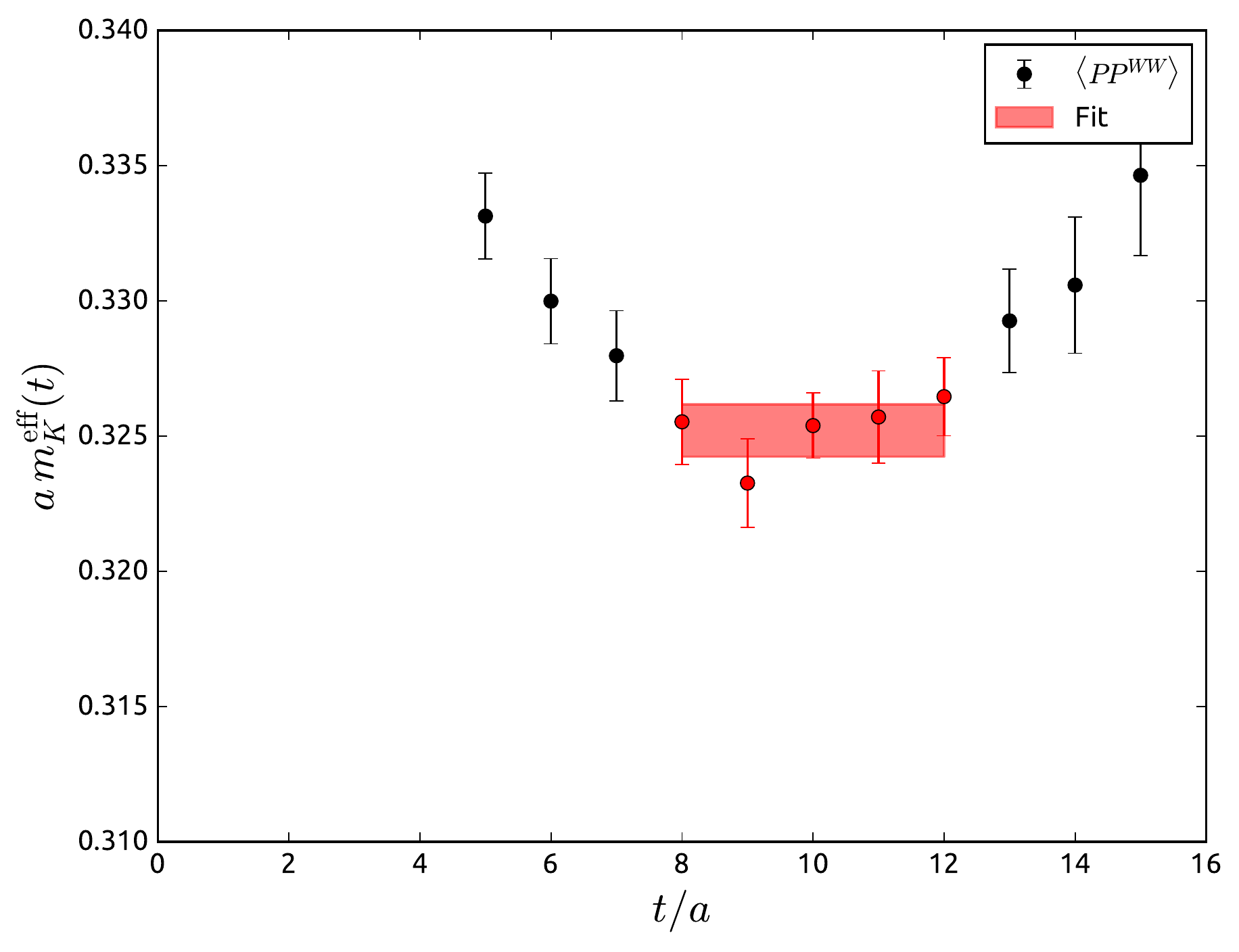}} \\
\subfloat{\includegraphics[width=0.48\linewidth]{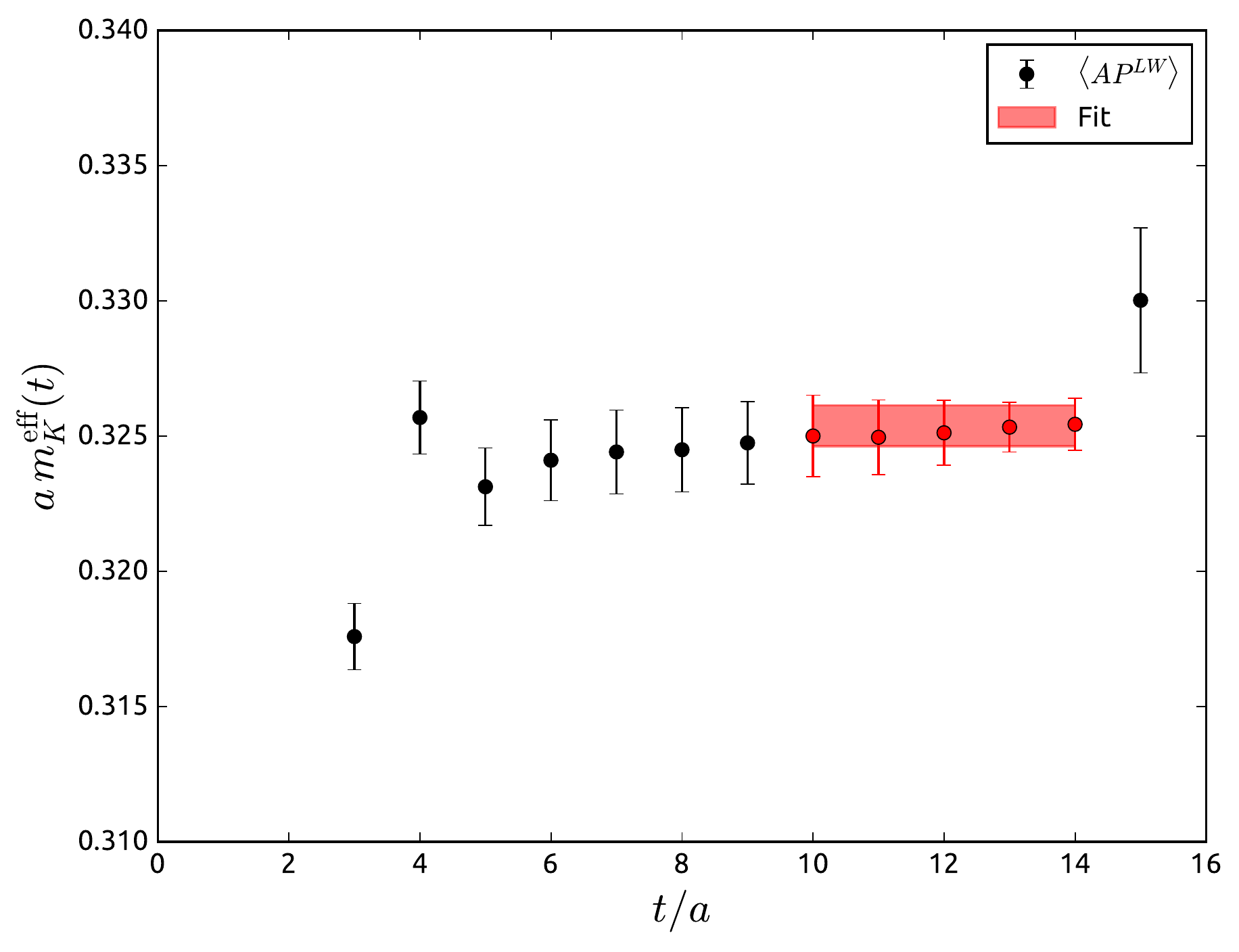}} \quad
\subfloat{\includegraphics[width=0.48\linewidth]{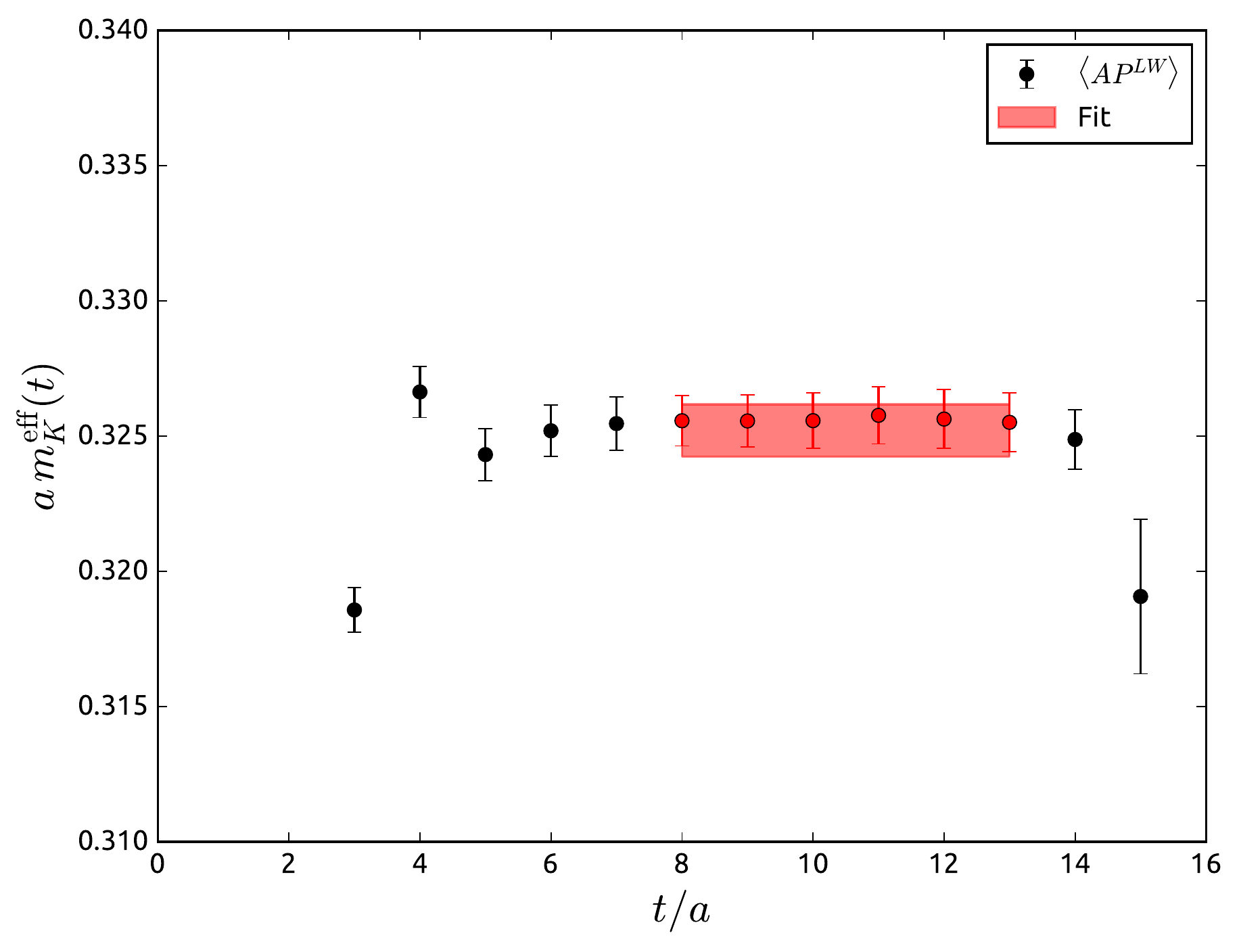}}
\caption{Effective kaon mass from a simultaneous fit to the $\langle PP^{LW} \rangle$ (top), $\langle PP^{WW} \rangle$ (middle), and $\langle AP^{LW} \rangle$ (bottom) correlation functions, as measured on the EOFA (left) and RHMC (right) 16I ensembles.} 
\end{figure}

\begin{figure}[!h]
\centering
\subfloat{\includegraphics[width=0.48\linewidth]{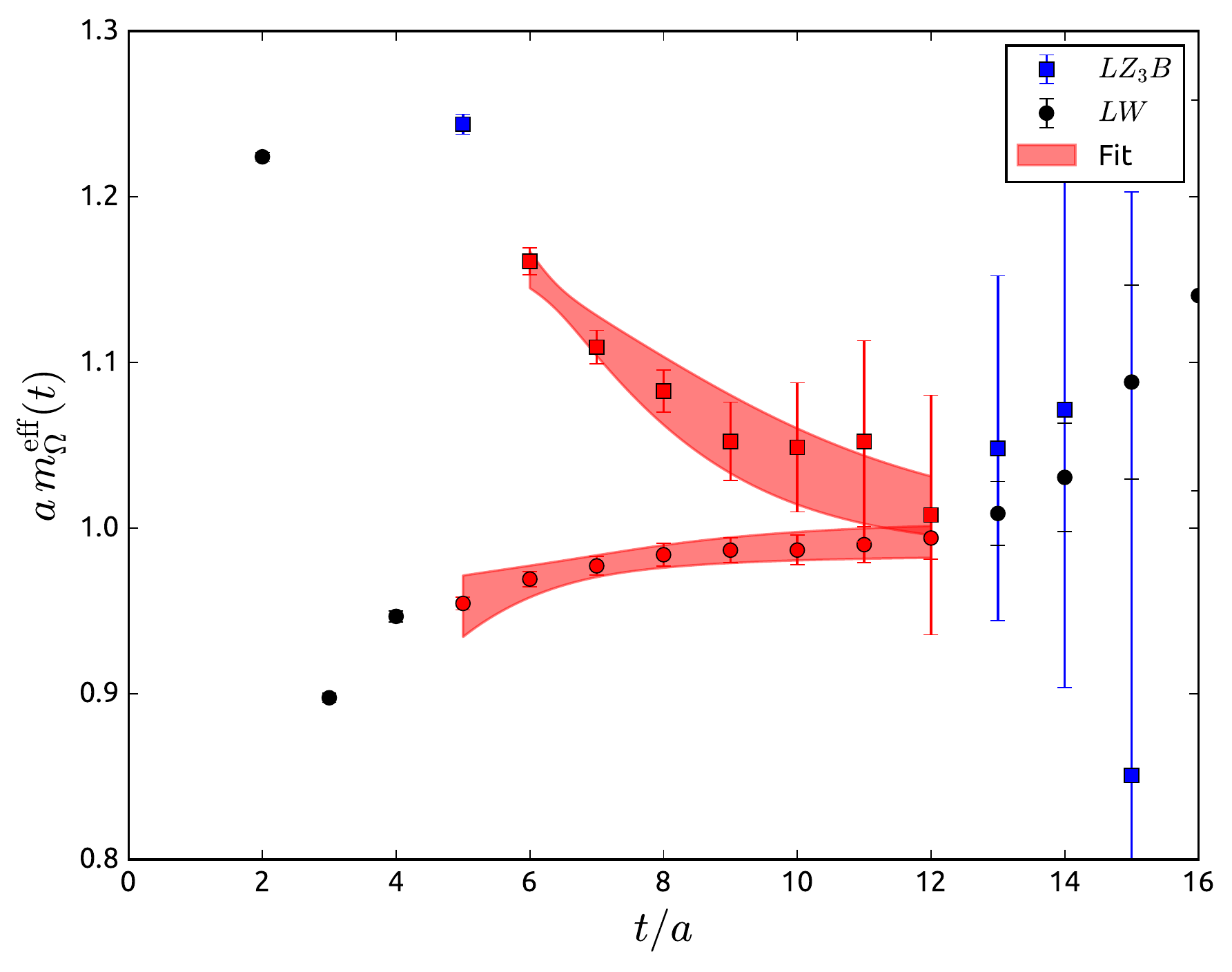}} \quad
\subfloat{\includegraphics[width=0.48\linewidth]{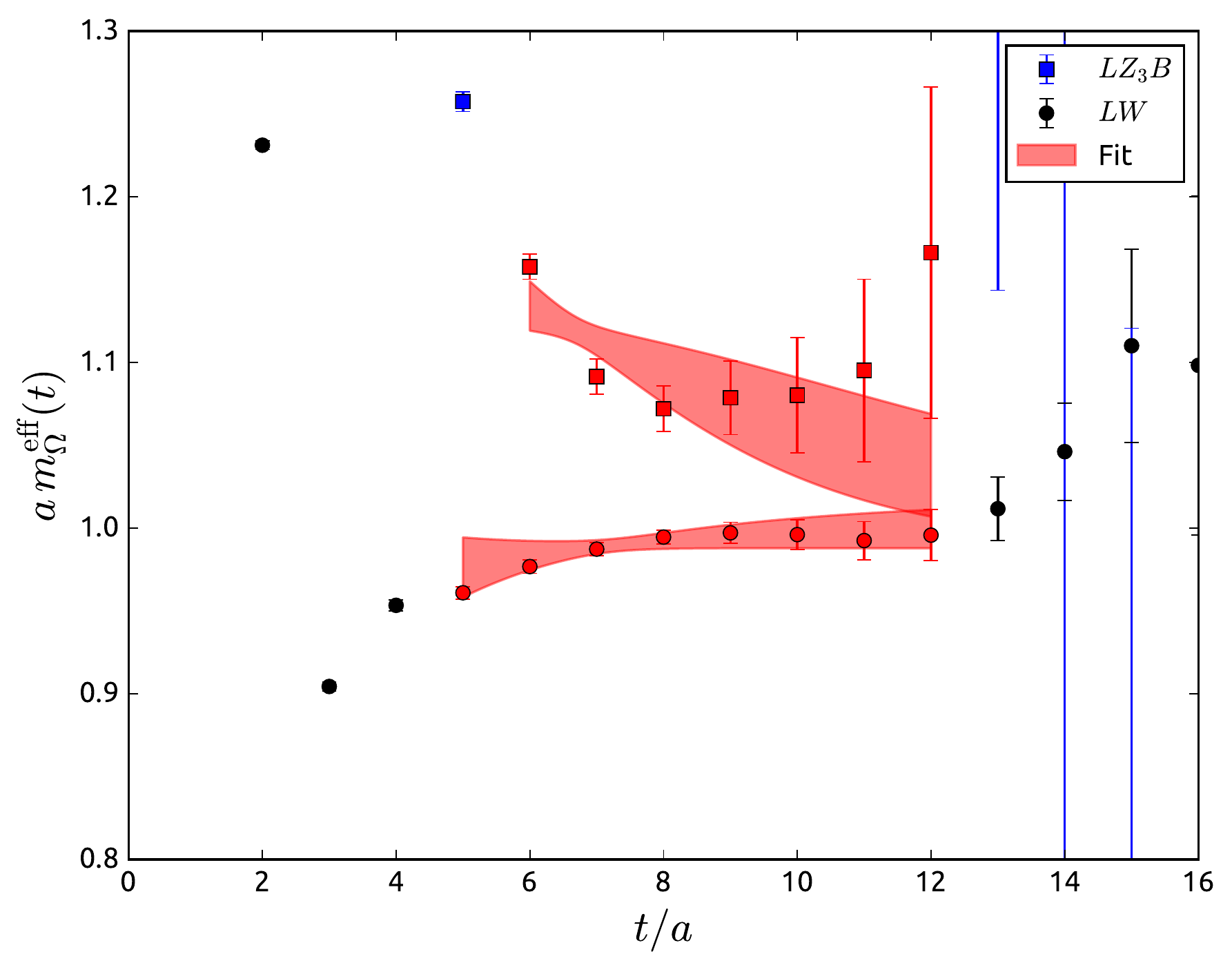}} \\
\caption{Effective $\Omega$ baryon mass from a simultaneous two-state fit to wall and $Z_{3}$ noise sources, as measured on the EOFA (left) and RHMC (right) 16I ensembles.} 
\end{figure}

\begin{figure}[!h]
\centering
\subfloat{\includegraphics[width=0.48\linewidth]{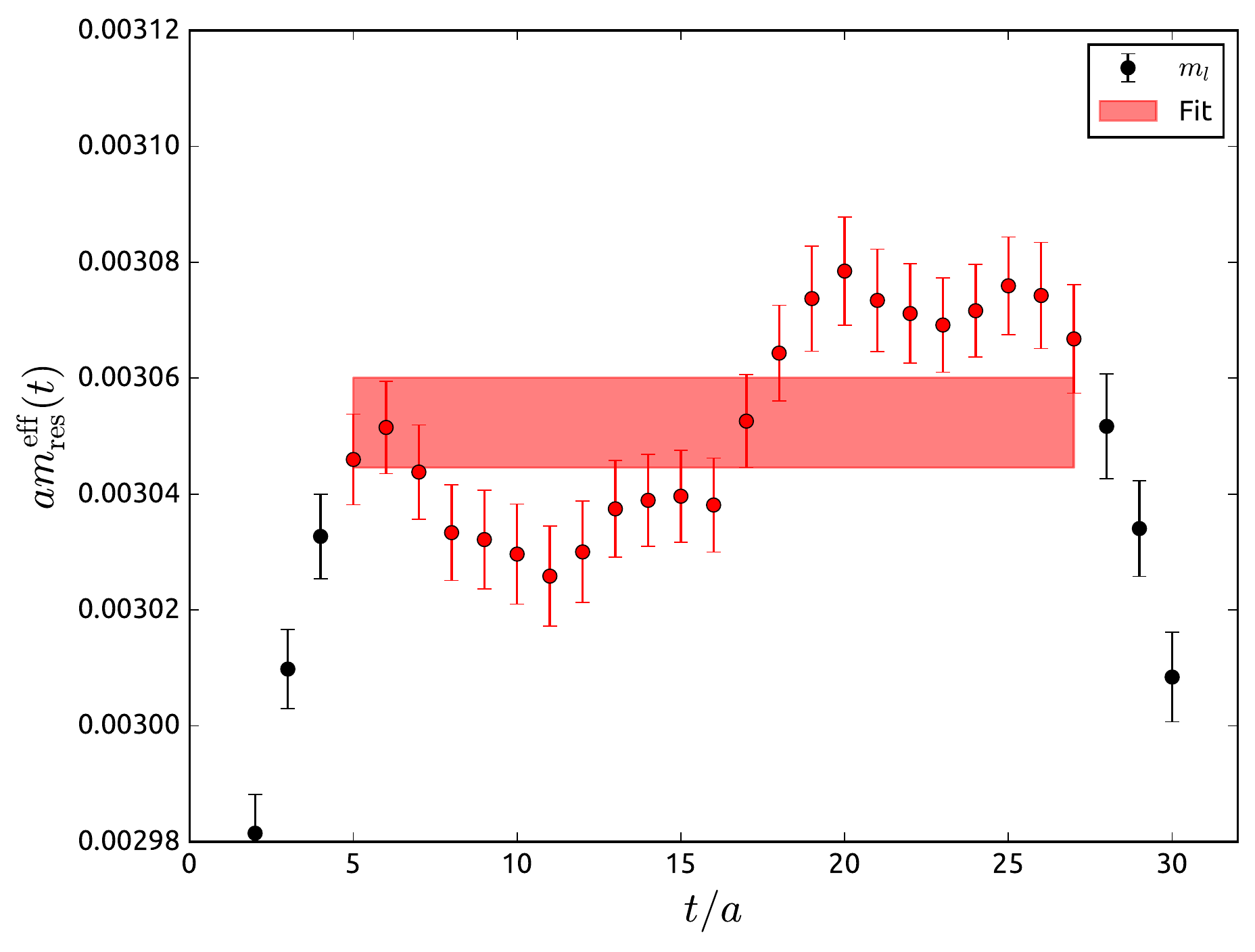}} \quad
\subfloat{\includegraphics[width=0.48\linewidth]{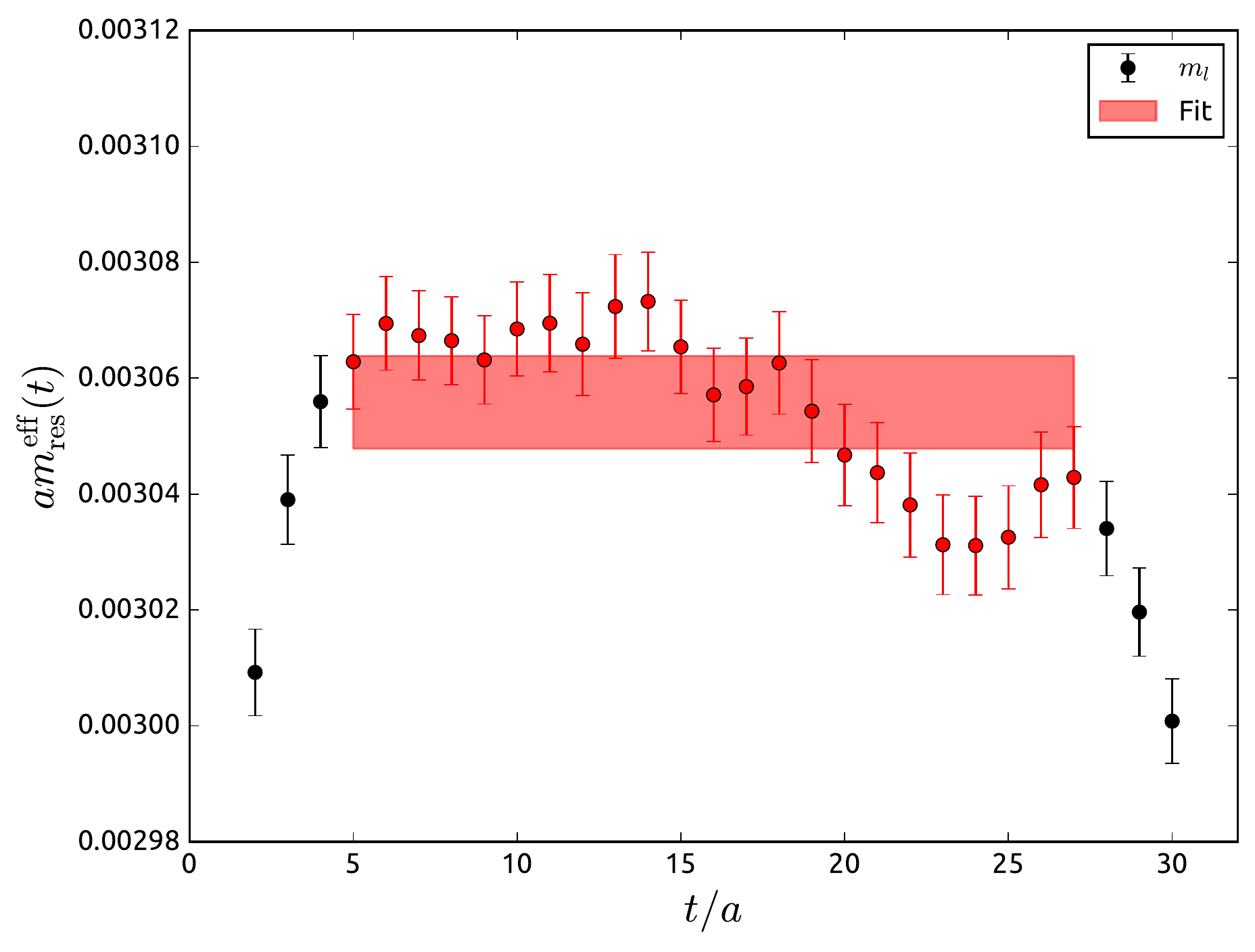}} \\
\caption{Effective $a m_{\rm res}'(m_{l})$, as measured on the EOFA (left) and RHMC (right) 16I ensembles.} 
\end{figure}

\begin{figure}[!h]
\centering
\subfloat{\includegraphics[width=0.48\linewidth]{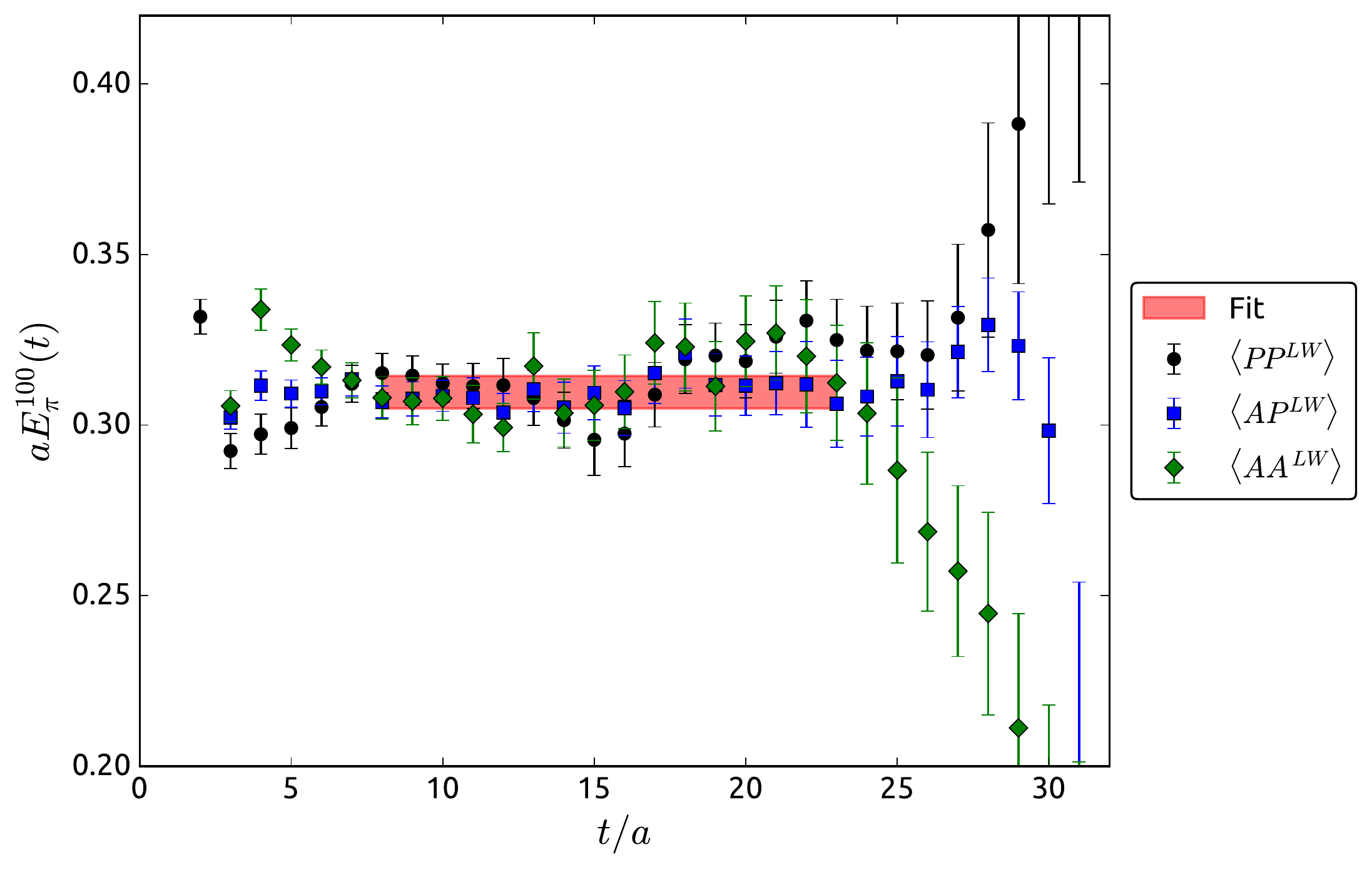}} \quad
\subfloat{\includegraphics[width=0.48\linewidth]{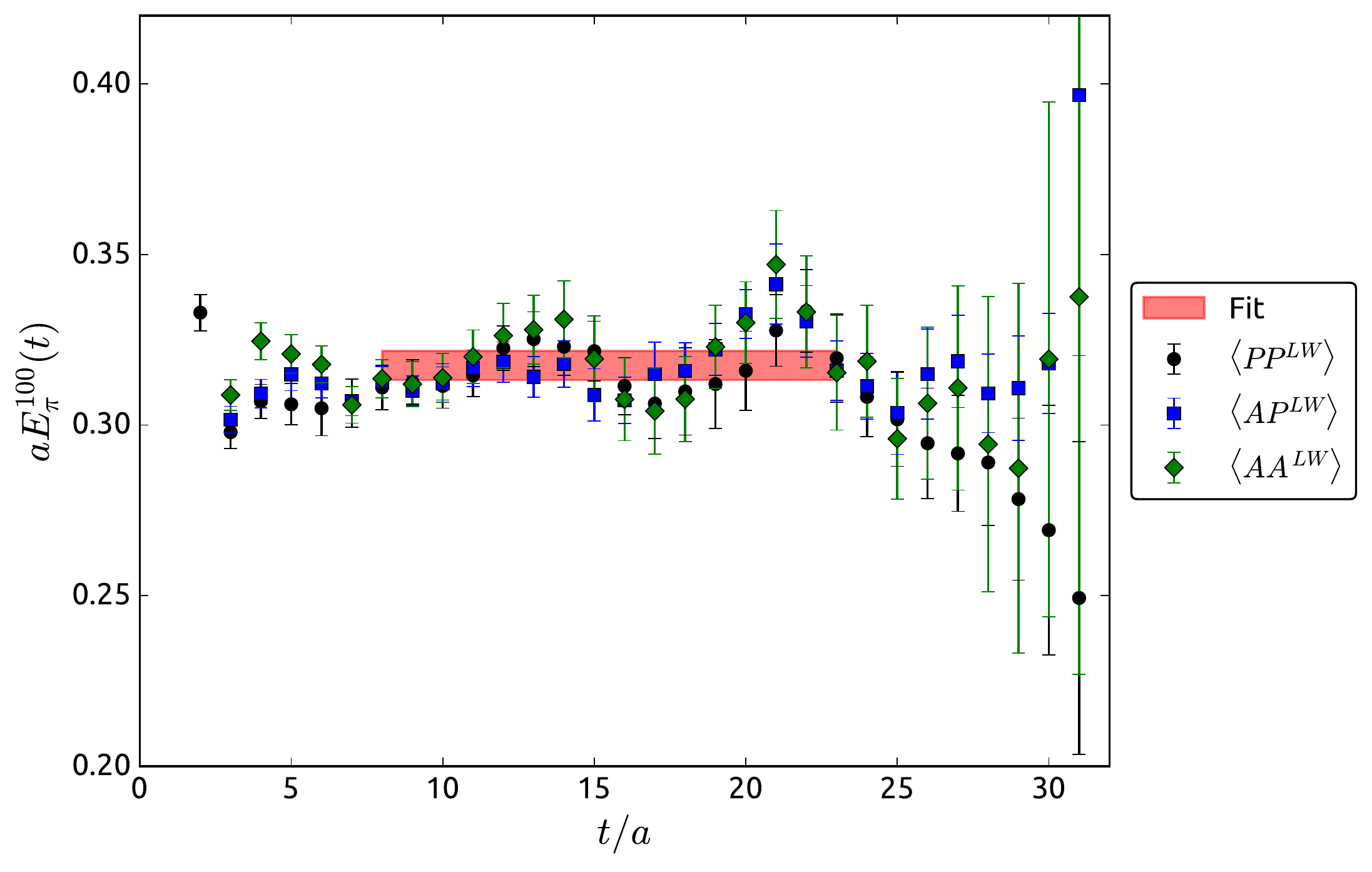}} \\
\subfloat{\includegraphics[width=0.48\linewidth]{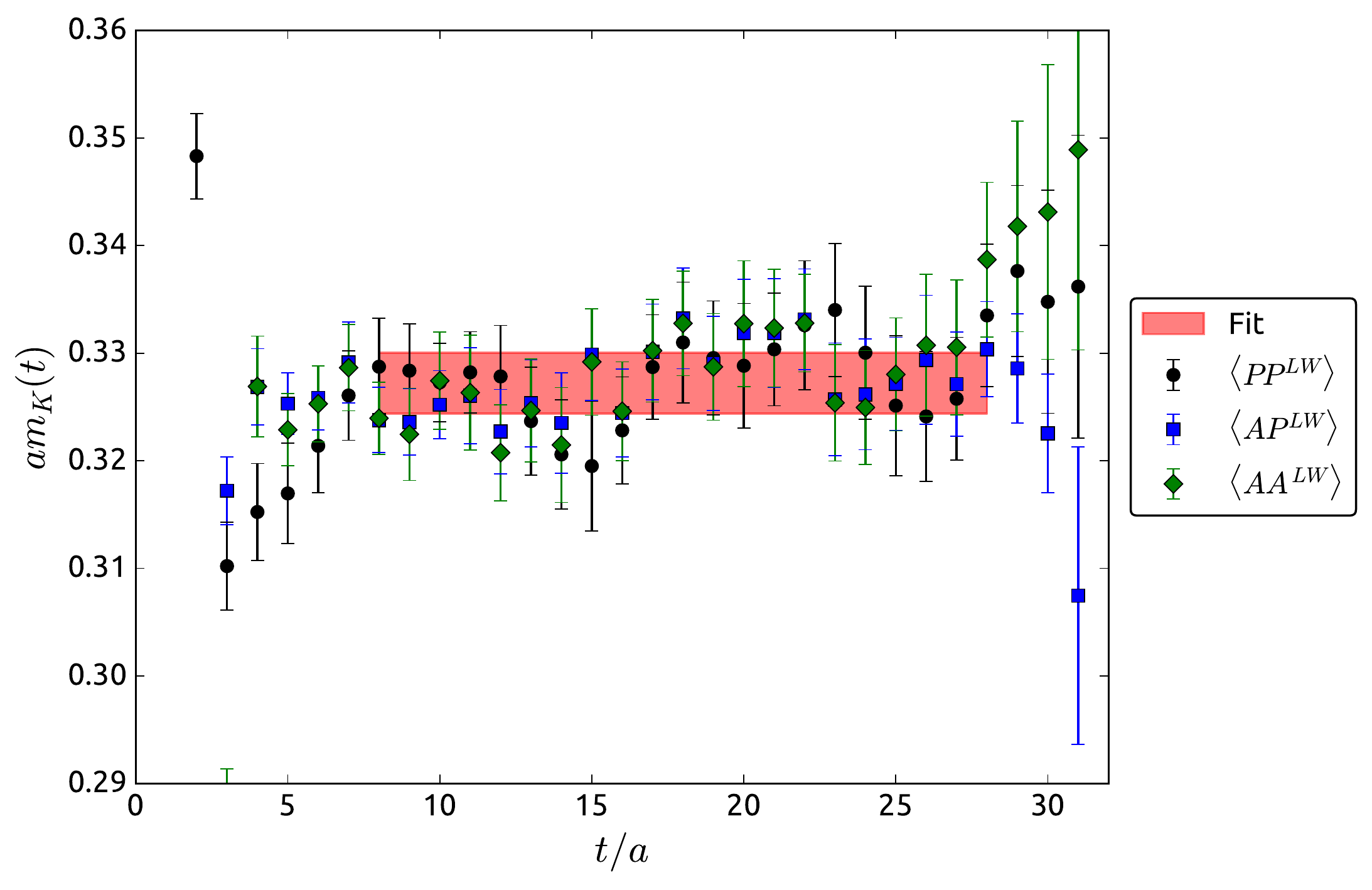}} \quad
\subfloat{\includegraphics[width=0.48\linewidth]{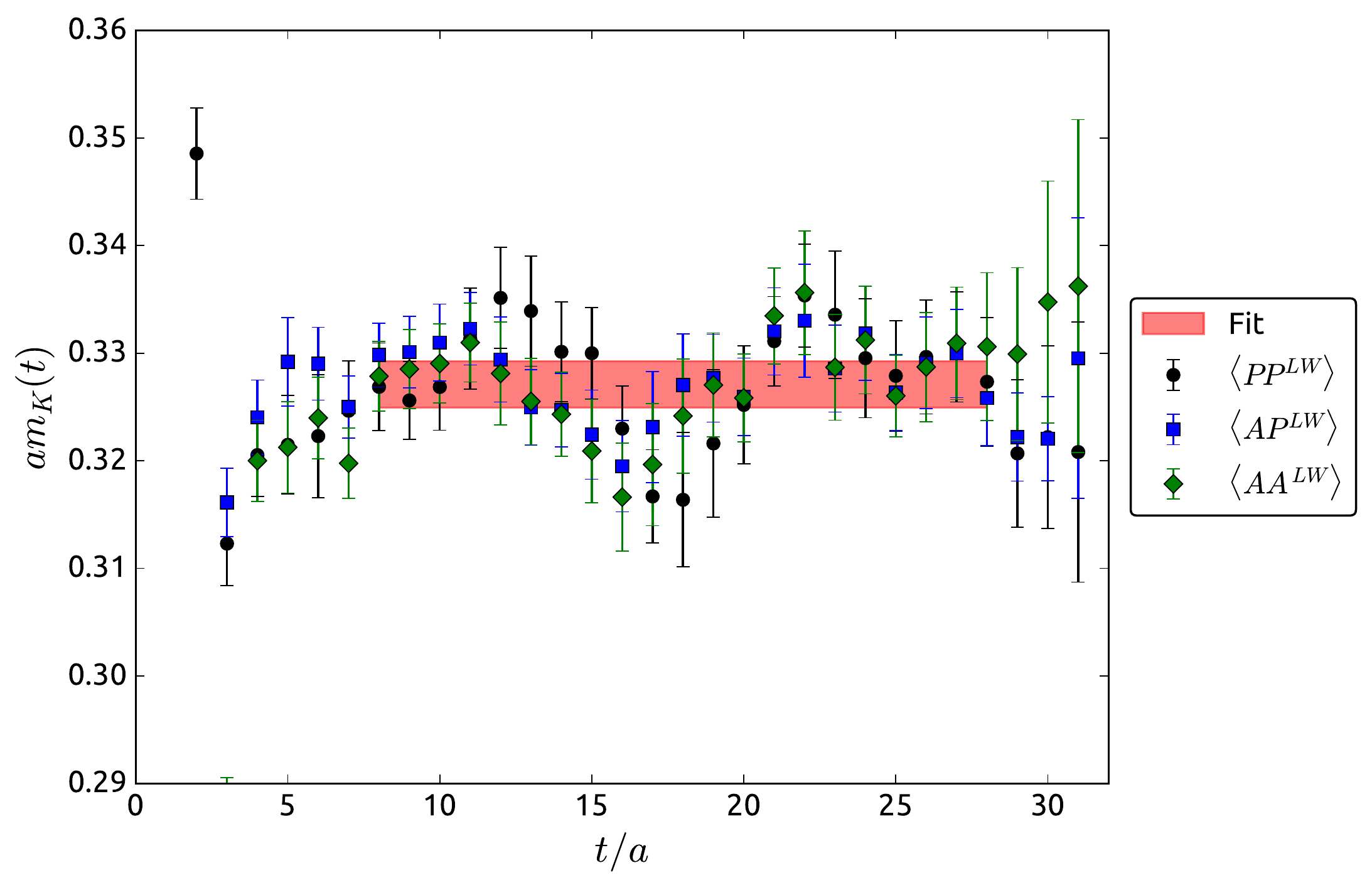}} \\
\subfloat{\includegraphics[width=0.48\linewidth]{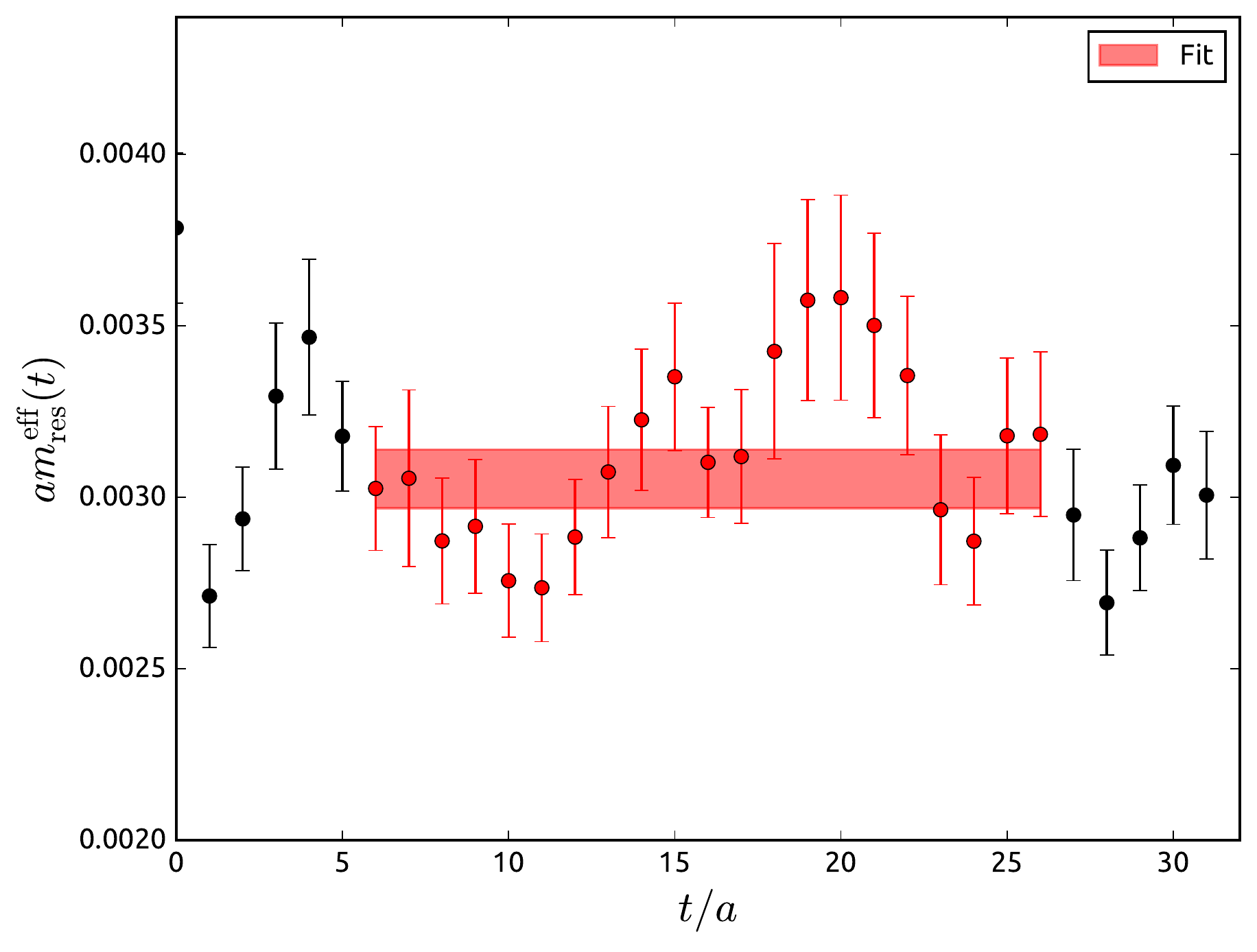}} \quad
\subfloat{\includegraphics[width=0.48\linewidth]{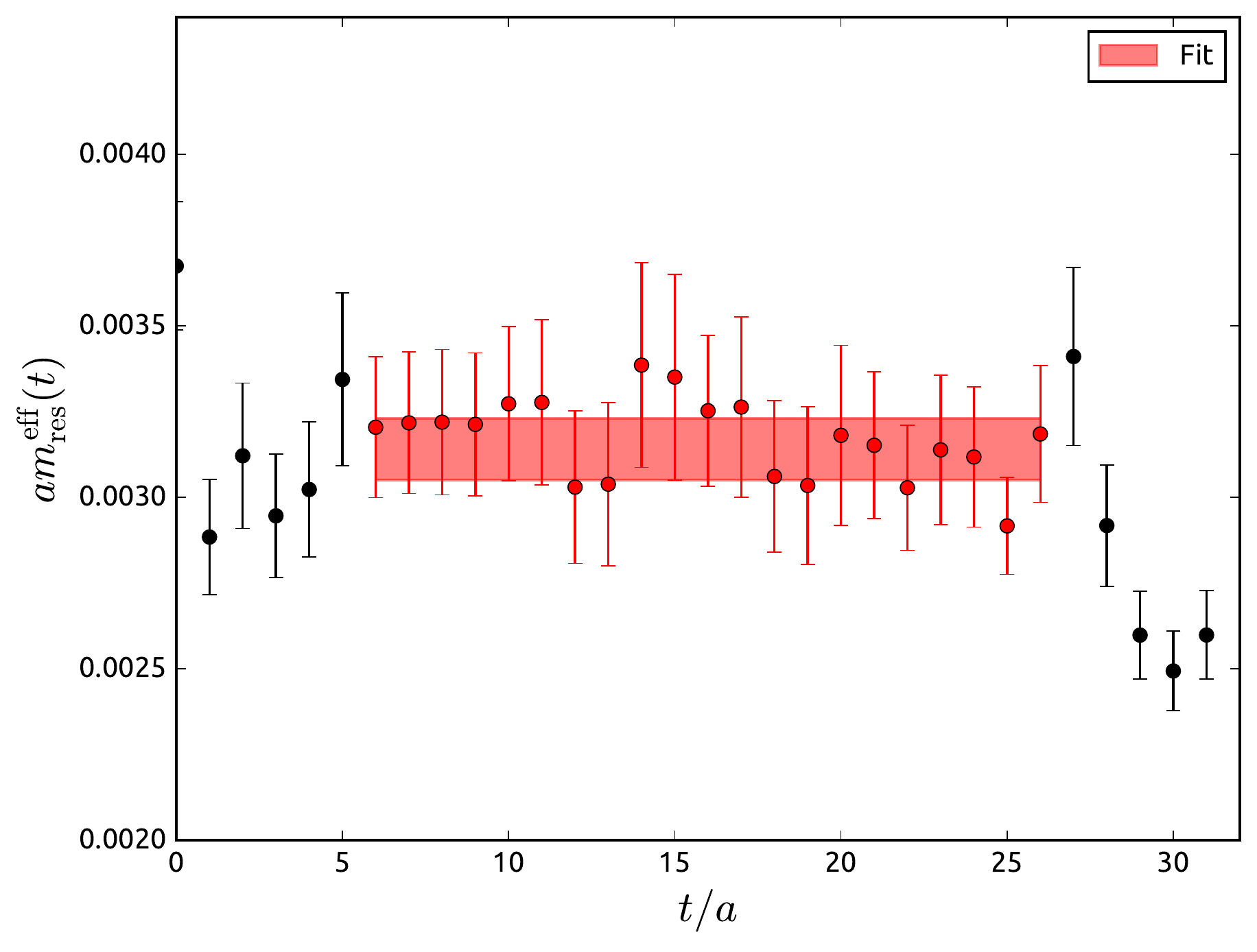}} \\
\caption{Effective ground state pion energy (top), kaon mass (middle), and $a m_{\rm res}'(m_{l})$ evaluated at the bare light quark mass, as measured on the EOFA (left) and RHMC (right) 16I-G ensembles.} 
\end{figure}

\begin{figure}[!h]
\centering
\subfloat{\includegraphics[width=0.48\linewidth]{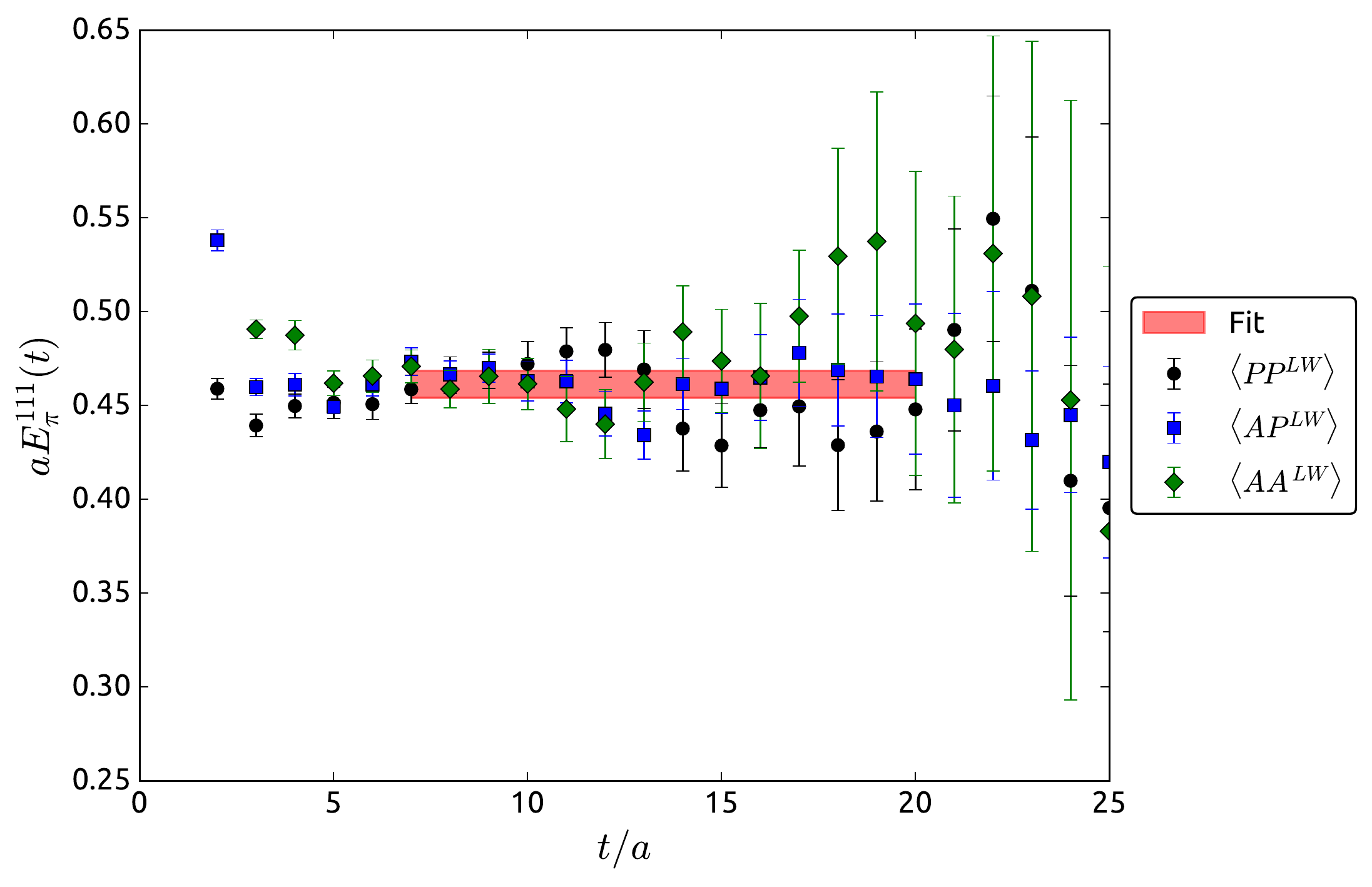}} \quad
\subfloat{\includegraphics[width=0.48\linewidth]{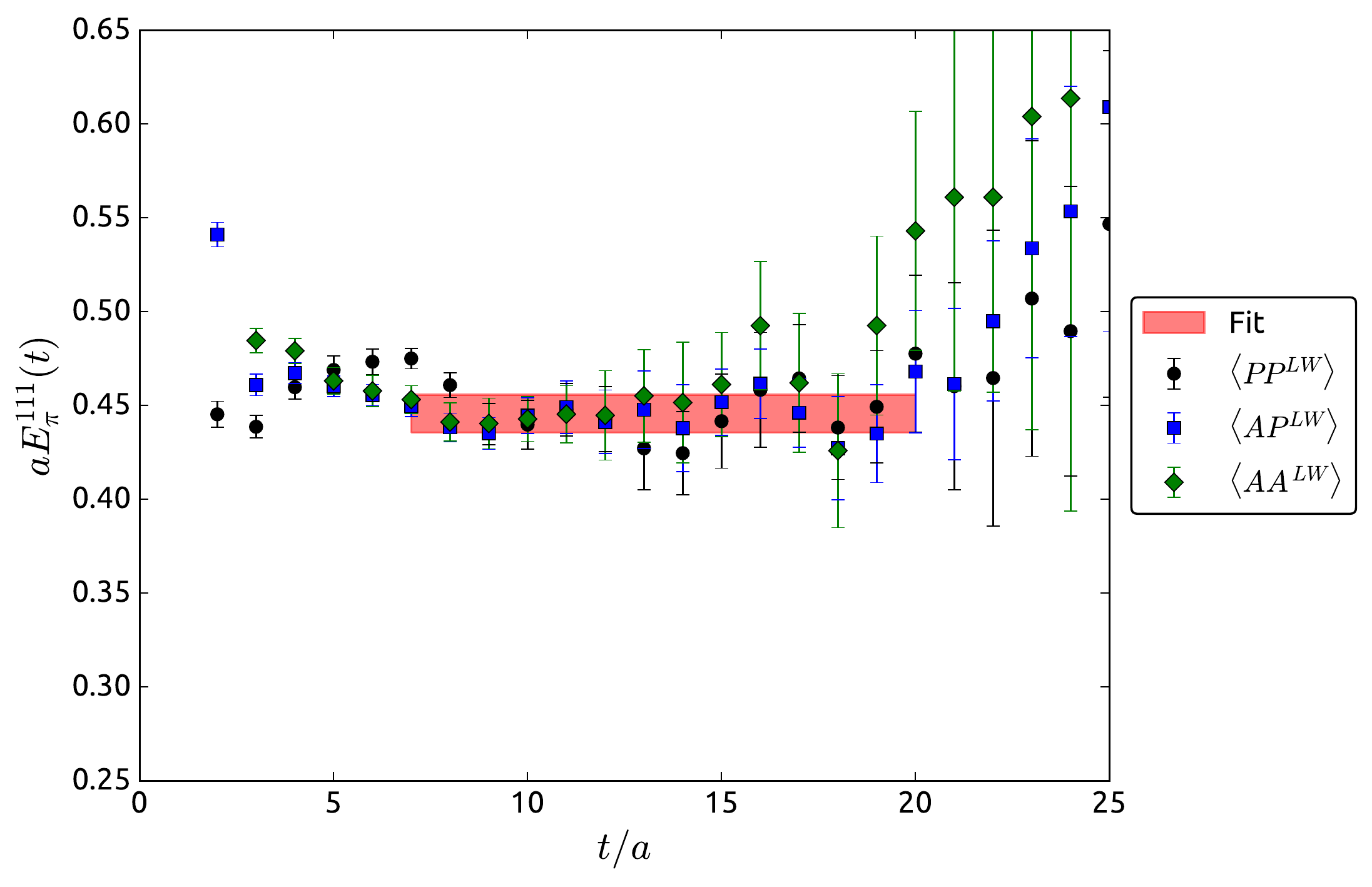}} \\
\subfloat{\includegraphics[width=0.48\linewidth]{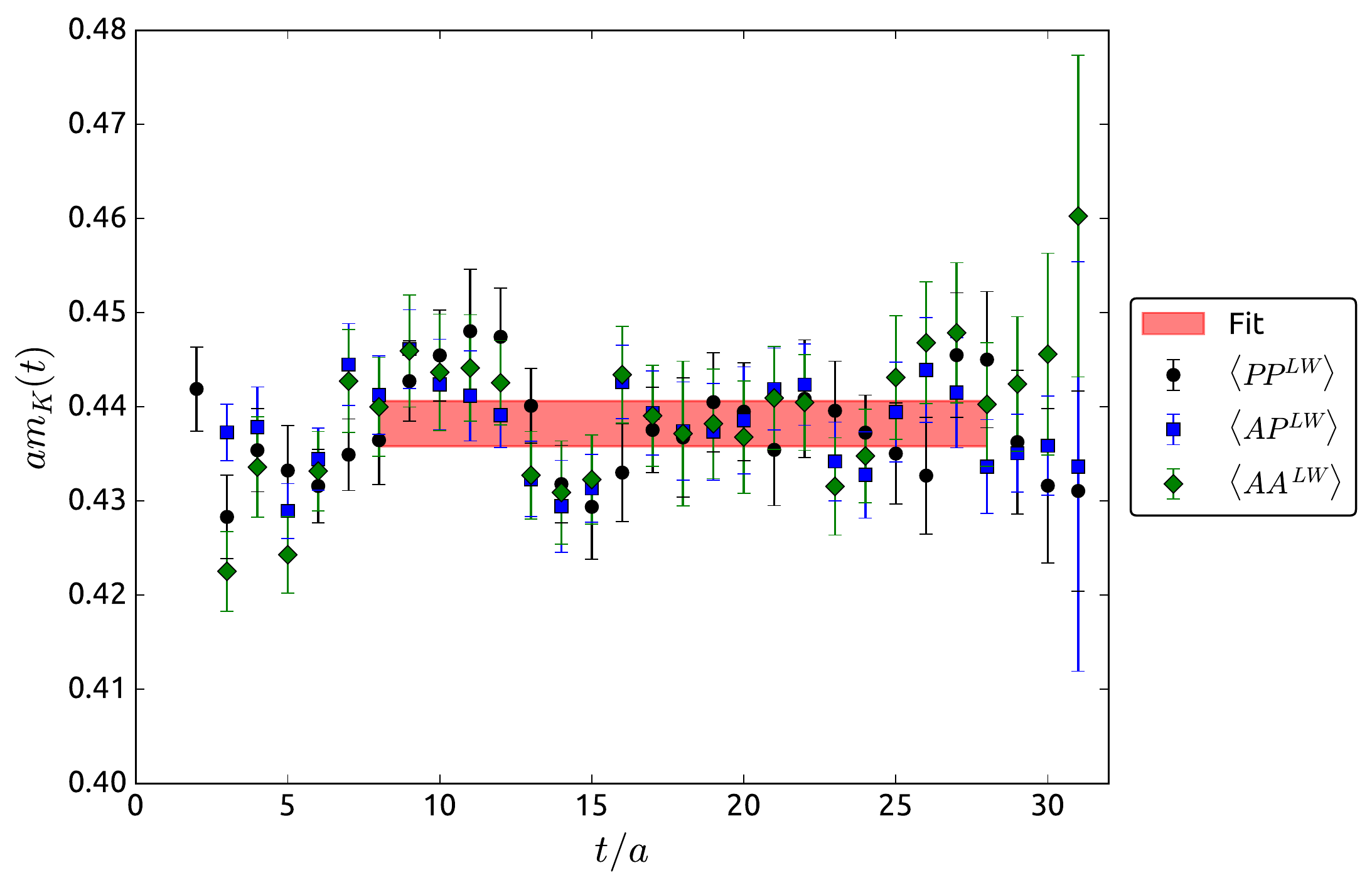}} \quad
\subfloat{\includegraphics[width=0.48\linewidth]{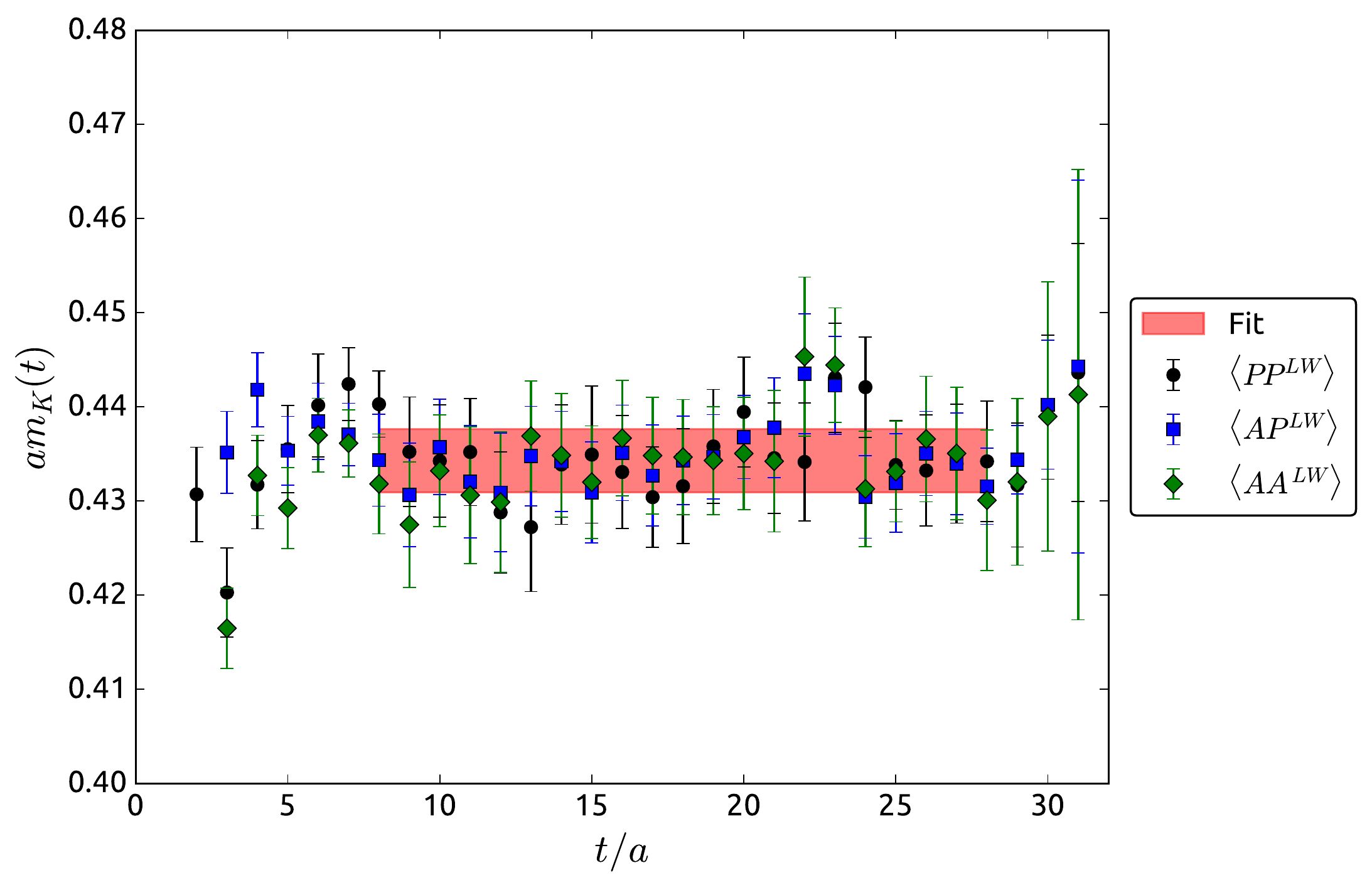}} \\
\subfloat{\includegraphics[width=0.48\linewidth]{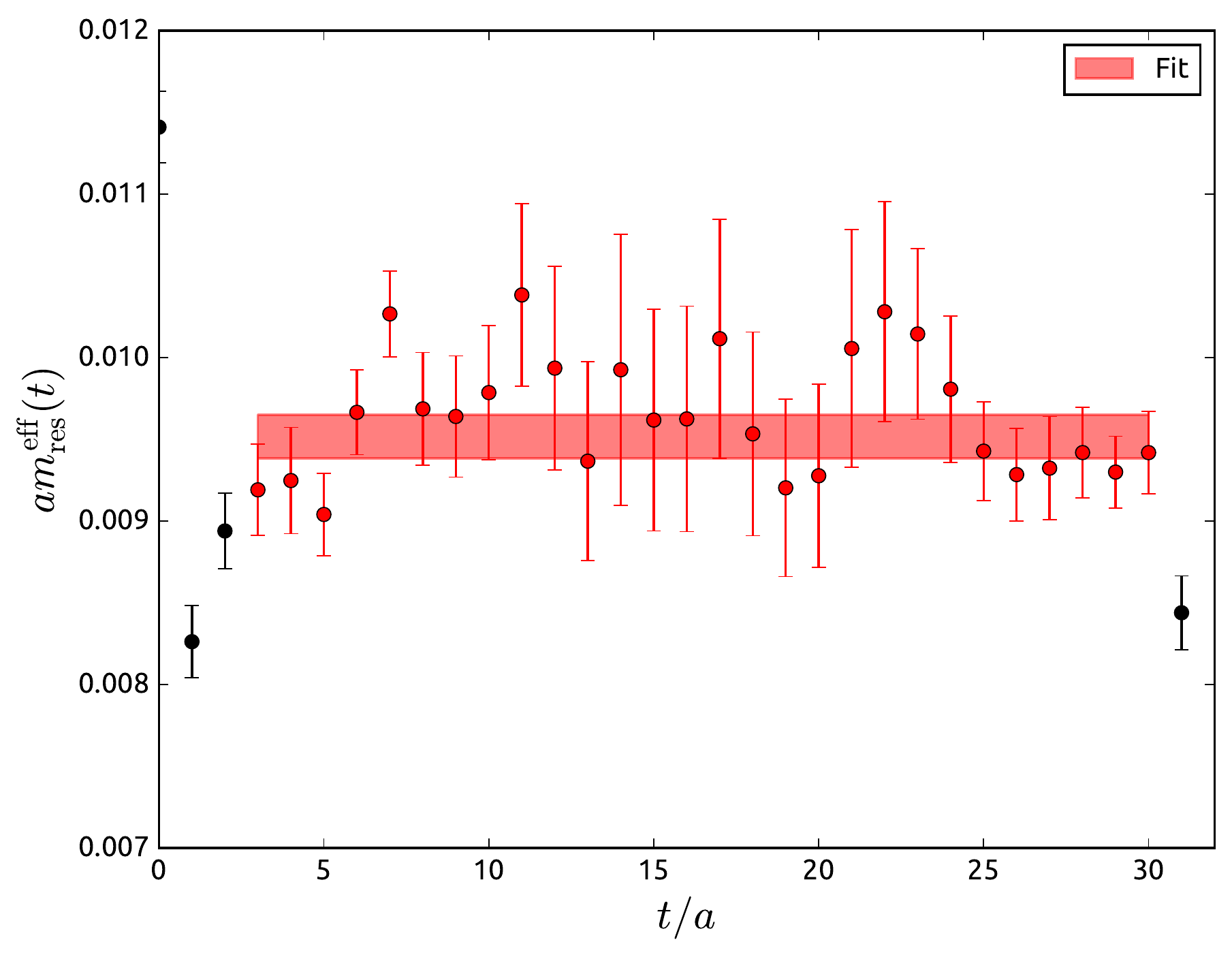}} \quad
\subfloat{\includegraphics[width=0.48\linewidth]{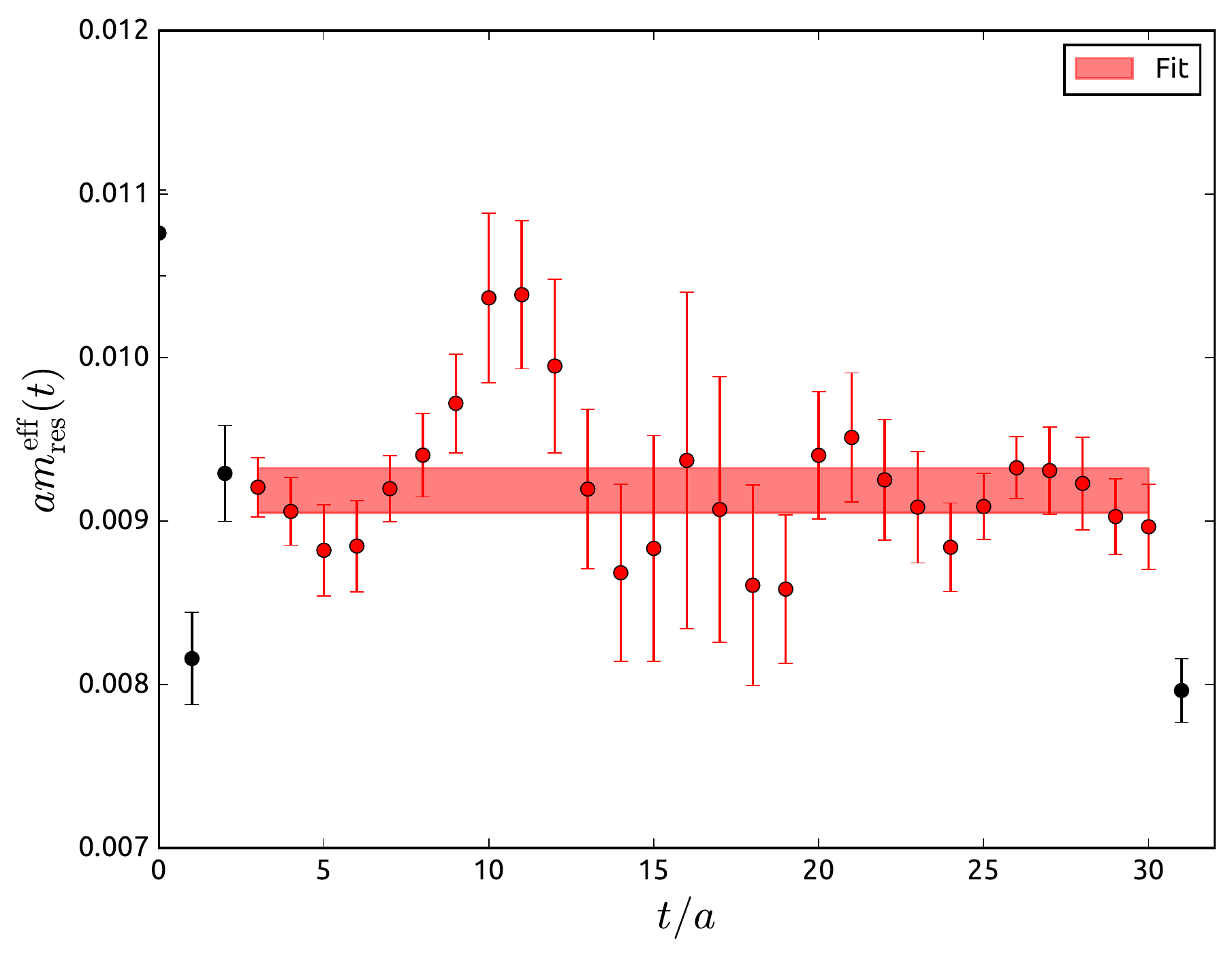}} \\
\caption{Effective ground state pion energy (top), kaon mass (middle), and $a m_{\rm res}$ evaluated at the bare light quark mass, as measured on the EOFA (left) and RHMC (right) 16ID-G ensembles.}
\end{figure}
\FloatBarrier
\FloatBarrier
	
\end{document}